\documentclass[12pt]{article}%
\usepackage[nosort]{cite}
\usepackage[usenames, dvipsnames]{xcolor}
\usepackage{graphicx}
\usepackage{multicol}
\usepackage{amsfonts}
\usepackage{amssymb}
\usepackage{amsmath}
\usepackage{heck}
\usepackage{afterpage}
\usepackage{setspace}
\usepackage{verbatim}
\usepackage{color}
\usepackage{longtable}
\usepackage{float}
\usepackage{subcaption}
\usepackage{epsfig}
\usepackage{enumerate}
\usepackage{epstopdf}
\usepackage[enableskew, vcentermath]{youngtab}
\usepackage{adjustbox}
\usepackage{multirow}
\usepackage{tikz}
\usepackage[margin=1in]{geometry}
\usepackage{titletoc}
\usepackage{hyperref}
\usepackage[percent]{overpic}
\usepackage{gensymb}
\usepackage{mathtools}%
\usepackage{tikz-cd}
\providecommand{\U}[1]{\protect\rule{.1in}{.1in}}
\pdfoutput=1
\newsavebox{\mysavebox}

\hypersetup{colorlinks,citecolor=black,filecolor=black,linkcolor=black,urlcolor=black}
\usetikzlibrary{decorations.markings}

\numberwithin{equation}{section}

\usetikzlibrary{chains}
\allowdisplaybreaks
\tikzset{node distance=2em, ch/.style={circle,draw,on chain,inner sep=2pt},chj/.style={ch,join},every path/.style={shorten >=4pt,shorten <=4pt},line width=1pt,baseline=-1ex}

\newcommand{\ba}{\begin{eqnarray}}
\newcommand{\ea}{\end{eqnarray}}

\newcommand{\be}{\begin{equation}}
\newcommand{\ee}{\end{equation}}

\tikzstyle{startstop} = [rectangle, rounded corners, minimum width=3cm, minimum height=1cm,text centered, draw=black, fill=blue!10]
\tikzstyle{startstop} = [rectangle, rounded corners, minimum width=3cm, minimum height=1cm,text centered, draw=black, fill=blue!10]
\tikzstyle{io} = [trapezium, trapezium left angle=70, trapezium right angle=110, minimum width=3cm, minimum height=1cm, text centered, draw=black, fill=blue!30]
\tikzstyle{process} = [rectangle, minimum width=3cm, minimum height=1cm, text centered, draw=black, fill=orange!30]
\tikzstyle{decision} = [diamond, minimum width=3cm, minimum height=1cm, text centered, draw=black, fill=green!30]
\tikzstyle{arrow} = [thick,->,>=stealth]
\tikzset{->-/.style={decoration={
  markings,
  mark=at position #1 with {\arrow[scale=2.4]{>}}},postaction={decorate}}}
\makeatletter \@addtoreset{equation}{section} \makeatother

\begin{document}

\date{May 2020}

\title{Geometric Approach to\\[4mm] 3D Interfaces at Strong Coupling}

\institution{PENN}{\centerline{Department of Physics and Astronomy, University of Pennsylvania, Philadelphia, PA 19104, USA}}

\authors{Markus Dierigl\footnote{e-mail: {\tt markusd@sas.upenn.edu}}, Jonathan J.\ Heckman\footnote{e-mail: {\tt jheckman@sas.upenn.edu}},\\[4mm]
Thomas B.\ Rochais\footnote{e-mail: {\tt thb@sas.upenn.edu}}, and Ethan Torres\footnote{e-mail: {\tt emtorres@sas.upenn.edu}}}

\abstract{We study 4D systems in which parameters of the
theory have position dependence in one spatial direction. In the limit where these parameters jump,
this can lead to 3D interfaces supporting localized degrees of freedom. A priori, this sort of position dependence
can occur at either weak or strong coupling. Demanding time-reversal invariance for $U(1)$ gauge theories
with a duality group $\Gamma \subset SL(2,\mathbb{Z})$ leads to interfaces at strong coupling
which are characterized by the real component of a modular curve specified by $\Gamma$. This provides a geometric
method for extracting the electric and magnetic charges of possible localized states. We illustrate these general
considerations by analyzing some 4D $\mathcal{N} = 2$ theories with 3D interfaces. These 4D systems can also be interpreted as
descending from a six-dimensional theory compactified on a three-manifold generated by a family of Riemann surfaces fibered over the real line.
We show more generally that 6D superconformal field theories compactified on such spaces also produce trapped matter by using the known
structure of anomalies in the resulting 4D bulk theories.}

\maketitle

\setcounter{tocdepth}{2}

\tableofcontents

\enlargethispage{\baselineskip}

\newpage

\section{Introduction} \label{sec:INTRO}

Insights from geometry and topology provide a non-trivial handle on many quantum systems,
even at strong coupling. In the context of high energy theory, this has typically
been applied in systems with supersymmetry. More generally,
however, one can hope that constraints on the topological structure of quantum fields
are enough to deduce many features of physics at long distance scales.

Indeed, there has recently been some progress in understanding 
some quantum field theories using constraints on the topological structure of
such systems. An example of this sort involves the effective field theory
associated with topological insulators \cite{Kane:2004bvs, Kane:2005zz, bernevig2006quantum, Moore:2006pjk, 2007Sci...318..766K, Fu:2007uya, PhysRevB.79.195322, Hasan:2010xy, 2011RvMP...83.1057Q, Hasan:2010hm, Ye:2017axd} in $3+1$ dimensions, which is one special type of symmetry-protected topological (SPT) phase of matter \cite{2010PhRvB..81f4439P, 2011PhRvB..83g5103F, 2011PhRvB..83g5102T, 2011PhRvB..83c5107C, 2011PhRvB..84p5139S, 2013PhRvB..87o5114C, 2008PhRvB..78s5424Q, 2008arXiv0810.2998E} with highly interesting surface behavior \cite{Vishwanath:2012tq, Chen:2013jha, Bonderson:2013pla, Wang:2013uky, 2014Sci...343..629W, Mross:2014gla, Wang:2014lca, Metlitski:2014xqa, Metlitski:2015bpa, Metlitski:2015eka, Fialkovsky:2019rum, Kurkov:2020jet}.
This phenomenon can be modeled in terms of the effective field theory
of a background $U(1)$ gauge theory with a position dependent $\theta$ angle \cite{Wilczek:1987mv, Qi:2008pi}.
Both $\theta = 0$ and $\theta = \pi$ preserve time-reversal symmetry, and demanding the system remain time-reversal invariant throughout
means that an interface between $\theta = 0$ and $\theta = \pi$ has trapped modes \cite{Jackiw:1975fn}. Indeed, this can
be explicitly verified by considering a 4D Dirac fermion with a  mass $m(x_\bot)$ which depends on a spatial direction of the
4D spacetime. A sign flip in $m$ leads to a trapped mode. There have been a
number of developments aimed at extending this analysis in various directions,
including new examples of dualities at weak coupling \cite{Seiberg:2016rsg}, as
well as possible strongly coupled phases for trapped edge modes \cite{Seiberg:2016gmd} and related dualities, see e.g.\ \cite{Hsin:2016blu, Karch:2016aux, Cordova:2017kue, Aitken:2017nfd, Gaiotto:2017tne, Benini:2018umh, Carl:2019bbf}.

In this paper we study a similar class of questions but in which we allow the system
to approach a regime of ``strong coupling in the bulk.'' This also means that we allow the $U(1)$ to be dynamical, but we will
assume that degrees of freedom charged under it are still quite heavy. We can, of course, still require that far away from the interface we are
at very weak coupling, but even this assumption can in principle be relaxed (though that would of course be more difficult to realize experimentally but might be relevant for materials that have magnetic excitations such as pyrochlores \cite{2009NatPh...5..258J}). Our aim will be to develop methods which apply in such situations as well.

The main theme running through our analysis will be to use methods from geometry to better understand the possible behavior of
localized modes. While much of our inspiration comes from the analysis of supersymmetric gauge theories in which these geometric structures descend from the extra-dimensional world of supersymmetric string compactifications, some aspects of our analysis
do not actually require the full machinery of these constructions. That being said, we will find it worthwhile to consider both low energy effective field theories in four dimensions, as well as compactification of six-dimensional superconformal field theories
as realized by string compactifications.

The first class of interfaces we study involve 4D $U(1)$ gauge theory with a complexified combination
of the gauge coupling $g$ and the theta angle:
\begin{equation}
\tau = \frac{4 \pi i}{g^2} + \frac{\theta}{2 \pi}.
\end{equation}
The main assumption we make is that our theory has a non-trivial set of duality transformations which act on this coupling as:
\begin{equation}
\tau \mapsto \frac{a \tau + b}{c \tau + d},
\end{equation}
for some $a,b,c,d$ integers such that $ad - bc = 1$. The most well-known case is that we just have a duality group $SL(2,\mathbb{Z}$) consisting
of all determinant one $2 \times 2$ matrices with integer entries, as associated with the famous electric-magnetic duality of Maxwell theory.
In systems with additional massive degrees of freedom, these duality groups can be smaller. Assuming this structure in the deep IR, we will be interested in the behavior of the 4D theory when $\tau(x_\bot)$ depends non-trivially on one of the spatial directions of the 4D spacetime.

In the case where the theory has an $SL(2,\mathbb{Z})$ duality group, there is a well-known correspondence between an equivalence class of $\tau$ and the geometry of a $T^2$ with complex structure $\tau$. One can think of this $T^2$ as the quotient $\mathbb{C} / \Lambda$ with $\Lambda = \omega^1 \mathbb{Z} \oplus \omega^2 \mathbb{Z}$ a two-dimensional lattice. In this case, the ratio $\omega^1 / \omega^2 = \tau$ dictates the ``shape'' of the $T^2$. In physical terms, $\Lambda$ is the lattice of electric and magnetic charges in the theory. Geometrically, we can replace $\tau(x_\bot)$ by a family of $T^2$'s which vary over a real line, building up a three-manifold with a boundary at $x_{\bot} \rightarrow \pm \infty$. Since there is a fixed choice of $T^2$ at both ends of the line, this $T^2$ comes with a distinguished marked point, and thus defines a one-dimensional family of elliptic curves.\footnote{An elliptic curve is a genus one curve with a marked point.}

We will be interested in a restricted class of 4D systems which enjoy time-reversal invariance in the bulk. This corresponds to a further condition of invariance of the physical theory under the mapping:
\begin{equation}
\tau \mapsto - \overline{\tau}.
\end{equation}
Geometrically, this corresponds to a further condition that the $j$-function of the elliptic curve is in fact a real number: $j \in \mathbb{R}$. This region splits into the familiar ``trivial phase'' with $\theta = 0$, the standard ``topological insulator phase'' with $\theta = \pi$ phase, and another ``strongly coupled phase'' in which $\vert \tau \vert = 1$. All other time-reversal invariant values of $\tau$ can be related to one of these three regions by an $SL(2,\mathbb{Z})$ transformation. As a point of nomenclature, we
note that this is somewhat of an abuse of terminology since in the topological insulator literature one views the $U(1)$ of the topological insulator as a global symmetry which is not broken (indeed it defines an SPT phase), and in which all excitations are gapped out. Part of the point of our analysis is to explore the effects of varying the gauge coupling as well as the theta angle. Hopefully the distinction will not be too distracting.

Viewed as a trajectory on the moduli space of elliptic curves, we thus see that an interface could a priori take two different routes between $\theta = 0$ and $\theta = \pi$. On the one hand, it could always remain at weak coupling. On the other hand, it could pass through a strongly coupled region. Asymptotically far away from the interface, both are a priori possible, but suggest very different possibilities for localized modes.  Singularities in this family of elliptic curves corresponds to the appearance of massless states. Since we are not assuming any supersymmetry, our knowledge of these states is somewhat limited, but we can, for example, deduce the electric and magnetic charge of states localized on the interface.

It can also happen that the duality group $\Gamma \subset SL(2,\mathbb{Z})$ is strictly smaller than that of the Maxwell theory. In this case, there are more possible phases, since the coset space $SL(2,\mathbb{Z}) / \Gamma$ is now non-trivial. Consequently, some values of $\tau$ related by an $SL(2,\mathbb{Z})$ duality transformation may now define different physical theories. The resulting moduli space of elliptic curves are specified by modular curves $X(\Gamma)$, and the geometry of these curves can be quite intricate. For our present purposes, we are interested in the subset of parameters which are time-reversal invariant. Thankfully, precisely this question has been studied in reference \cite{snowden2011real} which analyzes the real components of the modular curve, $X(\Gamma)_{\mathbb{R}}$. The key point for us is that $X(\Gamma)_{\mathbb{R}}$ consists of a collection of disjoint $S^1$'s. Each such $S^1$ itself breaks up into paths joined between ``cusps'' of the modular curve. These cusps are associated with the additional $SL(2,\mathbb{Z})$ images of the weak coupling point $\tau = i \infty$ which cannot be brought back to weak coupling via transformations in $\Gamma \subset SL(2,\mathbb{Z})$. Passing through such cusps is inevitable, and means that singularities in the family of elliptic curves are also dictated purely by topological considerations. For each such cusp, we can fix the associated electric and magnetic charge, thus indicating the corresponding charge of states localized on an interface.

We illustrate these general considerations with some concrete examples. As a first class, we consider some examples of 4D $\mathcal{N} = 2$ field theories in which the Seiberg-Witten curve has the topology of a $T^2$. As a second set of examples, we consider the compactification of a six-dimensional anti-chiral two-form on a family of elliptic curves. In this situation, we also present a general construction for realizing 4D $U(1)$ gauge theories with duality group given by the congruence subgroups $\Gamma_{0}(N), \Gamma_{1}(N),$ and $\Gamma(N)$.

As we have already mentioned, 3D interfaces appear in this geometric setting when the elliptic curve becomes singular. This raises the question
as to whether more singular transitions such as a change from a genus zero to a genus one curve could arise, and if so, what this would mean in
terms of the 4D effective field theory. Along these lines, we also consider a more general way to construct 3D interfaces from compactifying six-dimensional superconformal field theories on a three-manifold with boundaries. In this setting, we present explicit examples where the genus jumps as a function of $x_{\bot}$. By tracking the anomaly polynomial of the 4D theory before and after the jump, we deduce that the degrees of freedom on the two sides of a wall can be different. Such changes can be used to engineer more general examples of localized matter with a ``thickened interface.''

The rest of this paper is organized as follows. We begin in section \ref{sec:DUALITY} with a geometric characterization of
3D interfaces of a $U(1)$ gauge theory
with duality group $SL(2,\mathbb{Z})$. In section \ref{sec:MOREDUAL} we generalize this to cases where the duality group is $\Gamma \subset SL(2,\mathbb{Z})$ a proper subgroup. Section \ref{sec:NTWO} presents some explicit constructions based on 4D $\mathcal{N} = 2$ theories, and
section \ref{sec:6DCOMPACTIFY} presents examples based on compactification of the theory of a six-dimensional anti-chiral two-form.
We generalize these constructions in section \ref{sec:6DGENERAL} by considering compactifications of six-dimensional superconformal field theories on three-manifolds with boundary. We conclude in section \ref{sec:CONC}. Some additional details and examples are presented in the
Appendices.

\textbf{Note added 10/19/2020:} After our work appeared, a specific proposal for realizing QED-like systems at strong coupling was discussed in \cite{Pace:2020jiv} in the specific context of spin ice systems. This would provide an ideal setting for implementing a further study of the strong coupling phenomena indicated in this paper.

\section{Time-Reversal Invariance and Duality} \label{sec:DUALITY}

In this section we review some elements of the ``standard'' case of a 4D $U(1)$ gauge theory which
has an interface between two time-reversal invariant phases with $\theta = 0$ and $\theta = \pi$.
We will be interested in developing a geometric characterization of this sort of system with an eye
towards generalizing to strongly coupled examples.

Throughout this paper we will also confine our discussion to 4D theories on flat space $\mathbb{R}^{2,1} \times \mathbb{R}_{\bot}$.\footnote{Additionally, we will
ignore possible mixed gravitational/duality group anomalies which can appear on
some curved backgrounds \cite{Tachikawa:2017aux, Seiberg:2018ntt, Cordova:2019uob, Hsieh:2019iba, Cordova:2019jnf} as well as subtleties involving the spin-structure \cite{Rosenberg:2010ia, Metlitski:2013uqa, Metlitski:2015yqa}.
It would be interesting to extend the present considerations to these situations.}
We will, however, allow the coupling constants to depend on $x_{\bot}$, the local coordinate of $\mathbb{R}_{\bot}$.

The rest of this section is organized as follows. First, we introduce our conventions for time-reversal invariance, as well
$SL(2,\mathbb{Z})$ duality transformations. Using this, we identify different phases of parameter space which are time-reversal
invariant. Next, we study position dependent couplings which can generate an interface between these different phases.

\subsection{$U(1)$ Gauge Theory Revisited}

Consider an abelian gauge theory, with a possible coupling to some matter fields.
The corresponding Lagrangian density contains the terms:
\begin{equation}
\mathcal{L} = - \frac{1}{4 g^2} F_{\mu \nu} F^{\mu \nu} + \frac{\theta}{32 \pi^2} F_{\mu \nu} \widetilde{F}^{\mu \nu}  + \cdots \,,
\end{equation}
where the ``$\cdots$'' refers to contributions from all other matter fields. In terms of the electric and magnetic fields $\vec{E}$ and $\vec{B}$, we can also write this as:
\begin{equation}
\mathcal{L} = \frac{1}{2g^2} (\vec{E} \cdot \vec{E} - \vec{B} \cdot \vec{B}) - \frac{\theta}{8 \pi^2} \vec{E} \cdot \vec{B} + \cdots\,.
\end{equation}
It will be convenient to introduce the complexified coupling:
\begin{equation}
\tau = \frac{4 \pi i}{g^2} + \frac{\theta}{2 \pi}.
\end{equation}

Time reversal acts on the electric and magnetic fields as:
\begin{align}
\mathcal{T}: \quad \vec{E} \mapsto \vec{E} \,, \enspace \vec{B} \mapsto - \vec{B} \,.
\label{eq:tactemfields}
\end{align}
In terms of the original basis of fields, this has the effect of taking us to a new theory with the same gauge coupling, but with
$\theta_{\mathrm{new}} = -\theta_{\mathrm{old}}$. We can phrase this as a new choice of complexified gauge coupling:
\begin{equation}
\tau_{\mathrm{new}} = -\overline{\tau}_\mathrm{old}.
\end{equation}

We will be interested in values of the complexified coupling which can be identified with the old one via a duality transformation.
This takes us to a new basis of fields as well as dualized value of the coupling.
The most well-known situation is that our abelian gauge theory has an $SL(2,\mathbb{Z})$ duality group, which is the case for free Maxwell theory but also more interesting setups. We will shortly generalize this discussion to other duality groups.
Recall that the group $SL(2,\mathbb{Z})$ is defined as:
\begin{align}\label{generators}
SL(2,\mathbb{Z}) = \left\{ \begin{pmatrix} a & b \\ c & d \end{pmatrix}: a,b,c,d \in \mathbb{Z} \,, \enspace ad - bc = 1 \right\} \,.
\end{align}
Such duality transformations takes us to a new basis of electric and magnetic fields. Given a
state of electric charge $q_e$ and magnetic charge $q_m$, we introduce a two-component column vector which
transforms according to the rule:
\begin{equation}
\begin{pmatrix} q_e \\ q_m \end{pmatrix} \mapsto \begin{pmatrix} a & b \\ c & d \end{pmatrix} \begin{pmatrix} q_e \\ q_m \end{pmatrix} \,.
\end{equation}
For typographical purposes we shall also sometimes refer to this as a state having charge $(q_e,q_m)$, but we
stress that in our conventions this is to be viewed as a column vector, and not a row vector.
The Dirac pairing between two such charge vectors $\vec{q} \equiv q^{a}$ and $\vec{q^{\prime}} \equiv q^{\prime b}$ is:
\begin{equation}
\langle \vec{q} , \vec{q}^{\prime} \rangle = \epsilon_{ab} q^{a} q^{\prime b} = q_e q_m^{\prime} - q_{m} q_{e}^{\prime}.
\end{equation}
We can view a dyonic charge $(q_e , q_m)$ as coupling to a vector potential $A$ and its magnetic dual $A_D$ via the $SL(2,\mathbb{Z})$ invariant combination:
\begin{equation}
\epsilon_{ab} q^{a} \mathcal{A}^{b} = q_e A - q_m A_D,
\end{equation}
where we introduced the two-component vector $\mathcal{A}^{a}$ with
entries $\mathcal{A}^{1} = A_D$ and $\mathcal{A}^{2} = A$.
Under such a duality transformation, the complexified coupling also changes as:
\begin{equation}
\tau \mapsto \frac{a \tau + b}{c \tau + d}.
\end{equation}
Geometrically, the lattice of electric and magnetic charges can be written as:
\begin{equation}
\Lambda_{\tau} = \omega^1 \mathbb{Z} \oplus \omega^2 \mathbb{Z} = \tau \mathbb{Z} \oplus \mathbb{Z} \,
\end{equation}
where we can also view $\omega^{a}$ as a two-component column vector and the complex structure as $\tau = \omega^1 / \omega^2$.
Quotienting the complex plane $\mathbb{C}$ by this lattice results in an elliptic curve $E(\tau) = \mathbb{C} / \Lambda_{\tau}$. A pleasant feature of working with the elliptic curve is that $SL(2,\mathbb{Z})$ transformations leave the complex structure of the curve intact. This provides a geometric way to parameterize physically inequivalent $\tau$'s.

The group $SL(2,\mathbb{Z})$ is generated by the $T$ and $S$ transformations:
\begin{equation}
\begin{split}
T = \begin{pmatrix} 1 & 1 \\ 0 & 1 \end{pmatrix}:& \quad \tau \rightarrow \tau + 1 \,, \enspace \theta \rightarrow \theta + 2 \pi \,, \\
S = \begin{pmatrix} 0 & -1 \\ 1 & 0 \end{pmatrix}:& \quad \tau \rightarrow - \frac{1}{\tau} = - \frac{\overline{\tau}}{|\tau|^2} \,, \\
& \quad g^2 \rightarrow \Big( \Big( \frac{4 \pi}{g^2} \Big)^2 + \Big( \frac{\theta}{2 \pi} \Big)^2 \Big) g^2 \,, \enspace \theta \rightarrow - \Big( \Big( \frac{4 \pi}{g^2} \Big)^2 + \Big( \frac{\theta}{2 \pi} \Big)^2 \Big)^{-1} \theta \,
\end{split}
\end{equation}
observe that $\theta = -\pi$ can be mapped back to $\theta = +\pi$ under such a transformation. A priori, this gauge theory could be at strong or weak coupling, and have complicated interactions with other matter fields.

Assuming our theory enjoys an $SL(2,\mathbb{Z})$ duality group action, we need not work with the full set of
values of $\tau$, just the ones which are not identified by an $SL(2,\mathbb{Z})$ transformation. Implicit in this parameterization is that
when we label a theory, we allow ourselves to change to a dualized basis of fields.
Unitarity demands $\mathrm{Im} \tau  > 0$, so $\tau$ takes values in the upper half-plane $\mathbb{H}$. The quotient by $SL(2,\mathbb{Z})$
is known as the fundamental domain of $SL(2,\mathbb{Z})$, and we denote it as $Y = \mathbb{H} / SL(2,\mathbb{Z})$. Since we will also be interested in the very weakly coupled limit, we add on the ``point at infinity'' $\tau = i \infty$ as well as all of its $SL(2,\mathbb{Z})$ images (which are just rational numbers $a / c$ in the matrix presentation of line (\ref{generators})). Introducing the compactified upper half-plane:
\begin{equation}
\overline{\mathbb{H}} \equiv \mathbb{H} \cup \{ i \infty \} \cup \mathbb{Q},
\end{equation}
we can again consider the quotient space from an $SL(2,\mathbb{Z})$ action. This produces the compactified fundamental domain  which we denote as $X(\Gamma)$ with $\Gamma = SL(2,\mathbb{Z})$.

We will be interested in the space of couplings modulo such duality transformations. With this in mind, it is convenient to introduce an $SL(2,\mathbb{Z})$ invariant coordinate on the fundamental domain. This is simply the ``$j$-function'' of the parameter $\tau$.
The $j$-function is a modular form with $q$-expansion:
\begin{equation}
j = \frac{1}{q} + 744 + \cdots
\end{equation}
where $q = \exp(2 \pi i \tau)$. The $j$-function maps the fundamental domain $\overline{\mathbb{H}} / SL(2,\mathbb{Z})$ to the complex projective space $\mathbb{CP}^1$ with three distinguished points. This is the modular curve of the group $SL(2, \mathbb{Z})$. The three distinguished points are located at $\tau = i \infty, i, e^{2 \pi i/6}$, which are mapped to the points
\begin{align}
j (\tau) \underset{\tau \rightarrow i \infty}{\longrightarrow} \infty \,, \quad j (i) = 1728 \,, \quad j (e^{\pi i/3}) = 0 \,,
\end{align}
in the affine coordinate of $\mathbb{CP}^1$. For convenience we will use a rescaled version of the $j$-function defined by
\begin{align}
J (\tau) = \frac{j(\tau)}{1728} \,.
\end{align}

Having introduced a great deal of mathematical machinery, we now ask about which regions of our parameter space lead to a time-reversal invariant 4D theory. First of all, we can immediately identify a ``trivial phase'' with $\theta = 0$. This corresponds to the vertical line in the fundamental domain with $\tau \in i \mathbb{R}$. Additionally, we see that the region $\theta = \pi$ retains time-reversal invariance. We refer to this as the ``topological insulator'' phase. Using the $T$-generator of the $SL(2, \mathbb{Z})$ duality group one has
\begin{align}
(\theta = \pi) \overset{\mathcal{T}}{\longrightarrow} (\theta = - \pi) \overset{T}{\longrightarrow} (\theta = \pi) \,.
\end{align}
This means that utilizing the duality group, the value $\theta = \pi$ is also time-reversal invariant for arbitrary values of the gauge coupling $g$. In terms of the complex paremeter $\tau$, this region is given by $\tau \in \tfrac{1}{2} + i \mathbb{R}$. We will refer to a theory with $\theta = \pi$ as the ``topological insulator phase.'' This exhausts all possibilities for time-reversal invariance in regions of the moduli space that contain arbitrarily weak coupling, i.e.\ $g^2 \rightarrow 0$.

However, there is an additional phase that preserves time-reversal invariance at strong coupling.
In order to see that, assume $|\tau| = 1$ which means that we are at strong coupling. The $S$-generator of the $SL(2,\mathbb{Z})$ acts as
\begin{align}
\tau \rightarrow - \frac{\overline{\tau}}{|\tau|^2} = - \overline{\tau} \,,
\end{align}
i.e., exactly as $\mathcal{T}$! Therefore, there is a strongly coupled phase which preserves time-reversal
invariance for $|\tau| = 1$. We will refer to it as the ``strongly coupled phase''.

\begin{figure}[t!]
\centering
\includegraphics[width=0.7\textwidth]{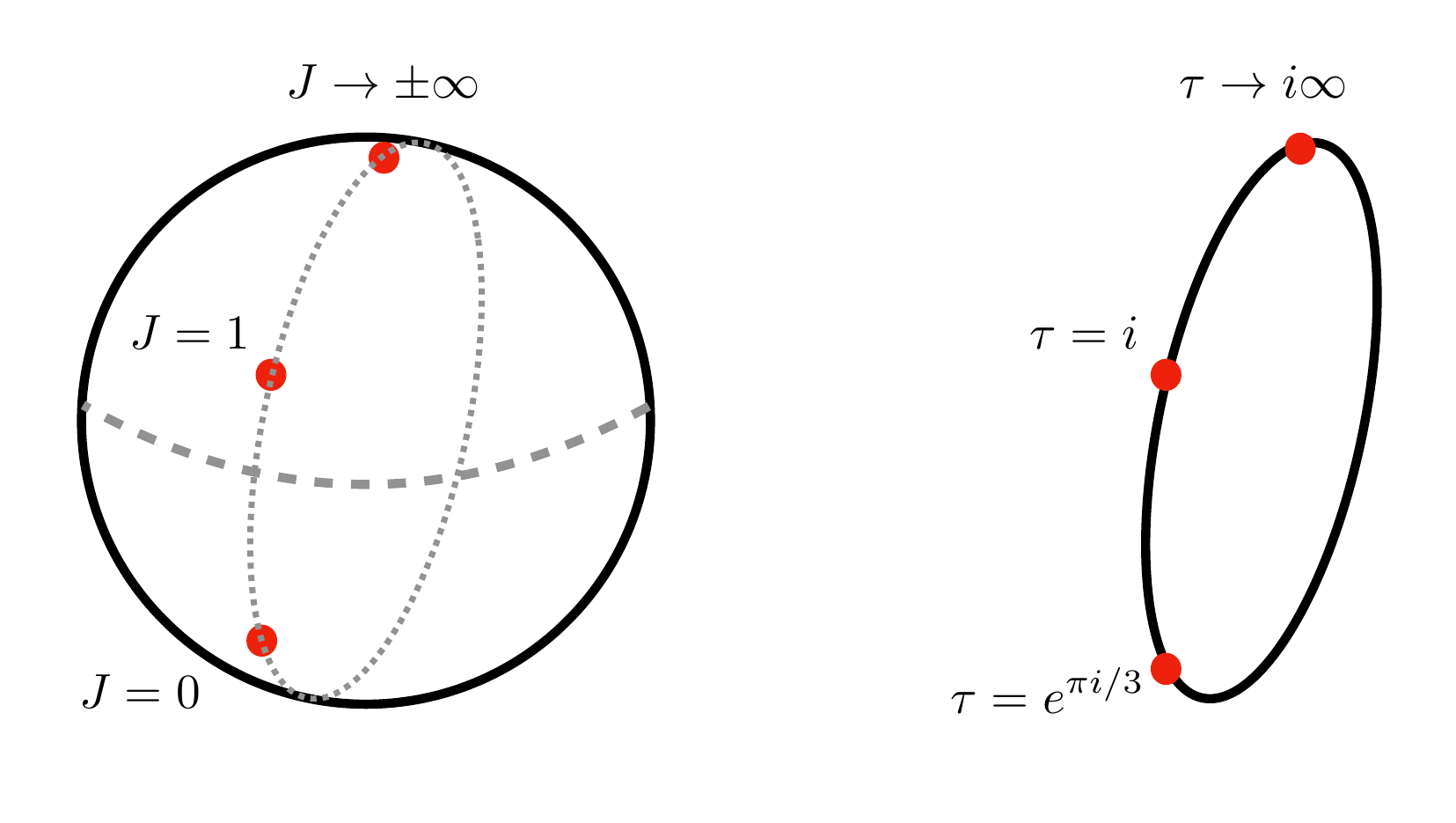}
\caption{Left: The image of the fundamental domain under $J$, with the marked points indicated as red dots. Right: The time-reversal subset $X(\Gamma)_{\mathbb{R}}$ of the modular curve $X (\Gamma)$ with $\Gamma = SL(2,\mathbb{Z})$.}
\label{fig:modcurve}
\end{figure}

The time-reversal invariant subspace indicated above is mapped as follows to the modular curve $X(\Gamma ) \sim \mathbb{CP}^1$
\begin{equation}
\begin{split}
\text{Trivial}:& \quad \tau = i \alpha \enspace \text{with} \enspace \alpha \in [1, \infty) \,, \quad 1 < J \,, \\
\text{Topological Insulator}:& \quad \tau = \tfrac{1}{2} + i \alpha \enspace \text{with} \enspace \alpha \in [\tfrac{\sqrt{3}}{2}, \infty) \,, \quad J < 0 \,, \\
\text{Strongly Coupled}:& \quad \tau = e^{i \alpha} \enspace \text{with} \enspace \alpha \in [\pi/3 , \pi/2] \,, \quad 0 \leq J \leq 1 \,.
\end{split}
\label{eq:tauregions}
\end{equation}
So we find that the image under $J(\tau)$ of the time-reversal invariant values of $\tau$ is the real line in $\mathbb{C}$, which is compactified to a circle in $\mathbb{CP}^1$. Since $J$ is a one-to-one map from the fundamental domain we see that all real values of $J$ correspond to time-reversal invariant values of $\tau$. That is to say, the time-reversal invariant subspace of a $U(1)$ gauge theory with duality group $SL(2,\mathbb{Z})$ is given by the real subspace of the corresponding modular curve denoted $X(\Gamma)_{\mathbb{R}}$.
Note further, that all three distinguished points are contained in the time-reversal invariant subset of $X(\Gamma)_{\mathbb{R}}$, see figure \ref{fig:modcurve} for a depiction.

We note that the above considerations have mainly focussed on the structure of the effective Lagrangian. A priori, it could happen that time-reversal invariance is spontaneously broken, as happens in some gauge theory examples (see e.g. \cite{Gaiotto:2017tne}). Here we
assume that time-reversal invariance is preserved by the system and explore the geometric and physical consequences.

\subsection{Localized Matter and Real Elliptic Curves}

In the previous subsection we reviewed some general features of 4D $U(1)$ gauge theory for a fixed value of the coupling
$\tau$. We now consider more general configurations in which the parameter $\tau(x_\bot)$ is a non-trivial function of position
in the 4D spacetime $\mathbb{R}^{2,1} \times \mathbb{R}_{\bot}$. In particular, we would like to understand what happens when we have an interface between two different time-reversal invariant phases. We argue that
the geometry of real elliptic curves provides a helpful tool in analyzing these situations.

On general grounds, demanding time-reversal invariance between phases of the system with different
values of the parameters means that we should expect states to be localized at
the region of transition (see e.g. \cite{Seiberg:2016rsg}).
To this end, we now
allow $\tau(x_{\bot})$ to be a non-trivial function of the position coordinate in our
4D spacetime $\mathbb{R}^{2,1} \times \mathbb{R}_{\bot}$. For each point
$x_{\bot} \in \mathbb{R}_{\bot}$, we get a value of $\tau$, and can also think about a 4D Lorentz
invariant theory with that particular value of the coupling. Indeed, in an interval of $\mathbb{R}_{\bot}$ where $\tau(x_{\bot})$
is constant, we just have a 4D theory compactified on an interval, and so we can still speak of the action of the duality group on the 4D basis of fields. So, for sufficiently adiabatic variations of the coupling, we can still fruitfully apply our 4D Lorentz invariant analysis.
On the other hand, we will also be interested in regions where there is a sharp jump in the profile of the coupling (sharp compared to all other length scales in the system). In such situations, we can expect new phenomena to be localized in the region where a jump occurs.

To a large extent, demanding time-reversal invariance for the system leads to the prediction that there are localized states
trapped at such an interface. Our discussion follows reference \cite{Seiberg:2016rsg}. Observe that if nothing is localized at the interface, the shift in $\theta$ angle from $\pi$ to $0$ at $x_\bot = 0$ would break time-reversal invariance. This can be seen by considering the $\theta$ term on a geometry with boundary
\begin{align}
\underset{x_{\bot} < 0}{\int} \frac{\theta}{8 \pi^2} F \wedge F = \underset{x_{\bot} < 0}{\int} \frac{\pi}{8 \pi^2} d (A \wedge F) = \frac{1}{8 \pi} A \wedge F \big|_{x_\bot = 0} \,.
\end{align}
This induces a half-integer quantized Chern-Simons term at the boundary which breaks time-reversal invariance.
Therefore, there have to be degrees of freedom living at the interface to compensate the variation with respect to time-reversal. One weakly coupled solution to the problem is a localized charged 3D Dirac fermion which compensates this variation by its parity anomaly \cite{Redlich:1983dv, Niemi:1983rq, AlvarezGaume:1984nf, Witten:2015aba, Tachikawa:2016cha, Cordova:2017kue, Kurkov:2017cdz, Cordova:2019wpi}, a version of the anomaly inflow mechanism \cite{Callan:1984sa}. Other weakly coupled options were discussed in \cite{Seiberg:2016rsg}, and some strongly coupled options were considered in reference \cite{Seiberg:2016gmd}.

In terms of the geometry of the modular curve $X(\Gamma)$ for the duality group $\Gamma = SL(2,\mathbb{Z})$, these weakly coupled completions correspond to motion in $X(\Gamma)_{\mathbb{R}}$ through the point at $\tau = i \infty$. The geometry of $X(\Gamma)_{\mathbb{R}}$ suggests an alternative route which might connect these two phases. Indeed, we can instead contemplate passing down through the strong coupling phase to reach the same value of the parameters. Observe that along this route, we need not pass through a cusp at all. Instead, we can pass through the strong coupling region with values $\tau = i$ and $\tau = \exp(2 \pi i / 6)$ at the ``bottom'' of the fundamental domain. In this case, one might be tempted to say that there is nothing localized, since there is a smooth interpolating in the value of $\tau$ which completely bypasses the cusp.

We now argue that even along this other trajectory, there are localized states. The main reason is that if we demand time-reversal invariance
for the system, then in the limit where there is a sharp jump across the $\vert \tau \vert = 1$ region, there must also be \textit{something} localized in this region. The one loophole in this argument is that it could happen that time-reversal invariance is somehow broken in this region. This, however, would be in conflict with the fact that after compactifying our 4D spacetime on a very large circle $S^{1}$, we see that there is a non-trivial winding number associated with maps $S^1 \rightarrow X(\Gamma)_{\mathbb{R}}$. This instead indicates that the pair of jumps $(\theta = 0) \rightarrow (\theta = \pi)$ and  $(\theta = \pi) \rightarrow (\theta = 2 \pi)$ retains time-reversal invariance.

To better understand what is happening in this region, we now study the geometry of the elliptic curve associated with the parameter $\tau$. Because correlation functions of the physical theory will depend on duality covariant expressions built out of $\tau$,
possible singularities associated with localized states will in general be associated with singularities in the geometry of the elliptic curve.

We geometrize the above statements by defining an auxiliary elliptic curve $E$ with complex structure modulus identified with the complexified coupling constant $\tau$. Any elliptic curve can be represented as a hypersurface in the weighted projective space $\mathbb{CP}_{[2,3,1]}^{2}$ via the coordinates $x$, $y$, and $z$. This leads to the so-called Weierstrass form of the elliptic curve:
\begin{align}
y^2 = x^3 + f x z^4 + g z^6 \,,
\end{align}
with complex coefficients $f$ and $g$. Away from the point $[x,y,z] = [1,1,0]$ we can use the $\mathbb{C}^*$-rescaling in order to set $z$ to $1$ and one obtains the standard form
\begin{align}
y^2 = x^3 + f x + g \,.
\label{eq:Weierstrass}
\end{align}
In this form the elliptic curve is given by a branched double-cover, with three branch points at the roots of the right hand side as well as a fourth root at infinity. For additional details on the geometry of elliptic curves, see Appendix \ref{app:ELLIPTIC}.

In terms of the parameter $\tau$, the coefficients $f$ and $g$ are associated with the Eisenstein series modular forms. We expect that
$f$ and $g$ depend non-trivially on the physical parameters of the system. This also holds for the discriminant:
\begin{equation}
\Delta = 4 f^3 + 27g^2.
\end{equation}
The $J$-function of the curve is given by the combination:
\begin{equation}
J = \frac{4 f^3}{4f^3 + 27g^2}.
\end{equation}

The appearance of this elliptic curve is quite familiar in a number of other contexts, including Seiberg-Witten theory, compactifications of 6D superconformal field theories on Riemann surfaces, as well as in the general approach to string vacua encapsulated by F-theory. In all of these cases, time-reversal invariance corresponds to a complex conjugation operation on the ``compactification coordinates'' $(x,y)$:
\begin{equation}
\mathcal{T}: (x,y) \mapsto (\overline{x}, \overline{y}).
\end{equation}
The special case of a time-reversal invariant Weierstrass model means we restrict to coefficients $f$ and $g$ which are real. Note that this is a strictly stronger condition than just demanding the $J$-function to be real. At least in supersymmetric settings, this is closely connected with the phase of BPS masses, and although we have less control in the non-supersymmetric setting, we expect a similar geometric condition to hold in this case as well. In section \ref{sec:NTWO} and Appendix \ref{app:FLAVA} we present some explicit $\mathcal{N} = 2$ examples illustrating these features, i.e., UV complete examples where $f$ and $g$ are purely real\footnote{Note that one could also consider models in which time-reversal invariance is restored in the deep IR, for which $f$ and $g$ can be complex numbers with correlated phases. In these cases, however, the mass parameters of the theory at high energies will break time reversal invariance in the UV.}.

Restricting $f$ and $g$ to be real means we are dealing with a real elliptic curve, namely the Weierstrass model makes sense over the real numbers. That being said, we will still view $x$ and $y$ as complex variables. This in turn leads to a constrained structure for the elliptic curve, especially as it moves through the different phases of $X(\Gamma)_{\mathbb{R}}$. To see this additional structure,
consider the factorization of the cubic in $x$:
\begin{equation}
x^3 + fx + g = \prod_{i = 1}^{3} (x - e_i),
\end{equation}
where the coefficients of the cubic are related to the roots as:
\begin{align}
0 & = e_1 + e_2 + e_3 \\
f & = e_1 e_2 + e_2 e_3 + e_3 e_1 \\
g & = -e_1 e_2 e_3\\
\Delta & = -\underset{i<j}{\prod}(e_i - e_j)^2.
\end{align}
The condition that $f$ and $g$ are real means that under complex conjugation, the roots $e_i$ must be permuted. There are two possibilities. Either all three roots are real, or one is real and the other two are complex conjugates. Without loss of generality, we can write these two cases as:
\begin{equation}
\begin{split}
\text{Case I}:& \quad e_1, e_2, e_3 \in \mathbb{R} \,, \\
\text{Case II}:& \quad e_1 \in \mathbb{R} \,, \enspace e_2 = \bar{e}_3 \,.
\end{split}
\end{equation}
Next, we want to relate the different configurations of the branch points to the time-reversal invariant values of $\tau$. The first comment is that from our explicit form of $f,g$ and $\Delta$, all of these quantities are real. In particular, the sign of the discriminant:
\begin{equation}
\Delta = - (e_1 - e_2)^2 (e_2 - e_3)^2 (e_3 - e_1)^2,
\end{equation}
tells us whether we are in Case I ($\Delta < 0$) or Case II ($\Delta > 0$). Since we also have:
\begin{equation}
J = \frac{4 f^3}{4 f^3 + 27 g^2} = \frac{4f^3}{\Delta},
\end{equation}
we conclude that when $f > 0$, we are in the regime of $0 \leq J \leq 1$, namely the strongly coupled phase.
If instead $f < 0$, then depending on the relative size of $4f^3$ and $27g^2$ we can get either sign of $\Delta$. Observe
that if $\Delta < 0$ and $f <0$ then, since $4f^3 + 27g^2 > 4f^3$ (recall $g^2$ is positive) we have $J > 1$, the ``trivial phase.'' If $\Delta >0$ and $f <0$ then we instead have $J <0$. Including the structure of the A- and B-cycles $\gamma_A$ and $\gamma_B$ of the elliptic curve, we see
there are three different phases of the time-reversal invariant contour specified by the following parameters:
\begin{itemize}
\item Trivial Phase: $J > 1 \Leftrightarrow \theta=0$ and $\tau = i \beta$ for $\beta>1$. There we have $\Delta < 0$, $f < 0$ and the roots $e_1 < e_3 < e_2$ are all real. The contours encircle $e_1$ to $e_3$ for $\gamma_B$ and $e_2$ to $e_3$ for $\gamma_A$.
\item Topological Insulator Phase: $J < 0 \Leftrightarrow \theta=\pi$. There we have $\Delta > 0$, $f < 0$ and the roots are such that $e_1 \in \mathbb{R}$, $e_2 = \bar{e}_3$, $\mathrm{Im}(e_2) > 0$. The contours encircle $e_1$ to $e_3$ for $\gamma_B$ and $e_2$ to $e_3$ for $\gamma_A$.
\item Strongly Coupled Phase: $0 \leq J \leq 1 \Leftrightarrow 0 \leq \theta \leq \pi$, $|\tau| = 1$. There we have $\Delta > 0$, $f \geq 0$ and the roots again satisfy $e_1 \in \mathbb{R}$, $e_2 = \bar{e}_3$, $\mathrm{Im}(e_2) > 0$. The contours encircle $e_1$ to $e_2$ for $\gamma_B$ and $e_1$ to $e_3$ for $\gamma_A$.
\end{itemize}
The different time-reversal invariant regions together with the signs of $f$, $g$, $\Delta$ are also indicated in figure \ref{fig:phasessigns}.
For some additional discussion, see Appendix \ref{app:ELLIPTIC}.

\begin{figure}[t!]
 \centering
 \includegraphics[width=.3\textwidth]{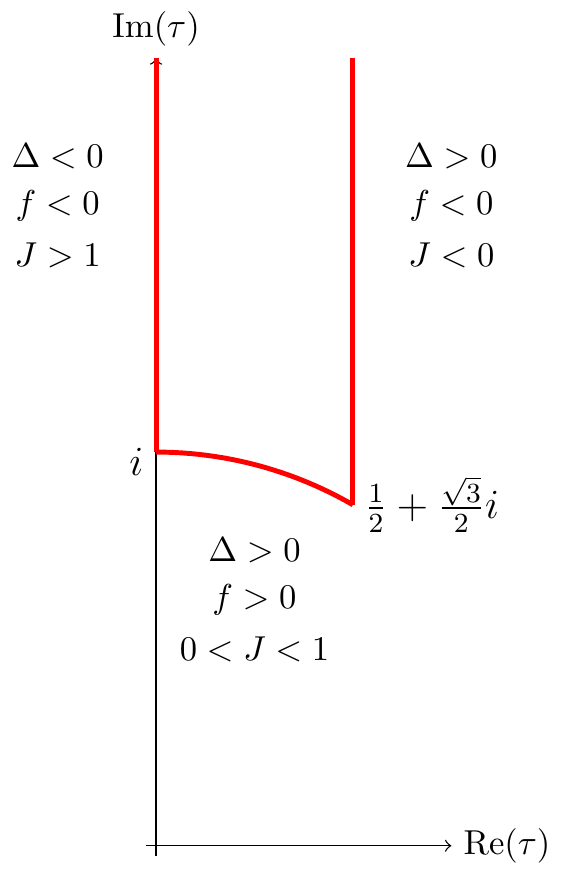}
 \caption{Values of the discriminant, $J(\tau)$ and $f$ for the elliptic curve as $\tau$ varies in its time-reversal invariant domain.}
 \label{fig:phasessigns}
\end{figure}

\begin{figure}[t!]
\centering
\includegraphics[width=\textwidth]{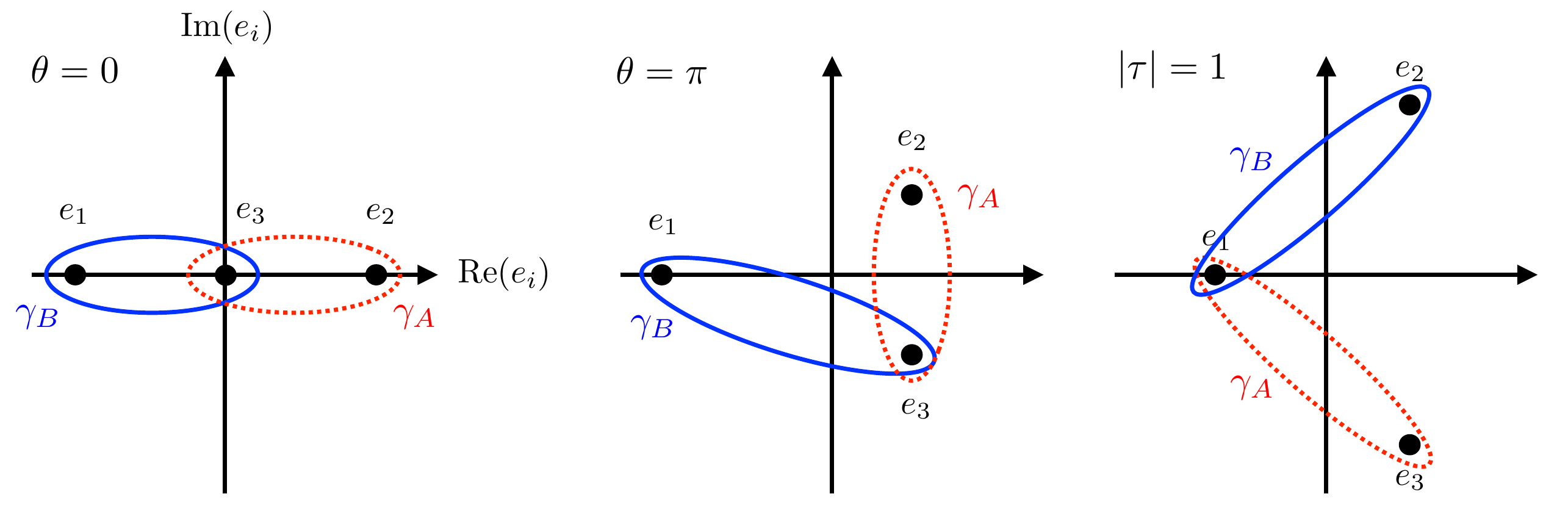}
\caption{Schematic view of the contours $\gamma_A$ (red dashed line) and $\gamma_B$ (blue solid line) for each phase.}
\label{fig:contours}
\end{figure}

\begin{figure}[t!]
\centering
\includegraphics[width=\textwidth]{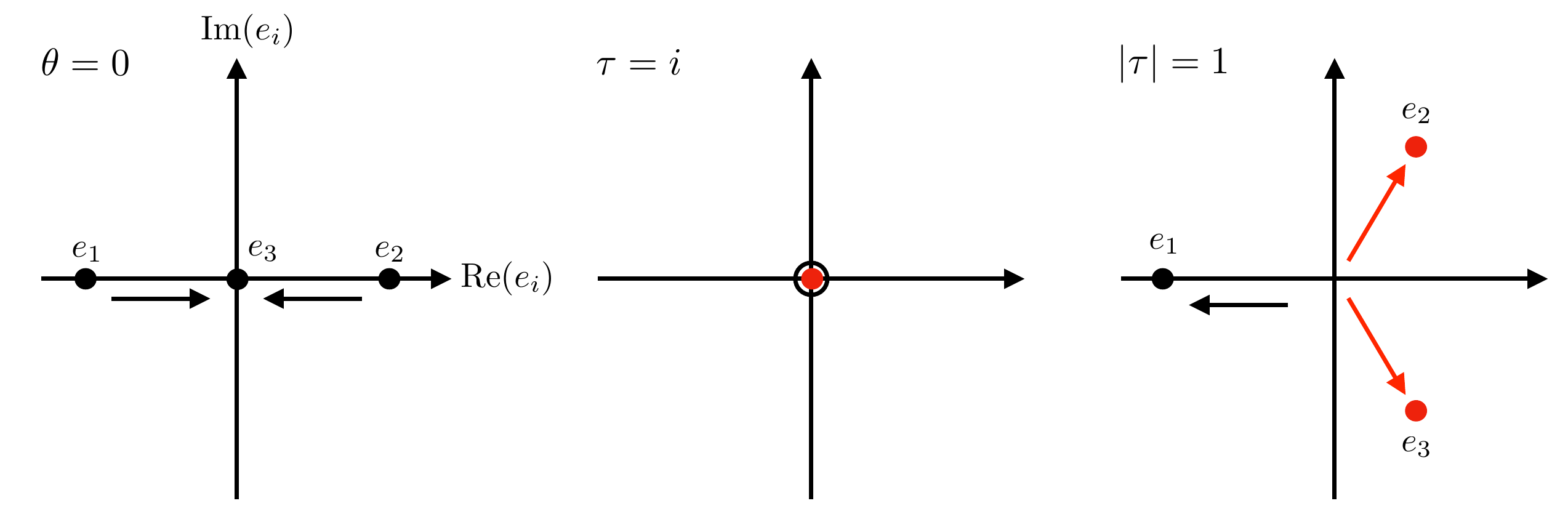}
\caption{Transition between the trivial and strongly coupled phase, with all three roots collapsing at $0$.}
\label{fig:transtrivstring}
\end{figure}

Finally, we want to ensure that we can move between the three different time-reversal invariant regions by adjusting the three roots $e_i$. As already indicated above one can transition between the phase with $| \tau | = 1$ and the topological insulator phase $\theta = \pi$ by moving two roots in the imaginary direction. Collapsing two conjugate roots on the real axis and then separating them as real roots along the real axis leads to the transition between the topological insulator phase and the trivial phase with $\theta = 0$. The last transition seems to happen when two of the roots go off to infinity, see figures \ref{fig:Jreal} and \ref{fig:Jcomp}. However, this transition can also happen at finite values of the roots, when all three roots collapse at $0$.
This last transition is depicted in figure \ref{fig:transtrivstring}. We see that the discriminant vanishes in the transition between the trivial and topological insulator phase as well as in the transition between the trivial and the strongly coupled phase.

Our analysis in terms of the real elliptic curve reveals that passing through a singularity in the elliptic curve also occurs when we move along the ``alternative contour'' connecting $\theta = 0$ and $\theta = \pi$. We take this to mean that there is also localized dynamics trapped at such an interface, in accord with general expectations from time-reversal invariance.

\section{Other Duality Groups}\label{sec:MOREDUAL}

In the previous section we presented some geometric tools to study 3D interfaces in 4D $U(1)$ gauge theory in the special case
where the duality group is $SL(2,\mathbb{Z})$. In systems with interacting degrees of freedom, one often encounters $U(1)$ gauge theories
where the duality group $\Gamma$ is a subgroup of $SL(2,\mathbb{Z})$. A common situation where this arises is in the
case where the $U(1)$ gauge theory has a non-trivial spectrum of line operators, which one can think of as various heavy non-dynamical states.

Our aim in this section will be to study interfaces with these smaller duality groups. Compared with the case of
$SL(2,\mathbb{Z})$ duality, we find a significantly richer set of possible interfaces. This is simply because there are now many different physically distinct field configurations which can no longer be related by a duality transformation under the smaller group. As before, we shall assume that time-reversal invariance is preserved, and in particular is not spontaneously broken by the vacuum.

For now, we assume that we have a $U(1)$ gauge theory where the duality group $\Gamma \subset SL(2,\mathbb{Z})$ is a finite index subgroup
of $SL(2,\mathbb{Z})$. Starting from the original lattice of electric and magnetic charges $\Lambda$, we can consider the orbits swept out by the group action $\Gamma$. This results in a refinement in the lattice $\Lambda_{\mathrm{refined}} \subset \Lambda_{\mathrm{orig}}$. This new lattice of electric and magnetic charges specifies a different elliptic curve $E = \mathbb{C} / \Lambda_{\mathrm{refined}}$. This new elliptic curve is related to the other by an isogeny; The complex structure is actually unchanged under this refinement, but additional data is now being specified by this choice.

The space of physically distinct values of $\tau$ as captured by the fundamental domain $X(\Gamma) = \overline{\mathbb{H}} / \Gamma$ is consequently bigger. In fact, for general $\Gamma \subset SL(2,\mathbb{Z})$, the resulting modular curve can be considerably more complicated than that obtained in the special case of $SL(2,\mathbb{Z})$ where we have the geometry of a $\mathbb{CP}^1$ with a single cusp at $i \infty$. For example, the genus of this new modular curve can be greater than zero. Additionally, the set of cusps is always bigger. Recall that the space of cusps is specified by taking the quotient of $\{i \infty \} \cup \mathbb{Q}$ by the group action specified by $\Gamma$. In terms of the electric and magnetic charge of a state, these rational numbers are specified by the ratio $q_e / q_m$ so that the ``purely electric'' cusp is at
$i \infty$. Observe that the value of $\tau$ at a cusp indicates either zero gauge coupling (as in the case of $\tau = i \infty$) or ``infinite coupling'' (as in the case of $\tau \in \mathbb{Q}$).

This also translates to a bigger set of values for $\tau$ which can lead to time-reversal invariant phases.
As before, these are obtained by focusing on the points of $X(\Gamma)$ which are invariant under the anti-holomorphic involution:
\begin{equation}
c_0 : \tau \mapsto - \overline{\tau}.
\end{equation}
Here, to aid the reader interested in comparing with reference \cite{snowden2011real} we have used that paper's notation. This operation is, of course, nothing but time-reversal conjugation!

We refer to the real locus of the modular curve as $X(\Gamma)_{\mathbb{R}}$:
\begin{align}
X (\Gamma)_{\mathbb{R}} = \{ \tau \in X (\Gamma): \enspace c_0 (\tau) = \tau \} = \{ \tau \in \overline{\mathbb{H}}: \enspace c_0(\tau) = \gamma \tau \enspace \text{with} \enspace \gamma \in \Gamma \} \,.
\end{align}
Thankfully this space has actually been studied in great detail in reference \cite{snowden2011real} for the congruence subgroups $\Gamma(N), \Gamma_1(N), \Gamma_0(N) \subset SL(2,\mathbb{Z})$ (see Appendix \ref{app:CONG} for details on the congruence subgroups). The results there hold for general congruence subgroups of $SL(2,\mathbb{Z})$. The topology of $X(\Gamma)_{\mathbb{R}}$ is a disjoint union of circles. Each such circle contains at least one cusp, but some cusps of $X(\Gamma)$ do not belong to any real component.\footnote{For example let $\Gamma=\Gamma_0(N)$, then $N=16$ is the lowest $N$ for which there are non-real cusps, and in this case there is one real component that crosses four real cusps, and two additional $\mathcal{T}$-violating cusps on the genus zero curve $X_0(16)$.} We refer to the cusps which are members of $X(\Gamma)_{\mathbb{R}}$ as ``real cusps.'' We note that the point at infinity is always a real cusp, and it specifies a distinguished $S^1$. Observe also that there are $S^1$'s which only involve cusps at ``infinite coupling.'' These are intrinsically strongly coupled regions of parameter space which are in some sense ``cut off'' from weak coupling.

Let us now turn to the structure of interfaces between time-reversal invariant phases. To build an interface, we allow $\tau(x_\bot)$ to be a non-trivial function of position in the 4D spacetime $\mathbb{R}^{2,1} \times \mathbb{R}_{\bot}$. As we move along one of the $S^{1}$'s of $X(\Gamma)_{\mathbb{R}}$ we encounter a cusp of electric and magnetic charge $(q_e , q_m)$ associated with the rational number $q_e / q_m \in \mathbb{Q}$. From all that we have said, we expect that the condition of time-reversal invariance
enforces the appearance of localized degrees of freedom at such an interface.

To better understand this, suppose we have such an interface located at $x_\bot = 0$. We can first specialize to the case $\Gamma=SL(2,\mathbb{Z})$. In this case all cusps $q_e / q_m \in \mathbb{Q}\cup \{ i \infty \}$ are dual to each other so it is enough to consider the electric duality frame where $(q_e,q_m)=(1,0)$. Crossing such a cusp at $x_\bot = 0$ involves having $g^2\rightarrow 0$ as $|x_\bot|\rightarrow 0$ while $\theta=0$ for $x_\bot < 0$ and $\theta=\pi$ for $x_\bot > 0$. This induces a localized Chern-Simons theory at level-$\frac{1}{2}$ on the interface. As noted in reference \cite{Seiberg:2016rsg}, the states trapped at the interface could exhibit a wide range of phenomena, including a charged, massless 3D Dirac fermion, or a system with non-trivial topological order.\footnote{We use this language since one is often interested in situations where the Maxwell theory arises as the IR limit of a more complicated 4d gauge theory.} If we do act by an $SL(2,\mathbb{Z})$ transformation to transform the cusp to a more general choice $(q_e,q_m)$, then we have that the putative localized states are charged under a dualized gauge potential $A_{(q_e, q_m)}$. In terms of the vector potentials for the electric field strength $F_{\mu \nu}$ and its magnetic dual counterpart $\widetilde{F}_{\mu \nu}$, we can write this as:
\begin{equation}
A_{(q_e, q_m)} = q_e A- q_m A_D.
\end{equation}
In other words, we can speak of localized dyonic states of electric charge $q_e$ and magnetic charge $q_m$!
Suppose now that we have a theory with smaller duality group $\Gamma$ a proper subgroup of $SL(2,\mathbb{Z})$.
We assume that we can supplement this theory by adding additional degrees of freedom to it so that in this enlarged theory,
$SL(2,\mathbb{Z})$ is the resulting duality group. This in turn means that in this bigger theory we can ask about the effects of an $SL(2,\mathbb{Z})$ transformation. In the original theory with the smaller duality group,
then, we learn that there can be states trapped at an interface with different electric and magnetic charges. Summarizing, we see that
if we encounter a cusp $q_e / q_m \in \mathbb{Q}$ in the original theory, the
localized degrees of freedom can be viewed as carrying an electric and magnetic charge $(q_e , q_m)$.

In section \ref{sec:DUALITY} we noted that there can be additional singularities other than those located at the cusps,
as associated to degeneration in the elliptic curve near the points $\tau = i$ and $\tau = \exp(2 \pi i / 6)$. These points are distinguished in the sense that they are fixed under some of the elements of $SL(2,\mathbb{Z})$ and are referred to as ``elliptic points'' of order two ($\tau = i$) and three ($\tau = \exp(2 \pi i / 6)$). It turns out that for most finite index subgroups $\Gamma \subset SL(2,\mathbb{Z})$ there are no elliptic points, but in the few cases when they are present we can expect localized matter to also be present,
at least when the associated elliptic curve degenerates in approaching such a point of the real moduli space.
In such situations, we expect states with non-zero Dirac pairing to be simultaneously localized.

\begin{figure}[t!]
\centering
\includegraphics[width=0.5\textwidth]{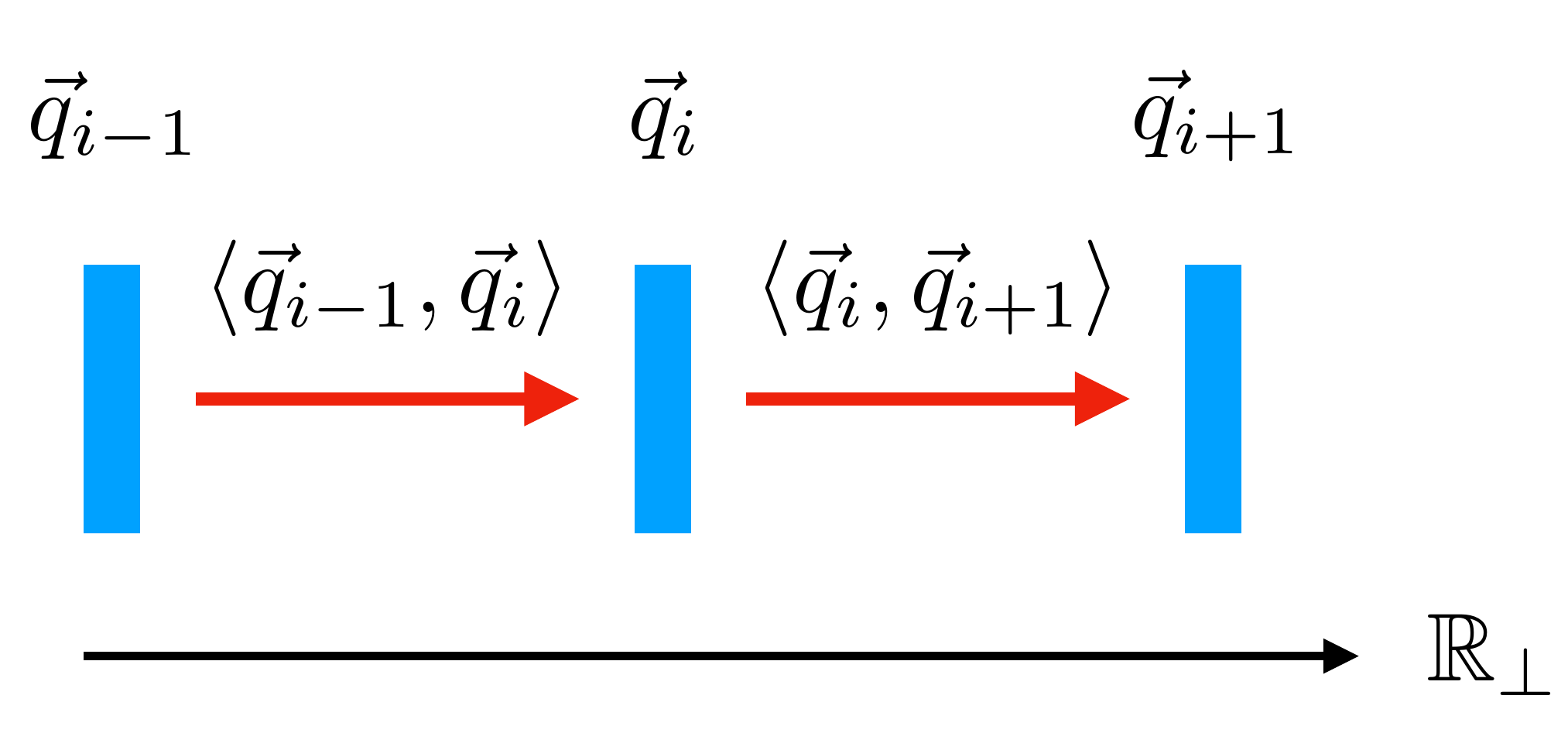}
\caption{Depiction of interfaces encountered in a trajectory through a component of $X(\Gamma)_{\mathbb{R}}$. Here,
each interface is associated with the $SL(2,\mathbb{Z})$ image of the cusp at weak coupling and therefore comes with excitations carrying an electric and magnetic charge which we denote as a two-component vector. States localized on neighboring walls have a non-zero Dirac
pairing, and this leads to a net angular momentum quantized in units of $\pm 1/2$ or $\pm 1$ between neighboring interfaces.}
\label{fig:SpinChain}
\end{figure}

We can also deduce the relative spin-statistics of the excitations on neighboring interfaces,
which also lead to a quantization of the angular momentum induced by the electro-magnetic field between the interfaces.
Although not stated in these physical terms, reference \cite{snowden2011real}
computes the Dirac pairing between neighboring interfaces.
Focusing on the generic situation where our interfaces are generated by cusps, it turns out that the excitations localized on
neighboring interfaces always have a
non-vanishing Dirac pairing equal to $\pm 1$ or $\pm 2$:
\begin{equation}
\langle \vec{q} , \vec{q} \, '  \rangle \in \{\pm 1, \pm 2 \}.
\end{equation}
Recall that the Dirac pairing between dyons specifies an intrinsic angular momentum in the system. What this pairing indicates is that there is
an intrinsic \textit{spin} quantized in units of $\pm 1/2$ or $\pm 1$ associated with regions of the 4D bulk. This is an additional topological feature of our 4D bulk, as controlled by the dynamics of the interface! See figure \ref{fig:SpinChain} for a depiction.

In the remainder of this section we illustrate these general considerations by focusing on some specific choices of duality groups. In particular, we leverage the results of reference \cite{snowden2011real} to obtain explicit information on the structure of 3D interfaces
in these systems. We consider the three most well-known congruence subgroups $\Gamma(N), \Gamma_{1}(N),$ and $\Gamma_{0}(N)$ which also show up frequently in the study of modular curves:
\begin{equation}
\begin{split}
\Gamma_0 (N) &= \bigg\{ \gamma \in SL(2,\mathbb{Z}): \gamma = \left( \begin{array}{c c} * & * \\ 0 & * \end{array} \right) \, \text{mod} \, N\bigg\} \,, \\
\Gamma_1 (N) &= \bigg\{ \gamma \in SL(2,\mathbb{Z}): \gamma = \left( \begin{array}{c c} 1 & * \\ 0 & 1 \end{array} \right) \, \text{mod} \, N\bigg\} \,, \\
\Gamma (N) &= \bigg\{ \gamma \in SL(2,\mathbb{Z}): \gamma = \left( \begin{array}{c c} 1 & 0 \\ 0 & 1 \end{array} \right) \, \text{mod} \, N\bigg\} \,,
\end{split}
\end{equation}
where $*$ denotes an arbitrary integer entry. Clearly, these subgroups satisfy
\begin{align}
\Gamma (N) \subset \Gamma_1 (N) \subset \Gamma_0 (N) \subset SL(2,\mathbb{Z}) \,,
\end{align}
and each is a finite index subgroup of $SL(2,\mathbb{Z})$.

For each of these choices, there is a corresponding modular curve $X(\Gamma)$ which we denote by $X(N)$ for $\Gamma = \Gamma(N)$, $X_1(N)$ for $\Gamma = \Gamma_1(N)$ and $X_0(N)$ for $\Gamma = \Gamma_0(N)$.  Further it is clear that in each case $X(\Gamma)_{\mathbb{R}}$ is non-trivial since one can always choose the fundamental domain in a way that it contains (part of) the imaginary axis, which is invariant under $c_0$. This subset of $X (\Gamma)_{\mathbb{R}}$ is the region with $\theta = 0$. Moreover, it is clear that some remnant of the standard $T$ generator in $SL(2,\mathbb{Z})$ survives:
\begin{align}
T \in \Gamma_0 (N), \Gamma_1 (N) \,, \quad T^N \in \Gamma (N) \,,
\end{align}
which means that for $\Gamma_0(N)$ and $\Gamma_1(N)$ there are regions in $X (\Gamma)_{\mathbb{R}}$ which correspond to $\theta = \pi$. For $\Gamma (N)$ the non-trivial time-reversal invariant value of $\theta$ is given by $N \pi$. Note, that these two regions meet in the weakly coupled cusp situated at $\tau = i \infty$, which is also contained in the set $X(\Gamma)_\mathbb{R}$.

Since we have already explained the significance of the time-reversal invariant components of these modular curves, we now review the graphical rules developed in \cite{snowden2011real} which enumerate which ($\Gamma$-equivalence classes of) cusps are on a given real component. These graphs were arrived at by a group-theoretic analysis of each $\Gamma$ which assigns a solid dot to a cusp, on open dot to an elliptic point, with a single line connecting two cusps if their Dirac pairing is $\pm 1$, and a double line if their Dirac pairing is $\pm 2$ which reference \cite{snowden2011real} refers to as a ``weight''. Similar considerations hold for lines which connect an elliptic point to a cusp, but in this case the pairing is trajectory dependent. In these cases, the elliptic point connects to a cusp, once with weight one, and once with weight two. We take this to mean that there are states with mutually non-local charges localized at the elliptic point. This is a phenomenon which is
known to occur in 4D $\mathcal{N} = 2$ theories \cite{Argyres:1995jj}.

\begin{figure}[t!]
  \centering
  \includegraphics[]{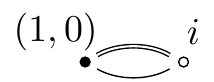}
  \caption{Real component for $X(1)$, namely the special case $\Gamma = SL(2,\mathbb{Z})$.
In the graph, cusps are denoted by solid dots and elliptic points are denoted by open dots.}%
  \label{fig:snowden2}%
\end{figure}

Each such line corresponds to a subset of points in $X(\Gamma)_{\mathbb{R}}$ satisfying:
\begin{align}
C_{\gamma}: \quad - \overline{\tau} = \gamma \tau \,.
\end{align}
for some conjugacy class $\gamma \in \Gamma$.
In general the subspace $X(\Gamma)_{\mathbb{R}}$ consists of the union of all these sets inside a single fundamental domain of the group $\Gamma$. For starters, we show the structure of this graph in figure \ref{fig:snowden2} in the case where $\Gamma = SL(2,\mathbb{Z})$.

\begin{figure}[t!]
  \centering
  \includegraphics[]{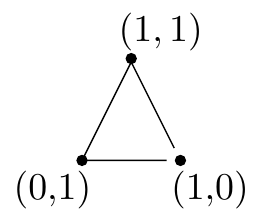}
  \includegraphics[]{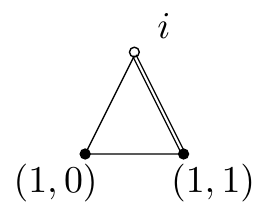}
 \caption{Real components of $X(2)$ (left) along with $X_0(2)$ and $X_1(2)$ (right).
On the right, the double line connecting $(1,1)$ to the elliptic point $\tau =i$
refers to the fact that if we follow a geodesic connecting
$(1,1)$ and $i$ we land on $(-1,1)$, and the Dirac pairing between $(1,1)$ and $(-1,1)$ is
$2$. Similarly, there is a single line connecting (1,0) and $i$ because the geodesic
through them lands on the cusp $(0,1)$, which has Dirac pairing $1$ with $(1,0)$.
In the graph, cusps are denoted by solid dots and elliptic points are denoted by open dots.
}
  \label{fig:snowden1}%
\end{figure}

As another example, consider the case of $X(2)$, for which there is one real component depicted in figure \ref{fig:snowden1} that (in a chosen duality frame) passes through the cusps $0, 1$, and $i \infty$. We represent this on the left side of figure \ref{fig:snowden1}. On the right side we depict the real component for $X_1(2)$ and $X_0(2)$ which passes through the cusps $1$ and $\infty$ and an order-2 elliptic point at $\tau=i$. Including $X(1)$, as shown in figure \ref{fig:snowden2}, we have actually exhausted all the cases where elliptic points can occur on a real component.

Having presented the general rules, we now summarize some of the important features of $X(\Gamma)_{\mathbb{R}}$ in the case of the aforementioned congruence subgroups. The statements we present amount to an adaptation of results given in \cite{snowden2011real}.

\subsubsection*{$\mathbf{X(N)}$}
Consider first the case where the duality group is $\Gamma = \Gamma(N) \subset SL(2,\mathbb{Z})$. In this case, the cusps are in the same $\Gamma(N)$-orbit if and only if $(a',b')\equiv \pm (a,b) \; \textnormal{mod} \; N$ and $\Gamma(N)$-equivalence classes of cusps are parametrized by pairs $\pm \frac{a}{b}$ of order-$N$ elements of $(\mathbb{Z}/N\mathbb{Z})^2$. To see the latter, note that we can reduce an element $(a,b)\in \mathbb{Z}^2$ modulo $N$ , which for $N>2$ is distinct from the modulo $N$ reduction of $(\pm a,b)$. Not every element of $(\mathbb{Z}/N\mathbb{Z})^2$ can be obtained from such a reduction though, since $\textnormal{gcd}(a,b)=1$. In particular $\textnormal{gcd}(a,b,N)=1$, which implies that at least either $a$ or $b$ must be an order-$N$ element of $\mathbb{Z}/N\mathbb{Z}$, making $(a,b)$ an order-$N$ element of $(\mathbb{Z}/N\mathbb{Z})^2$. The number of order-$N$ elements in $(\mathbb{Z}/N\mathbb{Z})^2$ is $N^2 \prod_{p|N}(1-1/p^2)$, where $p$ is a prime, but for $N>2$ we identify $(a,b) \, \textnormal{mod} \, N$ with $(-a,-b) \, \textnormal{mod} \, N$ since they represent the same cusp $\frac{a}{b}$, with similar considerations for the $-\frac{a}{b}$ cusp. Altogether we have
\begin{equation}
\textnormal{\# of cusps in fundamental domain} = \begin{cases}

\frac{1}{2} N^2 \underset{p|N}{\prod}(1-1/p^2) & N>2 \\
3 & N=2
\end{cases}
\end{equation}
for the total number of cusps.

\begin{figure}[t!]
  \centering
  \includegraphics[]{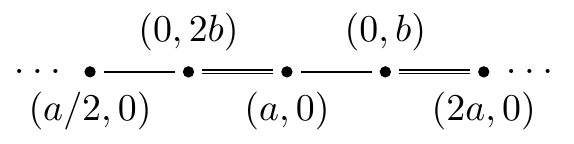}
  \includegraphics[]{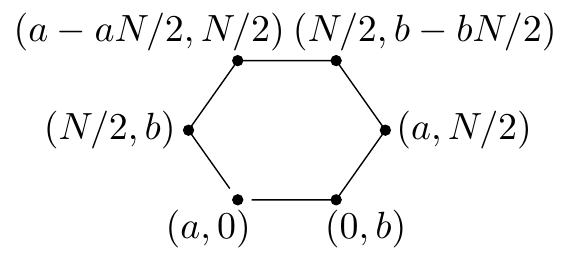}
  \includegraphics[]{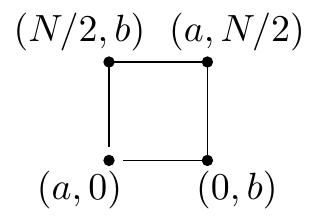}
 \caption{Real cusps/components (mod-$N$) for $r=0$ (left), $r=1$ (center), and $r\geq 2$ (right). Here $N=2^r N'$ for $N'$ odd. In all cases, $ab \equiv 1 \, \textnormal{mod} \, N$ and we take $\textnormal{gcd}(a,N) = 1$}%
  \label{fig:snowden3}%
\end{figure}

Turning next to the real cusps and components, we characterize the cases by the power $r$ in $N=2^r N'$ with $\mathrm{gcd}(2,N') = 1$ and we quote the results mainly without proof. The case $r=0$ is perhaps the most complicated, we have $\phi(N)$ real cusps\footnote{This is the Euler totient function which expresses how many numbers $m<N$ are coprime to $N$, or equivalently, the order of the multiplicative group $(\mathbb{Z}/N\mathbb{Z})^\times$. It can be expressed as $N\prod_{p|N}(1-\frac{1}{p})$.} spread across $\psi(N)$ real components.\footnote{Borrowing notation from \cite{snowden2011real}, $\psi(N)$ is defined as the order of the group $(\mathbb{Z}/N\mathbb{Z})^\times/ \langle -1,2 \rangle $ which has no known closed form expression.} The neighborhood of a cusp $(a,b)$ (taken mod $N$)  is shown on the left-hand side of figure \ref{fig:snowden3}.

The case $r=1$ ($N>2$) has $3\phi(N)$ real cusps spread evenly across $\frac{1}{2}\phi(N)$ real components, i.e. six cusps per component whose charges (mod $N$) are shown in figure \ref{fig:snowden3}. While the $r \geq 2$ cases have $2\phi(N)$ real cusps spread evenly across $\frac{1}{2}\phi(N)$ real components, i.e. four cusps per component.

\subsubsection*{$\mathbf{X_1(N)}$}
\begin{figure}[t!]
  \centering
  \includegraphics[]{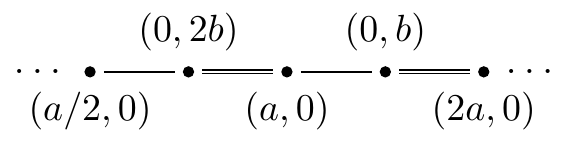}
 \caption{Real cusps/components (mod-$N$) for $X_1(N)$ ($N\neq 2,4$) for $r=0$. Here, $N=2^rN'$ with $\mathrm{gcd}(2,N') = 1$}%
  \label{fig:snowden4}%
\end{figure}

\begin{figure}[t!]
  \centering
  \includegraphics[width=50mm]{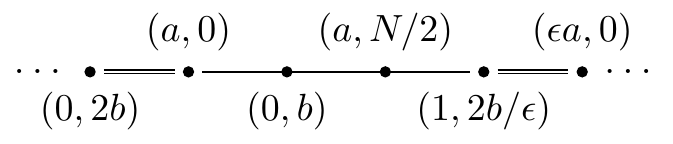}
 \caption{Real cusps/components (mod-$N$) for $X_1(N)$ ($N\neq 2,4$) for $r=1$. Here, $N=2^rN'$ with $\mathrm{gcd}(2,N') = 1$.
In the figure, $\epsilon \equiv 2+N/2$.}%
  \label{fig:snowden44}%
\end{figure}

\begin{figure}[t!]
  \centering
  \includegraphics[width=50mm]{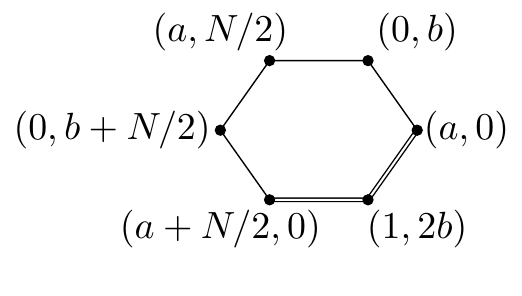}
 \caption{Real cusps/components (mod-$N$) for $X_1(N)$ ($N\neq 2,4$) for $r\geq 2$. Here, $N=2^rN'$ with $\mathrm{gcd}(2,N') = 1$.}%
  \label{fig:snowden444}%
\end{figure}

\begin{figure}[t!]
  \centering
  \includegraphics[]{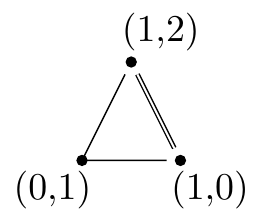}
 \caption{Real cusps/components for $X_1(4)$}%
  \label{fig:snowden5}%
\end{figure}

Consider next the case of the modular curve $X_1(N)$ as specified by the duality group $\Gamma_1(N) \subset SL(2,\mathbb{Z})$. In this case, the cusps are in the same $\Gamma_1(N)$ orbit if and only if $(a,b)\equiv \pm (a+jb,b) \; \textnormal{mod} \; N$ for some integer $j$. Equivalence classes can be parametrized by first fixing $a \; \textnormal{mod} \; \textnormal{gcd}(b,N)$, then enumerating pairs $\pm \frac{a}{b}$ of order-$N$ elements of $(\mathbb{Z}/N\mathbb{Z})^2$ under this restriction. The number of cusps (see e.g. \cite{diamond2006first}), is
\begin{equation}
\textnormal{\# of cusps in fundamental domain} = \begin{cases}

2 & N=2 \\
3 & N=4  \\
\frac{1}{2} \underset{d|N}{\sum} \phi(d)\phi(N/d)& N=3 \;  \textnormal{or} \; N>4
\end{cases}
\end{equation}
where $d$ is any divisor. Just like the $X(N)$ curves, the properties of the real cusps and components depend on the exponent $r$ in $N=2^rN'$ (with $\mathrm{gcd}(2,N') = 1$, and in fact the $r=0$ case is exactly the same for $X_1(N)$ and $X(N)$. For the $r=1$ case, there are $2\phi (N)$ real cusps and $\psi(N/2)$ real components (making the number of cusps per component more irregular than for the $X(N)$ curves), while the $r\geq2$ case has $\frac{3}{2}\phi(N)$ real cusps and $\frac{1}{4}\phi(N)$ real components arranged as in figures \ref{fig:snowden4}, \ref{fig:snowden44} and \ref{fig:snowden444}. There is an exception to this classification for $N=4$. The real structure of this case
is displayed in figure \ref{fig:snowden5}.

\subsubsection*{$\mathbf{X_0(N)}$}
\begin{figure}[t!]
  \centering
  \includegraphics[]{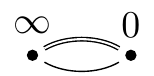}
  \caption{Real cusps/components for $X_0(N)$ when $N$ is odd.}%
  \label{fig:snowden6}%
\end{figure}

Finally, consider the case of the modular curve $X_0(N)$ as associated with the duality group $\Gamma_0(N) \subset SL(2,\mathbb{Z})$.
The cusps in this case are in the same $\Gamma_0(N)$ orbit if and only if $(ya,b)\equiv \pm (a+jb,yb) \; \textnormal{mod} \; N$ for some integers $j$ and $y$ such that $\mathrm{gcd}(y,N)=1$. Conveniently, it turns out that equivalence class of cusps can be described simply as elements of $\mathbb{P}^1(\mathbb{Z}/N\mathbb{Z})$ and we can represent the mod-$N$ charges of cusps as $[a:b]$. The total number of cusps is then
\begin{equation}
\textnormal{\# of cusps in fundamental domain} =  \sum_{d|N} \phi(\textnormal{gcd}(d,N/d))
\end{equation}
for any $N$. For $r=0$ ($N$ odd), let $k$ be the number of distinct prime factors of $N$, then there are $2^{k-1}$ real components all of the form shown in figure \ref{fig:snowden6}.

\begin{figure}[t!]
\centering
  \includegraphics[]{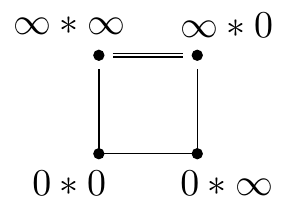}
  \includegraphics[]{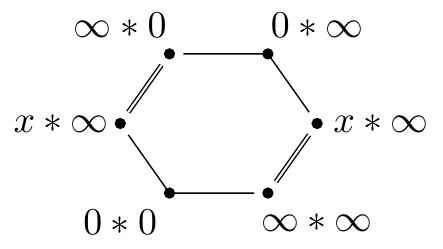}
 \caption{The real cusps/components for $X_0(N)$ when $r=1$ (left) and $r=2$ (right) where the $*$ notation refers to the decomposition $\mathbb{P}^1(\mathbb{Z}/N\mathbb{Z})=\mathbb{P}^1(\mathbb{Z}/2^r\mathbb{Z})\times \mathbb{P}^1(\mathbb{Z}/N'\mathbb{Z})$ since we do not want to conflate this with the parentheses notation $(\cdot, \cdot)$ used to label the electric and magnetic charges.
Here we define $x \equiv [1:2]$, viewed as an element of $\mathbb{P}^1(\mathbb{Z} / 2^r \mathbb{Z})$.}%
  \label{fig:snowden7}%
\end{figure}

\begin{figure}[t!]
\centering
  \includegraphics[]{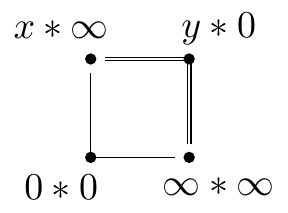}
  \includegraphics[]{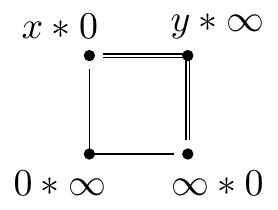}
 \caption{The real cusps/components for $X_0(N)$ when $r\geq 3$, where we have two flavors of components (an equal number of each).
the $*$ notation refers to the decomposition $\mathbb{P}^1(\mathbb{Z}/N\mathbb{Z})=\mathbb{P}^1(\mathbb{Z}/2^r\mathbb{Z})\times \mathbb{P}^1(\mathbb{Z}/N'\mathbb{Z})$ since we did not want to confuse with the parentheses $(,)$ for the electric and magnetic charges.
Here we defined $x \equiv [1:2]$ and $y \equiv [1:2^{r-1}]$ viewed as elements of $\mathbb{P}^1(\mathbb{Z} / 2^r \mathbb{Z})$.}%
  \label{fig:snowden8}%
\end{figure}

The behavior for even $N$ is again governed by the number of distinct odd prime factors $k$. For $r = 1$, $r = 2$, and $r \geq 3$, there are respectively $2^{k+1}$, $3\cdot 2^k$ , and $2^{k+2}$ real cusps and $2^{k-1}$, $2^{k-1}$, and $2^k$ real components. See figures \ref{fig:snowden7} and \ref{fig:snowden8} for the corresponding real components of the modular curves.

\section{$\mathcal{N} = 2$ Examples} \label{sec:NTWO}

To illustrate some of these general considerations, we now present some examples based on $\mathcal{N} = 2$ supersymmetry.
Recall that a helpful way to study such theories involves the geometry of the Seiberg-Witten curve \cite{Seiberg:1994rs, Seiberg:1994aj}.

We begin by considering a class of 4D $\mathcal{N} = 2$ superconformal field theories obtained from a D3-brane probing a stack of seven-branes with and ADE gauge group. This determines a flavor symmetry on the 4D worldvolume theory of the D3-brane \cite{Banks:1996nj,
Minahan:1996fg, Minahan:1996cj, Noguchi:1999xq}.
In these cases, there is a one-dimensional Coulomb branch, specified by a complex coordinate $u$, and mass parameters $m$ in the adjoint representation of the seven-brane gauge group. The Seiberg-Witten curves for this class of examples can all be written as:
\begin{equation}
y^2 = x^3 + f(u,m)x + g(u,m),
\end{equation}
where the $f$'s and $g$'s are polynomials in the Coulomb branch parameters and the $m$'s. These polynomials in the $m$'s are constructed
from Casimir invariants of the associated flavor symmetry. In the string compactification geometry, time-reversal invariance corresponds to a complex conjugation operation on the elliptic curve itself. We get a time-reversal invariant system by demanding the Weierstrass coefficients $f$ and $g$ are real. Observe that in a suitable basis of fields, we can simply demand that the $u$'s and $m$'s are all real. This corresponds to a situation in which any mass terms being switched on preserves time-reversal invariance
along the flow from the UV fixed point to the IR, namely where the Seiberg-Witten curve description is valid.

We obtain examples of interfaces by allowing position dependent mass terms
$m(x_{\bot})$. One can also contemplate giving a position dependent value to $u$, though in this case we need to consider
the spacetime dependence for a dynamical field. Switching on a $\mathcal{N} = 1$ superpotential deformation as well as possible supersymmetry breaking mass terms, we can also produce theories in the IR which only have a $U(1)$ gauge field remaining. This strategy was used, for example in \cite{Tachikawa:2016xvs} to analyze some examples of SPTs with non-abelian gauge dynamics.

Assuming we vary the mass parameters $m$ adiabatically, we can continue to use 4D $\mathcal{N} = 2$ supersymmetry to look for the appearance of localized states. In the F-theory realization of these systems as obtained from D3-branes probing a stack of seven-branes, this corresponds to moving the seven-branes around in the $\mathbb{R}_{\bot}$ direction of the 4D spacetime. In the vicinity of some of these seven-branes, however, we can continue to use a 4D analysis. In particular, the location of these seven-branes will occur at some locations $u = u_{\ast}$ in the original Coulomb branch parameter.

Now, the appearance of massless states occurs when the discriminant $\Delta$ vanishes to some order in the variable $(u - u_{\ast})$. In fact, for elliptically-fibered K3 spaces there is a Kodaira classification\footnote{Which also classifies possible codimension one singularities for higher-dimensional elliptically fibered Calabi-Yaus.} of possible singularities \cite{kodaira},
as controlled by the order of vanishing for:
\begin{align}
f & \sim (u - u_{\ast})^{\mathrm{ord}(f)}\\
g & \sim (u - u_{\ast})^{\mathrm{ord}(g)}\\
\Delta & \sim (u - u_{\ast})^{\mathrm{ord}(\Delta)}.
\end{align}
These tell us about the appearance of flavor enhancements, as well as the appearance of massless states, including the associated electric and magnetic charges. In Appendix \ref{app:FLAVA} we consider in detail the special case of $SU(2)$ gauge theory with four hypermultiplets in the fundamental representation of $SU(2)$. In particular, we calculate the periods and the appearance of massless states for a specific choice of mass parameters.

The case of a cusp corresponds to an $I_N$ singular fiber (associated with an $SU(N)$ flavor symmetry), in which $\mathrm{ord}(f) = \mathrm{ord}(g) = 0$, and $\mathrm{ord}(\Delta) = N$. Observe that in the vicinity of such a point, we have:
\begin{equation}
\tau \sim \frac{N}{2 \pi i} \log(u - u_{\ast}),
\end{equation}
indicating a jump of $\theta$ by $2 \pi N$ as we cross this sort of singularity.

The Kodaira classification also shows that we can expect mutually non-local states to be trapped at an interface. For example, a $III^{\ast}$ singular fiber (associated with an $E_7$ flavor symmetry) corresponds to the special case where $\mathrm{ord}(f) = 3$, $\mathrm{ord}(g) \geq 5$ and $\mathrm{ord}(\Delta) = 9$. In this case, we also note that the $J$-function has a well-defined limit, even though the elliptic curve becomes degenerate in this region. The specific value is $J = 1$, as associated with $\tau = i$.

We can also get trapped matter at the other elliptic point of $\Gamma = SL(2,\mathbb{Z})$, namely $\tau = \exp(2 \pi i / 6)$, as associated with $J = 0$. This occurs, for example, with a $II^{\ast}$ singularity (associated with an $E_8$ flavor symmetry), in which $\mathrm{ord}(f) \geq 4$, $\mathrm{ord}(g) = 5$, and $\mathrm{ord}(\Delta) = 10$. In the non-supersymmetric setting we have less analytic control over the ways in which $f,g$, and $\Delta$ might vanish.

Our discussion so far has focused on the case where the $U(1)$ gauge theory on the Coulomb branch enjoys an $SL(2,\mathbb{Z})$ duality group, as directly inherited from the F-theory realization of these systems.\footnote{Strictly speaking one should speak of the $\mathbb{Z} / 2 \mathbb{Z}$ extension of $SL(2,\mathbb{Z})$, as in reference \cite{Pantev:2016nze}. We will not dwell on this issue here.} We get examples with smaller duality groups by holding fixed some of the mass parameters of the system. For example, the ADE series of superconformal field theories just introduced can also be engineered by taking M5-branes wrapped on a $\mathbb{CP}^{1}$ with punctures \cite{Gaiotto:2009we}. These punctures dictate the behavior of mass parameters in the 4D effective field theory. In this formulation, the mapping class group of the curve determines the structure of the duality group. Doing so, we can engineer smaller duality groups. As an example, for $SU(2)$ gauge theory with four flavors, we have two M5-branes wrapped on a sphere with four punctures. In this case, taking some mass parameters held fixed to equal values can produce a smaller duality group such as $\Gamma_{0}(2)$.

We can also consider examples which have a smaller duality group right from the start. As an example of this sort, consider pure $\mathfrak{su}(2)$ gauge theory. Here, we have no mass parameters, so we will consider varying the Coulomb branch parameter $u$
as a function of $x_{\bot}$ with the implicit assumption that we have introduced a suitable $\mathcal{N} = 1$ superpotential deformation to generate jumps in the value of $\tau$ in a given interface region.

Consider first the limit where no superpotential deformation has been switched on.
Following \cite{Seiberg:1994rs, Seiberg:1994aj}, the $\mathcal{N} = 2$ vector multiplet contains a scalar field in the adjoint representation $\phi$. Non-zero values of this scalar move the theory onto the Coulomb branch. In the following we use the gauge invariant combination:
\begin{align}
u = \tfrac{1}{2} \text{tr} ( \phi^2 ) \,.
\end{align}

The Seiberg-Witten curve of the system is given by
\begin{align}
y^2 = (x - u) (x - \Lambda^2) (x + \Lambda^2) \,,
\end{align}
which can be brought to Weierstrass form by a coordinate transformation on $x$. The weakly coupled $U(1)$ gauge theory arises for $|u| \rightarrow \infty$ in which case the gauge coupling goes to zero. Other interesting limits are described by the limits $u \rightarrow \pm \Lambda^2$, which are at strong coupling. At these points one finds light magnetically charged states.

By moving around the moduli space parameterized by $u$ one finds the following monodromy actions in $SL(2,\mathbb{Z})$ on the auxiliary elliptic curve:
\begin{align}
\gamma_+ = \begin{pmatrix} 1 & 0 \\ -2 & 1 \end{pmatrix} \,, \quad  \gamma_- = \begin{pmatrix} -1 & 2 \\ -2 & 3 \end{pmatrix}   \,.
\end{align}
These do not generate the full $SL(2, \mathbb{Z})$ but instead a congruence subgroup given by $\Gamma(2)$.\footnote{Here we do not dwell on the distinctions between $\Gamma(2) \subset SL(2,\mathbb{Z})$ and $P\Gamma(2) \subset PSL(2,\mathbb{Z})$.}

\begin{figure}[t!]
\centering
\includegraphics[width=0.6\textwidth]{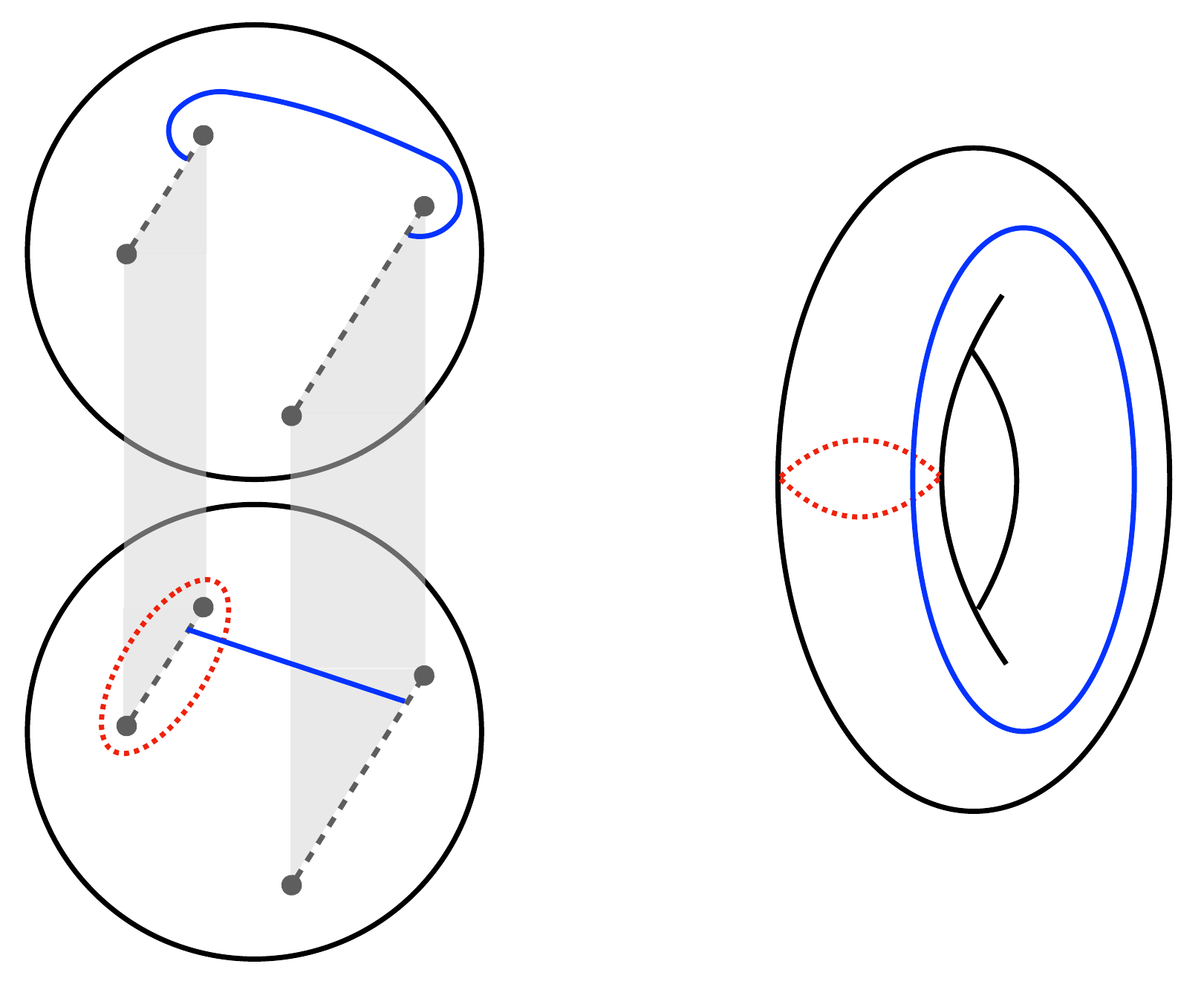}
\caption{Schematic description of the torus as double cover of $\mathbb{CP}^1$.}
\label{fig:doublecov}
\end{figure}

Instead of using the usual Weierstrass form one can also describe the Seiberg-Witten curve in terms of a branched double cover of $\mathbb{CP}^1$, parameterized by the complex coordinate $z$. For a schematic description of the relation between the torus and the double cover of $\mathbb{CP}^1$, see figure \ref{fig:doublecov}.
One possible parametrization is given in \cite{Tachikawa:2013kta} and reads as:
\begin{align}
\Lambda^2 z + \frac{\Lambda^2}{z} = x^2 - u \,.
\end{align}
In terms of these variables the Seiberg-Witten differential reads
\begin{align}
\lambda = x \frac{dz}{z} \,.
\end{align}
The UV curve is given by the $\mathbb{CP}^1$ in combination with the four branch points connected by two branch cuts.

\begin{figure}[t!]
\centering
\includegraphics[width=0.5\textwidth]{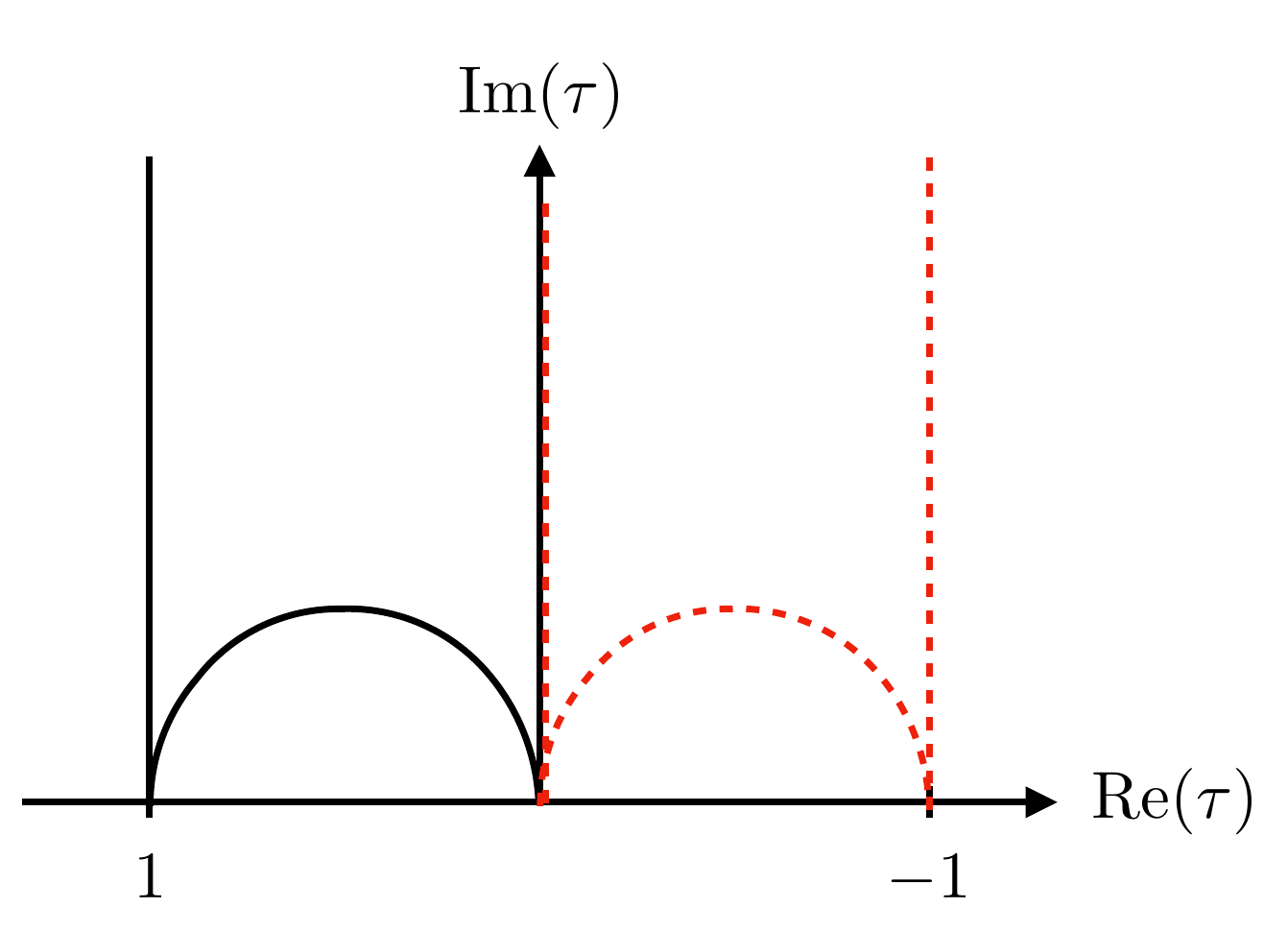}
\caption{Fundamental domain of $\Gamma (2)$ on the upper half plane as well as its time-reversal invariant subset $X(2)_{\mathbb{R}}$.}
\label{fig:G2phases}
\end{figure}

The pure gauge theory describes an elliptic curve, with moduli space given by $X(2)$. The fundamental domain as well as its time-reversal invariant subset are depicted in figure \ref{fig:G2phases}.
It contains three distinct cusps at $\tau \in \{ 0, 1, i \infty \}$ and is topologically a $\mathbb{CP}^1$ with the cusps marking three points. In this case the time-reversal invariant subset $X(2)_{\mathbb{R}}$ contains all three cusps.

Let us see what the three cusps correspond to in terms of data extracted from
the Seiberg-Witten curve. The equivalent of the $j$-function in the case
of $\Gamma = \Gamma(2)$ is its so-called Hauptmodul, defined by
\begin{align}
\lambda (\tau) = \bigg( \frac{\Theta_2 (\tau)}{\Theta_3 (\tau)} \bigg)^4 \,,
\end{align}
where the $\Theta$'s denote theta functions, the explicit form of which we will not need.
This yields a map $\lambda: X(2) \rightarrow \mathbb{CP}^1$. The values at the cusps are
\begin{align}
\lambda (0) = 1 \,, \quad \lambda (1) = \infty \,, \quad \lambda (i \infty) = 0 \,.
\end{align}
Taking the original form of the Seiberg-Witten curve, we expect cusps at the collision of two of the branch points, i.e.\
\begin{align}
u = \Lambda^2 \,, \quad u = - \Lambda^2 \,, \quad u \rightarrow \infty \,.
\end{align}
For the two strongly coupled cusps at $u = \pm \Lambda^2$, which are associated to $\tau = 0$ and $\tau = 1$, we know that we get either a massless monopole or dyon.

Next, we assume a suitable superpotential deformation has been switched on which produces a domain wall solution with multiple
kinks which passes through the different cusps. Our expectation is that the wall will now carry a charge as dictated by the sort of cusp encountered. The cusp at weak coupling corresponds to $u \rightarrow \infty$ and at first poses a puzzle. In the limit of large $u$ the theory becomes classical and one has the identification $a \sim \sqrt{u}$. Therefore, the $\mathfrak{su}(2)$ gauge algebra is broken to $U(1)$ at a very high scale and the supermultiplets containing the electrically charged $W$-bosons are very massive with
\begin{align}
m_{W} \sim a \rightarrow \infty \,.
\end{align}
Therefore, even though there is a cusp, one naively does not expect any light modes. That being said, building an interface that is very thin
relative to the mass scale, the corresponding energy scales are very high and the classical description in terms of a weakly coupled $\mathfrak{su}(2)$ gauge theory remains valid throughout the system. In this sense there actually are massless $W$ bosons and the $\mathfrak{su}(2)$
is restored.

Assuming the presence of light electric states of charge $q_e$ on the interfaces associated to the cusp at $\tau \rightarrow i \infty$, we can use coset representatives in order to investigate the other cusps at strong coupling. For this we choose
\begin{equation}
\begin{split}
\alpha_1 = \begin{pmatrix} 0 & -1 \\ 1 & 0 \end{pmatrix}:& \quad \tau = i \infty \enspace \mapsto \enspace \tau = 0 \,, \\
\alpha_2 = \begin{pmatrix} 1 & -1 \\ 1 & 0 \end{pmatrix}:& \quad \tau = i \infty \enspace \mapsto \enspace \tau = 1 \,.
\end{split}
\end{equation}
Then we can find the action on the charges of states as:
\begin{align}
\alpha_1 \begin{pmatrix} q \\ 0 \end{pmatrix} = \begin{pmatrix} 0 \\ q \end{pmatrix} \,, \quad \alpha_2 \begin{pmatrix} q \\ 0 \end{pmatrix} = \begin{pmatrix} q \\ q \end{pmatrix} \,,
\end{align}
which suggests the presence of massless purely magnetically charged and dyonic states, respectively. These are exactly the states associated to the monopole and dyon point for the pure gauge Seiberg-Witten theory! This can be precisely matched to the behavior of the elliptic $\lambda$-function in terms of the three branch points
\begin{align}
\lambda = \frac{2 \Lambda^2}{u + \Lambda^2} \,.
\end{align}
For $u \rightarrow \Lambda^2$, which is the monopole point one obtains $\lambda = 1$ which corresponds to $\tau = 0$. Similarly, for $u \rightarrow - \Lambda^2$, the dyon point, one has $\lambda \rightarrow  \infty$, i.e.\ $\tau = 1$.

\section{Examples via Compactification} \label{sec:6DCOMPACTIFY}

In this section we present a construction of 4D $U(1)$ gauge theories
with duality groups $\Gamma = \Gamma_{0}(N), \Gamma_{1}(N), \Gamma(N)$
by compactifying the theory of an anti-chiral two-form in six spacetime dimensions.
We view this theory as an edge mode coupled to a bulk 7D Chern-Simons theory.
This provides us with a geometric way to visualize much of the structure associated with
the spectrum of states and line operators in these 4D theories.

Using this, we can build 3D interfaces by just taking this 6D theory and compactifying on a three-manifold $M_{3}$ given by a family of elliptic curves fibered over the line $\mathbb{R}_{\bot}$ of the 4D spacetime $\mathbb{R}^{2,1} \times \mathbb{R}_{\bot}$. In this picture, singularities of the fibration indicate the locations of 3D interfaces.

This section is organized as follows. We begin by discussing the spectrum of charged states and line operators
for the different choices of duality groups. Much of this discussion follows what is presented in
reference \cite{Aharony:2013hda}. After this, we turn to the realization of this structure
via compactification of an anti-chiral two-form. In particular, we show that the level of the
associated 7D Chern-Simons theory provides a general way to control the set of possible duality groups.

\subsection{Line Operators and Charges \label{sec:lineops}}

A $U(1)$ gauge group is always specified together with a charge quantization condition. This quantization condition is not necessarily correlated with the presence of dynamical degrees of freedom with the corresponding charges. Instead it can be described by the set of genuine line operators.

For an abelian $U(1)$ gauge theory without any charged particles this defines a lattice of charges which are mutually local, i.e.\ they are consistent with the Dirac quantization condition, that enters in the definition of a general line operator. An electric line operator is given by
\begin{align}
\mathcal{O}^{(q_e , 0)}_{L} = \exp \Big( i q_e \underset{L}{\int} A \Big) \,,
\end{align}
where $A$ denotes the electric gauge field, and $L$ denotes a line in the 4D spacetime to integrate over.
The corresponding purely magnetically charged line operator can be given in terms of the dual gauge field $A_D$, and reads:
\begin{align}
\mathcal{O}^{(0, q_m)}_{L} = \exp \Big( - i q_m \underset{L}{\int} A_D \Big) \,.
\end{align}
In general, one can also define dyonic line operators $\mathcal{O}^{(q_e,q_m)}_{L}$, that carry both electric and magnetic charges.
For consistency, $q_e$ and $q_m$ have to be in the charge lattice defined by Dirac quantization. Moreover, these operators are charged with respect to global one-form symmetries \cite{Banks:2010zn, Gaiotto:2014kfa}. In the case of pure $U(1)$ gauge theory there are two global $U(1)$ one-form symmetries. The electric one-form symmetry acts by shifting $A$ by a flat $U(1)$ connection, the magnetic one acts accordingly on the dual gauge field $A_D$.

In the presence of dynamical charges the one-form symmetries are broken explicitly. However, if the dynamical charges only fill out a sublattice of the allowed charge lattice, discrete one-form symmetries remain. One example which will be relevant in the following is the case where the dynamical charges are of the form
\begin{align}
(q_e, q_m)_{\text{dyn}} = (N k, l) \,, \enspace \text{with} \enspace k,l \in \mathbb{Z} \,,
\end{align}
where without loss of generality we normalized the charges in a way that the full charge lattice is given by $\mathbb{Z} \times \mathbb{Z}$, i.e.\ integer charges. In this case the full magnetic one-form symmetry is broken. The electric one-form symmetry is only broken to a discrete subgroup, namely $\mathbb{Z} / N \mathbb{Z}$, with the charge carried by the line operators
\begin{align}
\mathcal{O}_L^{(r,0)} = \exp \Big( i r \underset{L}{\int} A \Big) \,, \enspace \text{with} \enspace r \in \{ 1, \dots, N - 1 \} \,.
\label{eq:ellinetors}
\end{align}
Note that line operators of the form discussed are objects in the theory which are also present at very low energies. The same is not necessarily true for dynamical charged particles, which can be integrated out below their mass scale.

On general grounds, the line operators transform non-trivially under duality, so to fully specify the action of the duality group we need to take this into account. To present explicit examples associated with different duality groups, we now turn to a 6D realization of these structures, starting first with $SL(2,\mathbb{Z})$.

\subsection{Geometrizing Duality}

One way of making this connection between line operators, charged states, and the congruence subgroups more apparent is to describe the $U(1)$ theory as a compactification of an anti-chiral two-form potential $B$ compactified on a torus, see e.g.\ \cite{Tachikawa:2013hya, Lawrie:2018jut, Eckhard:2019jgg, Garcia-Etxebarria:2019cnb}. At a classical level, we
can think of this as being specified by a three-form field strength $H$ subject to the condition:
\begin{equation} \label{6Dselfduality}
\ast_{6D} H = - H.
\end{equation}
The two-form potential couples to anti-chiral strings via integration of the pull-back of $B$ to the worldsheet of the string. It is well-known that the compactification of this theory on a $T^2$ produces a $U(1)$ gauge theory with complexified gauge coupling $\tau$ controlled by the complex structure of the $T^2$. Letting $\gamma_A$ and $\gamma_B$ denote the A- and B-cycles of this $T^2$, we observe that wrapping a string on the one-cycle $q_e \gamma_A + q_m \gamma_B$ results in a 4D point particle of electric and magnetic charge $(q_e,q_m)$. The celebrated S-duality of Maxwell theory corresponds to interchanging the A- and B-cycles of this
torus.

We would like to understand the structure of line operators and dynamical operators in the associated quantum theory.
To give a proper account, we of course need to quantize this 6D theory. This is somewhat subtle because the self-duality condition of equation \eqref{6Dselfduality} clashes with the condition that such fluxes should be quantized. As noted in \cite{Witten:1998wy, Belov:2006jd, Monnier:2017klz, Heckman:2017uxe}, the proper way to handle this sort of situation is to view the 6D theory as an edge mode coupled to a 7D Chern-Simons theory with three-form potential $C$ and action:
\begin{equation}
S_{7D} = \frac{k}{4 \pi i} \underset{M_7}{\int} C \wedge dC.
\end{equation}
with $M_7$ a seven-manifold with 6D boundary $M_{6} = \partial M_{7}$, e.g.\ \cite{Hsieh:2020jpj}. There are some subtleties in fully defining this 7D theory. For example, the analog of spin structure for a 3D Chern-Simons theory involves specifying a Wu structure (see e.g. \cite{Monnier:2017klz, Monnier:2018nfs}). Since we will primarily work on spaces with no metric curvature, most of these issues have little impact on the general statements we make. The boundary condition for the three-form potential is:
\begin{equation}
C \vert_{\partial M_7} = - \ast_{6D} C \vert_{\partial M_7}.
\end{equation}
This is the analog of the same condition one would impose for a bulk 3D Chern-Simons theory coupled to a chiral boson. In this bulk 7D theory we have a three-form potential, so our system couples to two-branes. Given a three-chain which ends on a two-cycle in the 6D spacetime, we obtain a two-dimensional string of the 6D theory. Much as in 3D Chern-Simons theory, the level $k \in \mathbb{Z}$ must be quantized. This is just to ensure that the phase factor $\exp(iS)$ remains well-defined under large gauge transformations of the three-form potential.

The analog of a line operator in this setting is specified by integrating the three-form potential over a three-chain.
Calling such a three-chain $\Sigma$, these operators take the form:
\begin{equation}
\mathcal{O}^{Q}_{\Sigma} = \exp \Big( i Q \underset{\Sigma}{\int} C \Big).
\end{equation}
If we were to quantize this theory with ``time'' indicated by the direction perpendicular to a 6D Euclidean slice, we
would obtain a non-trivial braid relation between these operators (see e.g. \cite{Witten:1998wy, DelZotto:2015isa}) given by:
\begin{equation}
\mathcal{O}^{Q}_{\Sigma} \mathcal{O}^{Q^{\prime}}_{\Sigma^{\prime}} = \exp \left(\frac{2 \pi i}{k} Q Q^{\prime} \Sigma \cdot \Sigma^{\prime} \right)\mathcal{O}^{Q^{\prime}}_{\Sigma^{\prime}} \mathcal{O}^{Q}_{\Sigma}.
\label{eq:corrfunc}
\end{equation}
In the case where the 6D slice is instead Lorentzian, this this fixes a Dirac pairing between strings
of the 6D theory \cite{Deser:1997se}. This Dirac pairing descends to the expected one in 4D.
Now, the important point for us is that we are interested in the spectrum of
line operators which commute, namely those which have integer valued Dirac pairing.
The main thing we will need to track is the level $k$ of the anti-chiral two-form $B$.

Let us now turn to the compactification of a level $k$ anti-chiral two-form on an elliptic curve $E$ with complex structure
$\tau$.  We will be interested in the
periods of the $B$-field on a two-cycle of the 6D spacetime $\mathbb{R}^{3,1} \times E$ of the form:
\begin{equation}
L \times (q_e \gamma_A + q_m \gamma_B).
\end{equation}
First of all, we see that the intersection pairing from the closed path on the elliptic curve amounts to the Dirac pairing which is invariant with respect to $SL(2, \mathbb{Z})$ transformations. Moreover, correlation functions are only sensitive to charges $(q_e,q_m)$ modulo $k \mathbb{Z}$. This naturally draws a connection to the classification of congruence subgroups acting in a particular way on operators specified by their electric and magnetic charges modulo $\mathbb{Z} / k \mathbb{Z} \times \mathbb{Z} / k \mathbb{Z}$, which we want to explain next.

First, let $k = N^2$ be a square of an integer $N$. Then one possible solution to the constraint that two genuine line operators have to commute is given by
\begin{align}
q_e \in N \mathbb{Z} + \tfrac{1}{N} m r \,, \quad q_m \in N \mathbb{Z} \,, \quad \text{with} \enspace r \in  \{ 0, 1, \dots, N-1 \}
\end{align}
which fills out a $\mathbb{Z} / N \mathbb{Z} \times \mathbb{Z} / N \mathbb{Z}$, a subset of $\mathbb{Z} / k \mathbb{Z} \times \mathbb{Z} / k \mathbb{Z}$.
\begin{figure}[t!]
\centering
\includegraphics[width=0.3\textwidth]{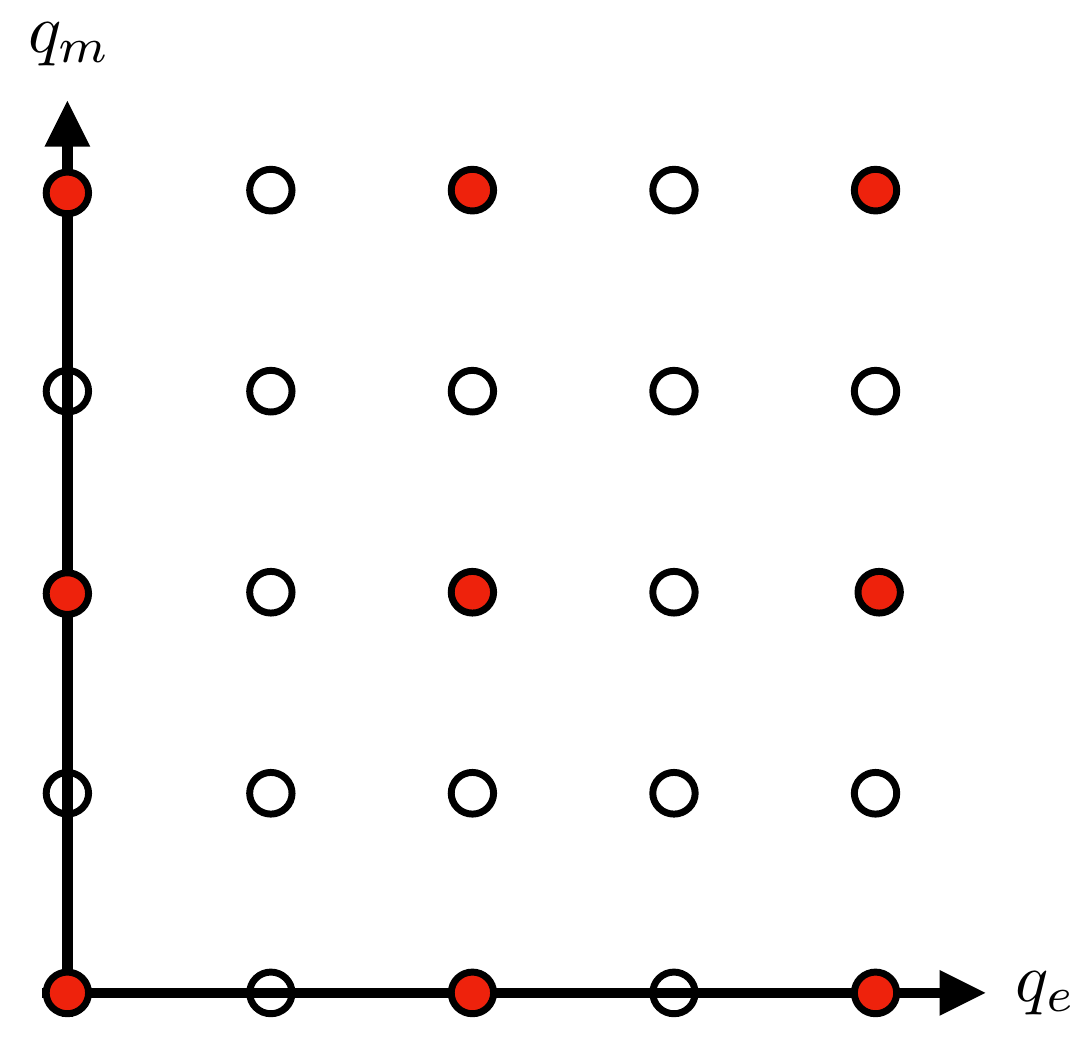}
\caption{Possible sublattice of commuting dynamical charges for $k = N^2 = 4$, corresponding to the case $\Gamma = \Gamma(2)$.}
\label{fig:chlatG4}
\end{figure}
Further demanding that $N$ times the charge has to be a trivial charge in $\mathbb{Z} / k \mathbb{Z} \times \mathbb{Z} / k \mathbb{Z}$ fixes $r$ to zero and one obtains the sublattice depicted in figure \ref{fig:chlatG4} for $k = N^2 = 4$. The charges of the genuine line operator are therefore labeled by elements of $\mathbb{Z} / N \mathbb{Z} \times \mathbb{Z} / N \mathbb{Z}$. Restricting the duality group to a subgroup keeping these operators invariant mod $k$ will lead to the congruence subgroup defined by $\Gamma (N)$.

For general $k$ such a sublattice is not accessible, but one always can define the charges to satisfy $q_e \in \mathbb{Z}$ and $q_m \in k \mathbb{Z}$, which naturally lead to a maximal set of charges with mutually local line operators.
\begin{figure}[t!]
\centering
\includegraphics[width=\textwidth]{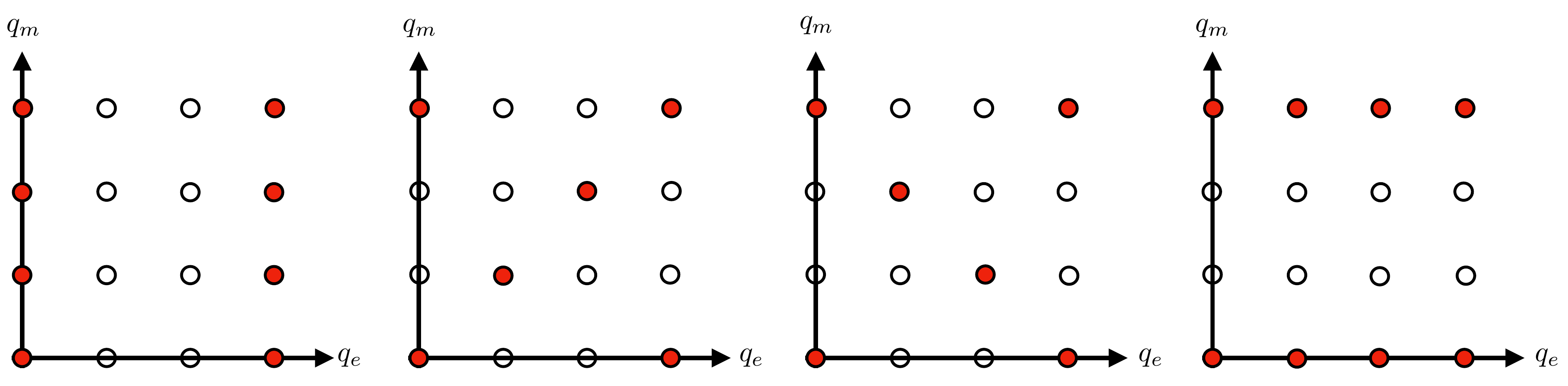}
\caption{Possible spectrum of genuine line operators for $k = 3$. Here, the duality group is taken to be either $\Gamma_0(3)$ or $\Gamma_1(3)$. In the case of $\Gamma_1(3)$, a torsional point (and its multiples) is fixed, while in the case of $\Gamma_0(3)$ only the zero element is fixed.}
\label{fig:chlatG1_3}
\end{figure}
Since the Dirac pairing is invariant with respect to the action of $SL(2, \mathbb{Z})$ one can also use the transformed spectrum of charges. In figure \ref{fig:chlatG1_3} we show the different possible choices for $k = 3$. Demanding invariance of the chosen spectrum of genuine line operators under the duality group then leads to the congruence subgroups $\Gamma_1 (k)$ and $\Gamma_0 (k)$, or a conjugate by a coset representative. In the case of $\Gamma_1 (k)$ one requires the invariance of each line operator individually. In the case of $\Gamma_0 (k)$ one allows an action on the line operators keeping the full spectrum fixed.

These congruence subgroups in connection with a specification of line operators also appear in the context of non-abelian gauge symmetries. There, the line operators specify the explicit realization of the gauge group as opposed to the gauge algebra \cite{Aharony:2013hda, Gaiotto:2014kfa, Garcia-Etxebarria:2019cnb}. In these cases the one-form symmetry is related to the center of the gauge group and mixed anomalies with time-reversal invariance can lead to interesting insights concerning the phase structure of four-dimensional theories as well as their possible interfaces \cite{Gaiotto:2017tne, Gaiotto:2017yup, Hsin:2018vcg}.

\subsection{The Jacobian Curve}

There is also a close connection with the Jacobian of the elliptic curve given as:
\begin{align}
\mathcal{J} (E) = H^1 (E, \mathbb{R}) / H^1 (E, \mathbb{Z}) \simeq \widetilde{E}
\end{align}
which itself is an elliptic curve with the origin defined as the vanishing gauge field. In physical terms, the Jacobian
specifies non-trivial flat fields on the torus $E$. In fact, the complex structure of this elliptic curve as specified by a parameter $\widetilde{\tau}$ is determined by the complex structure $\tau$ of the elliptic curve $E$; they are in fact the same.

With the basis of $H^1 (E, \mathbb{Z})$ given by $\{ \sigma_A, \sigma_B \}$ defining the lattice of $\widetilde{E}$, the relevant forms are given by $\alpha \sigma_A + \beta \sigma_B$, with $\alpha, \beta \in [0,1)$. Now we can specify the subset of $\mathcal{J}(E)$ which is trivial on the physical states, by which we mean that
\begin{align}
\underset{q_e \gamma_A + q_m \gamma_B}{\int}( \alpha \sigma_A + \beta \sigma_B ) \in \mathbb{Z} \,.
\end{align}
The structure specified by the level of the anti-chiral two-form
thus determines a corresponding level in the elliptic curve $\widetilde{E}$. This level structure is associated with the appearance of
torsional points in $\widetilde{E}$. Recall that these are obtained by viewing the curve $\widetilde{E} = \mathbb{C} / \widetilde{\Lambda}$ as a group. An $N$-torsional point $P$ in this group is one for which $N[P]$ is just the zero element of this additive group. In terms of the lattice $\widetilde{\Lambda} = \widetilde{\omega}^1 \mathbb{Z} \oplus \widetilde{\omega}^2 \mathbb{Z} \subset \mathbb{C}$, these $N$-torsion points can be written as:
\begin{equation}
\widetilde{E}(N) = a \frac{\widetilde{\omega}^1}{N} + b \frac{\widetilde{\omega}^2}{N} \,\,\, \text{for} \,\,\, a,b = 0,...,N-1.
\end{equation}

For the example above $(q_e,q_m) \in N \mathbb{Z} \times \mathbb{Z}$ this is given by the elements
\begin{align}
\big\{ \big( \tfrac{r}{N} + k \big) \sigma_A + l \sigma_B \big\} \,, \quad \text{with \ } k,l \in \mathbb{Z} \enspace \text{and} \enspace r \in \{0, 1, \dots, N-1 \} \,.
\end{align}
We see that up to lattice vectors this defines a set of $N$-torsion points on the Jacobian $\widetilde{E}$. In general, one can get the full set of $N$-torsion points by demanding that a dynamical state has charge $(q_e,q_m) \in N \mathbb{Z} \times N \mathbb{Z}$. An $SL(2,\mathbb{Z})$ action on the line operators can then be perceived as an action on the torsion points in the dual curve $\widetilde{E}$.

Invariance of (a subset of) the spectrum of line operators therefore restricts the duality group to a subgroup of $SL(2, \mathbb{Z})$.
One way to think about this is to start with the original lattice of electric and magnetic charges $\Lambda$, along with the corresponding elliptic curve $\widetilde{E}$. We can consider a non-zero holomorphic map to another complex torus $\widetilde{E^{\prime}}$ along with its corresponding defining lattice $\Lambda^{\prime}$. Such mappings are known as isogenies and in general correspond to either rescalings of the original lattice via the multiplication map $\Lambda \rightarrow N \Lambda$ or involve picking an order $N$ cyclic subgroup $C \subset \widetilde{E}[N] = \mathbb{Z} / N\mathbb{Z} \times \mathbb{Z} / N\mathbb{Z}$  and constructing a new lattice out of the cosets. All isogenies can be obtained from these two basic operations (see e.g. \cite{diamond2006first}), and they serve to define different lattices of electric and magnetic charges. We now turn to the three congruence subgroups $\Gamma(N), \Gamma_1(N)$, and $\Gamma_0(N)$, which are obtained as follows.

\subsubsection*{$\mathbf{\Gamma (N)}$}

For the congruence subgroup $\Gamma (N)$ the full set of line operators classified by the lattice $\mathbb{Z} / N \mathbb{Z} \times \mathbb{Z} / N \mathbb{Z}$ remains invariant. In terms of the Jacobian, that means that the full set of torsion points in $\widetilde{E}(N)$ is invariant up to lattice vectors. Specifically, the line operators are given by
\begin{align}
\mathcal{O}_{L}^{(r,s)} = \exp \Big( i \underset{L \times (r \gamma_A + s \gamma_B)}{\int} B\Big) = \exp \Big( i r \underset{L}{\int} A - i s \underset{L}{\int} A_D \Big) \,, \quad \text{with} \enspace r,s \in \mathbb{Z} / N \mathbb{Z}\,,
\end{align}
which are invariant under $\Gamma (N)$ up to the addition of a worldline of a dynamical particle. In the four-dimensional description this is a theory with dynamical electric and magnetic charges that are a multiple of $N$.

\subsubsection*{$\mathbf{\Gamma}_1 (N)$}

For the congruence subgroup $\Gamma_1(N)$ we fix an $N$-torsion point of $\widetilde{E}(N)$. This leads to the invariance of a full $\mathbb{Z} / N \mathbb{Z}$ subgroup of $\widetilde{E}(N)$ by the linearity of the modular transformation. With the help of an $SL(2, \mathbb{Z})$ element which is not in $\Gamma_1 (N)$ we can always map this torsion point to be $\tfrac{1}{N} \sigma_A$. We see that $\Gamma_1 (N)$ leaves invariant the line operators defined by
\begin{align}
\mathcal{O}^{(r,0)} =  \exp \Big( i \underset{L \times (r \gamma_A)}{\int} B \Big) = \exp \Big( i r \underset{L}{\int} A \Big) \,, \quad \text{with} \enspace r \in \mathbb{Z} / N \mathbb{Z}\,.
\label{eq:G1lines}
\end{align}
In the compactified theory this means that only dynamical electric charges which are a multiple of $N$ are present.
There can be other realizations of this choice which differ by the action of a coset representative.

\subsubsection*{$\mathbf{\Gamma}_0 (N)$}

Finally, in $\Gamma_0 (N)$ one has a set of elements generating a $\mathbb{Z} / N \mathbb{Z}$ subgroup of $\widetilde{E}(N)$ which stays invariant. The individual elements, however, can be transformed among each other. Again, we can use a coset representative in order to map the $\mathbb{Z} / N \mathbb{Z}$ subgroup to $\big\{ \tfrac{r}{N} \sigma_A \big\}$, which translates to the same line operators as in \eqref{eq:G1lines}. The transformation of the individual elements among each other defines an action on the line operators. For example if $\gamma \in \Gamma_0 (N)$ acts as
\begin{align}
\tfrac{r}{N} \sigma_A \mapsto \tfrac{r'}{N} \sigma_A \,,
\end{align}
up to lattice vectors, the induced action on the line operators reads
\begin{align}
\mathcal{O}_L^{(r, 0)} \rightarrow \mathcal{O}_L^{(r', 0)} \,.
\end{align}
In the four-dimensional effective action, we see that $\Gamma_0 (N)$, describes a theory with dynamical electric charges being a multiple of $N$ together with an action on the line operators $\mathcal{O}_{L}^{(r,0)}$.

\subsection{Generalization to Other Riemann Surfaces}
\label{subsec:genRiemann}

The generalization to higher-genus Riemann surfaces is straightforward from what we said above. Compactifying a 6D anti-selfdual tensor on a genus $g$ Riemann surface $C_g$ leads to $g$ abelian $U(1)$ gauge fields in four dimensions. Whereas the mapping class group of higher-genus realizations is highly complicated and these surfaces do not have a generic way to add points, the interpretation using the Jacobian is still applicable. The Jacobian of the Riemann surface is:
\begin{align}
\mathcal{J} (C_g) = H^1 (C_g, \mathbb{R}) / H^1 (C_g, \mathbb{Z}) \simeq \widetilde{T}^{2g} \,,
\end{align}
and on the torus $\widetilde{T}^{2g}$ we can define $N$-torsion elements as harmonic one-forms with
\begin{align}
N \sigma \in H^1 (\Sigma_g, \mathbb{Z}) \,,
\end{align}
which we denote by $\mathcal{J}_N (C_g)$. For the case of $C_g = E$ this lead to the identification of the congruence subgroups of $SL(2,\mathbb{Z})$ via the action on the torsion elements in $\widetilde{T}^2 = \widetilde{E}$.

For a general Riemann surface we can restrict the actions of the duality group, i.e.\ the mapping class group in such a way that the integral over a basis of one-cycles for all or a subset of torsion elements modulo $N$ has a well-defined behavior. It either remains fixed or it allows for an action on the set of torsion elements. Since now the set  of torsion elements in $\mathcal{J}_N(C_g)$ are defined by $( \mathbb{Z} / N \mathbb{Z} )^{2g}$ it is also conceivable that mixed version of the possibilities above are realized. For example, a certain $\mathbb{Z} / N \mathbb{Z}$ subgroup can be held fixed element by element and another subgroup might be held fixed up to an action on the individual elements. This leads to a generalization of congruence subgroups in the context of the mapping class groups of higher genus Riemann surfaces.

\section{More General Interfaces at Strong Coupling} \label{sec:6DGENERAL}

In the previous sections we used time-reversal invariance in 4D $U(1)$ gauge theories to produce examples of 3D interfaces
at strong coupling, and we also presented some explicit examples realizing these features.

A common theme in these constructions is the appearance of a six-dimensional field theory. In the case of the compactification of an anti-chiral
two-form, this is manifest from the start. In the case of our $\mathcal{N} = 2$ theories, this follows from the class $\mathcal{S}$ construction based on compactification of a 6D $\mathcal{N} = (2,0)$ superconformal field theory on a Riemann surface (see e.g. \cite{Witten:1997sc, Gaiotto:2009we}). In both these cases, the geometry of the interface can thus be understood in terms of compactification on a three-manifold with boundary, constructed from a family of Riemann surfaces fibered over the real line. Returning to the analysis of the previous sections, we have been considering singularities in the associated elliptic curve with real coefficients, deducing the appearance of localized matter from singular fibers. This method of construction relies heavily on the special features of time-reversal invariance, in tandem with the structure of congruence subgroups of $SL(2,\mathbb{Z})$.

In this section we present another method for generating interfaces at strong coupling. Instead of relying on the additional structure of time-reversal invariance we will instead consider compactification of higher-dimensional field theories on families of Riemann surfaces. The main theme here will be to identify the appearance of singularities in the associated fibers as a diagnostic for tracking the appearance of localized matter. We focus on the case of compactification of six-dimensional superconformal field theories on three-manifolds with boundary. There has recently been significant progress in understanding the construction and study of such 6D SCFTs (see e.g. \cite{Heckman:2013pva, Ohmori:2014kda, Heckman:2015bfa, Cordova:2015fha} and \cite{Heckman:2018jxk} for a recent review), and in particular the compactification of such theories to various lower-dimensional systems \cite{Morrison:2016nrt, Apruzzi:2016nfr, Razamat:2016dpl, Apruzzi:2018oge}. Notably, however, compactifications of 6D SCFTs on three-manifolds has mainly focussed on the special case of $\mathcal{N} = (2,0)$ theories as in references \cite{Cecotti:2011iy, Dimofte:2011py}. From this perspective, the present study provides a general starting point for building 3D field theories associated with the degrees of freedom localized on an interface.

The main idea will be to first consider a 4D $\mathcal{N} = 1$ theory as obtained from compactification of a 6D SCFT on a Riemann surface. This sort of compactification involves a choice of background metric on the Riemann surface, and can also be supplemented by switching on various flavor symmetry fluxes. All of these choices lead to a wide range of possible 4D theories. In many cases, these compactifications are expected
to produce a 4D $\mathcal{N} = 1$ SCFT \cite{Morrison:2016nrt, Razamat:2016dpl, Apruzzi:2018oge}, but there are also situations where such a compactification instead leads to a trivial fixed point in the IR (either fully gapped or with just free fields) \cite{Apruzzi:2018oge}. Assuming we can switch on some choice of background fields in the 6D theory, the 4D theory inherits some of its symmetries as well their anomalies from the 6D theory.

To build a 3D interface, we can next consider a family of Riemann surfaces, each equipped with a set of flavor symmetry fluxes. Fibering over a real line $\mathbb{R}_{\bot}$ we can vary both the metric and the fluxes. In fact, by allowing for singular fibrations and gauge field configurations, we can allow both the genus and the Chern classes of these fluxes to jump as we move along $\mathbb{R}_{\bot}$. This is problematic when viewed as a motion inside the moduli space of genus $g$ Riemann surfaces with $n$ marked points (such as $\overline{\mathcal{M}}_{g,n}$, the Deligne-Mumford compactification of the moduli space), but is not problematic when viewed in terms of the geometry of the total space. Indeed, we can construct an interface by gluing together piecewise constant profiles for the metric and fluxes such that when interpreted as a 4D theory, the anomalies are bigger in an interior region. We view this as building an interface with non-zero thickness. In the singular limit where the interior region degenerates to zero thickness, we have a sharp interface.

The rest of this section is organized as follows. First, we set up the relevant mathematical bordism problem and show that there are no obstructions to constructing an interpolating profile of the sort needed to build a thick interface. We then illustrate these considerations with a few examples. We consider the special case of a 6D hypermultiplet compactified on a three-manifold with boundary, and then turn to the more general structure of compactifications of interacting 6D SCFTs.

\subsection{Cobordism Considerations}

To construct more general examples of 3D interfaces, we now discuss the general cobordism problem for our compactification.
Consider $Q$ a cobordism between two Riemann surfaces $C^L$ and $C^R$. A cobordism always has the structure of a fibration\footnote{To suit our needs, what we refer to as a cobordism here is actually a noncompact manifold gotten by deleting the boundary components of a cobordism (which is a compact manifold with boundary) so that $C^R$ and $C^L$ lie ``at infinity". The fibration structure is usually presented in the math literature as being over [0,1], but we use $\mathbb{R}_t$ for our physical purposes.} over $\mathbb{R}_{\bot}$ where the fiber may become singular, change its topology, and have multiple components. This is equivalent to the well-known statement that there always exists a smooth Morse function, $f$, on a cobordism with $f^{-1}(-\infty)=C^L$ and $f^{-1}(+\infty)=C^R$, which induces a codimension-one foliation which is singular at the critical points of $f$ \cite{milnor}. Further, we choose a metric on $Q$ that is in the conformal class of a metric that gives the same volume to each of the Morse fibers. We emphasize that while the fibers may become singular at given values of $x_\bot$, the smoothness of the compactification manifold $Q$ suggests we should be careful about our expectation of localized states since this is merely a coordinate singularity.

To understand what happens, first note that the second oriented cobordism group, $\Omega_2^{SO}$, is trivial for the reason that we can take any oriented three-manifold and cut out two disjoint oriented Riemann surfaces of any genus out of it. The fibration structure will depend on a choice of Morse function and will in general consist of several jumps in the genus of the fiber along with the possibility of the fiber being a disjoint union of Riemann surfaces. To eliminate certain pathologies, we will assume that this Morse function saturates the Morse inequalities from now on, and our choice of three-manifolds will force the fiber to always be connected.

As a warmup let us take our three-manifold to be an $S^3$. If we then cut out two $S^2$'s this is topologically $S^2\times \mathbb{R}_{\bot}$, so the fibration structure in this case is clear. If we instead cut out two tori, then the fibers of the fibration will jump in the following manner along $\mathbb{R}_{\bot}$:
\begin{equation}
  g=1 \quad \vert \quad g=0 \quad \vert \quad g=1.
\end{equation}
To generate thickened 3D interfaces, we will actually be interested in situations where the genus is bigger in the interior. The reason is that as a rule of thumb, compactifications of 6D SCFTs on higher genus spaces tend to produce 4D theories with more degrees of freedom. With this in mind, the typical situation of interest will be:
\begin{equation}
  g_L \quad \vert \quad g_{\text{mid}} \quad \vert \quad g_R, \,\,\, \text{with} \,\,\, g_{L},g_{R} < g_{\text{mid}}.
\end{equation}

Focusing on the case where the genus increases inside the interface, we accomplish this by cutting out Riemann surfaces with genera $g_{L,R}$ out of the suspension\footnote{Given a topological space X, the suspension is defined as $\Sigma X \coloneqq X\times [0,1]/\{ (x,0)\sim (y,0) \; \text{and} \; (x,1)\sim (y,1)\}$. This has the important property that $\Sigma S^2 \simeq_{Top.} S^3$ and we note that while normally $\Sigma$ is called the reduced suspension by mathematicians, we favor this symbol here for aesthetic purposes.} of a Riemann surface $\Sigma C_{g_{\mathrm{mid}}}$ such that $g_L,g_R < g_{\text{mid}}$. The 3D theory living on the interface can be equivalently studied as either the compactification of a 6D SCFT on $\Sigma C_{g_\mathrm{mid}} \; \; \textnormal{with lower genus ``punctures''}$ or (from the fibration point-of-view) as the compactification of the 4D theory associated to $C_{g_{\text{mid}}}$ on an interval with appropriate boundary conditions.

\begin{figure}[t!]
\centering
\includegraphics[width=0.24\textwidth]{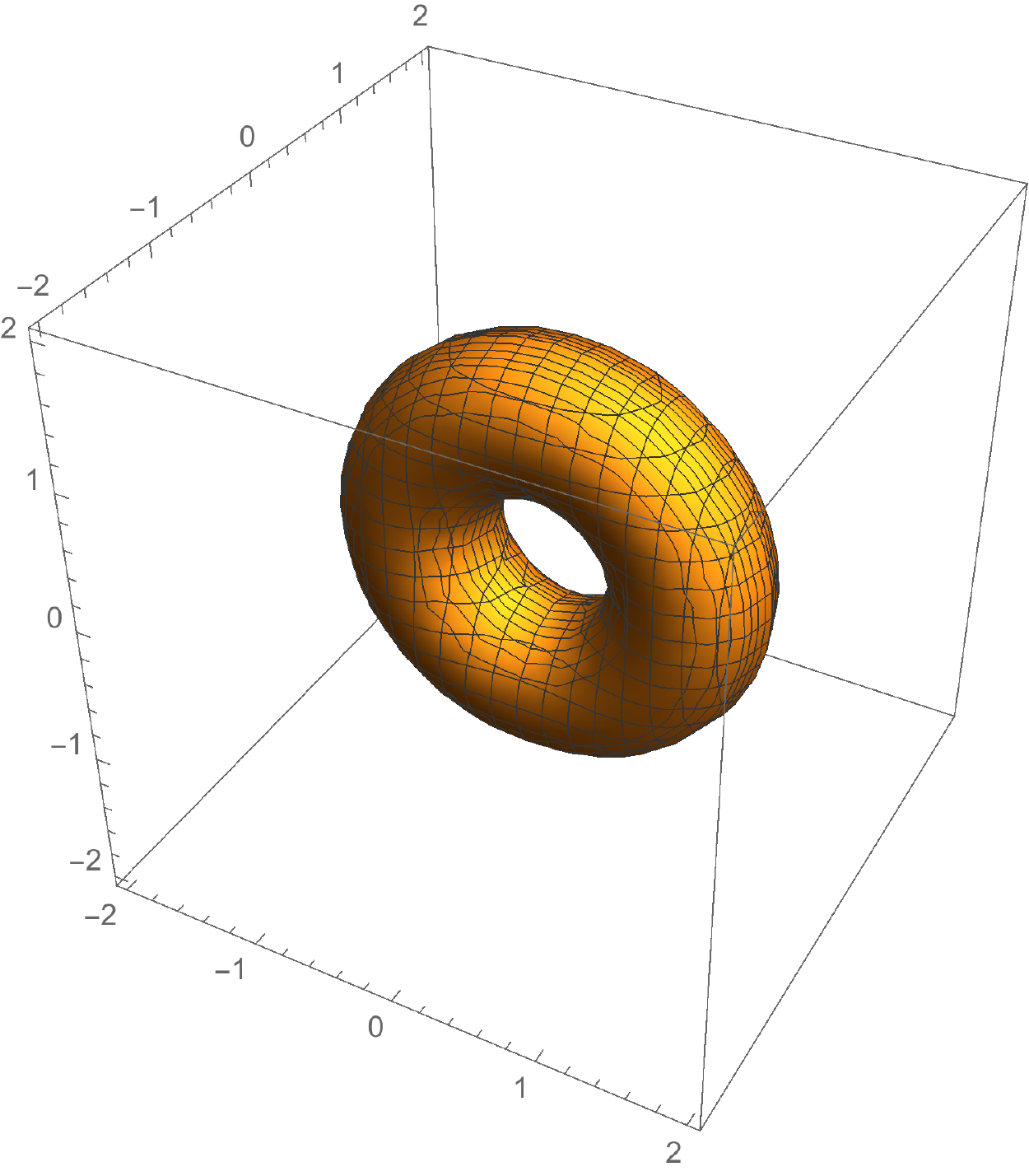}
\includegraphics[width=0.24\textwidth]{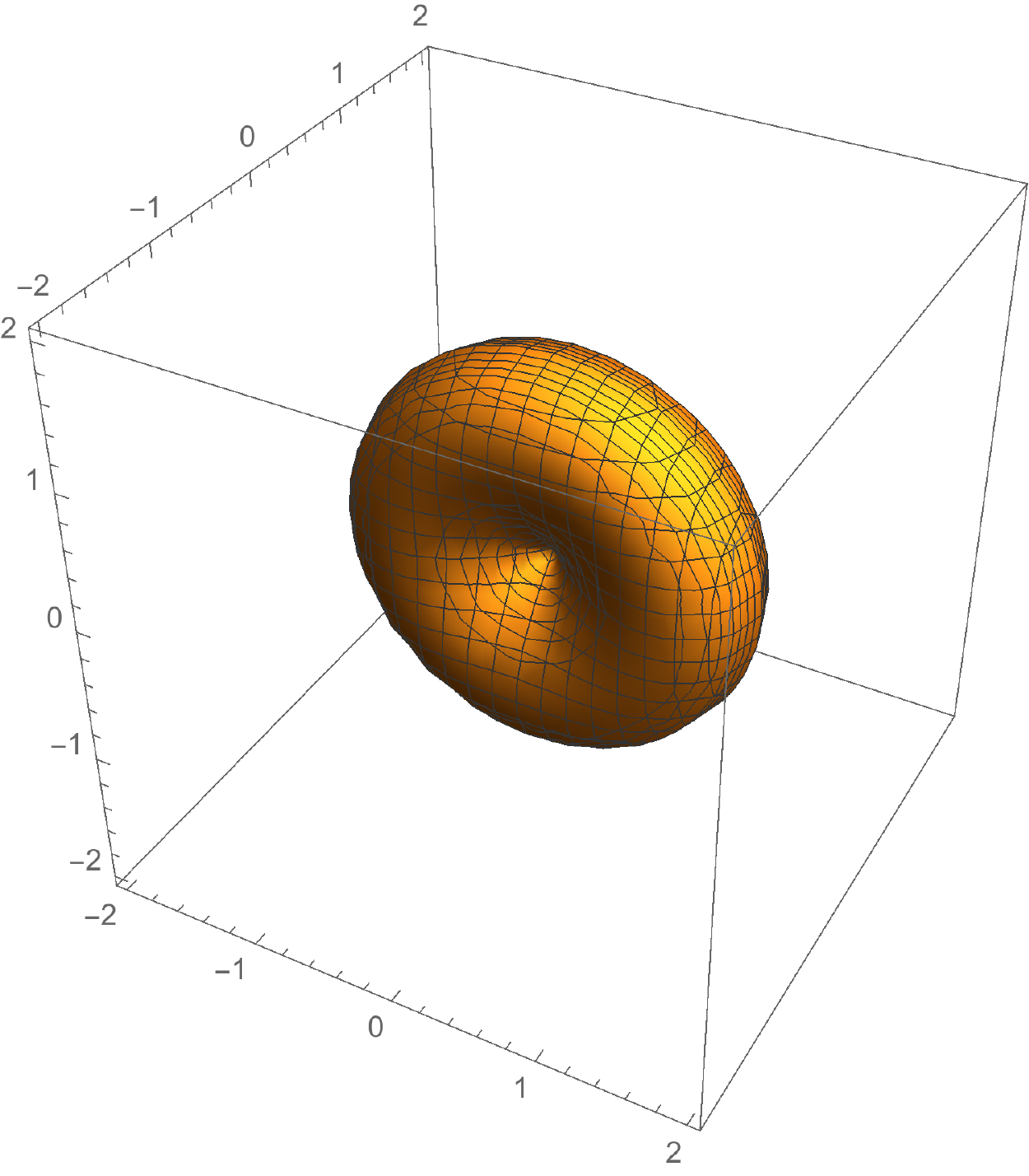}
\includegraphics[width=0.24\textwidth]{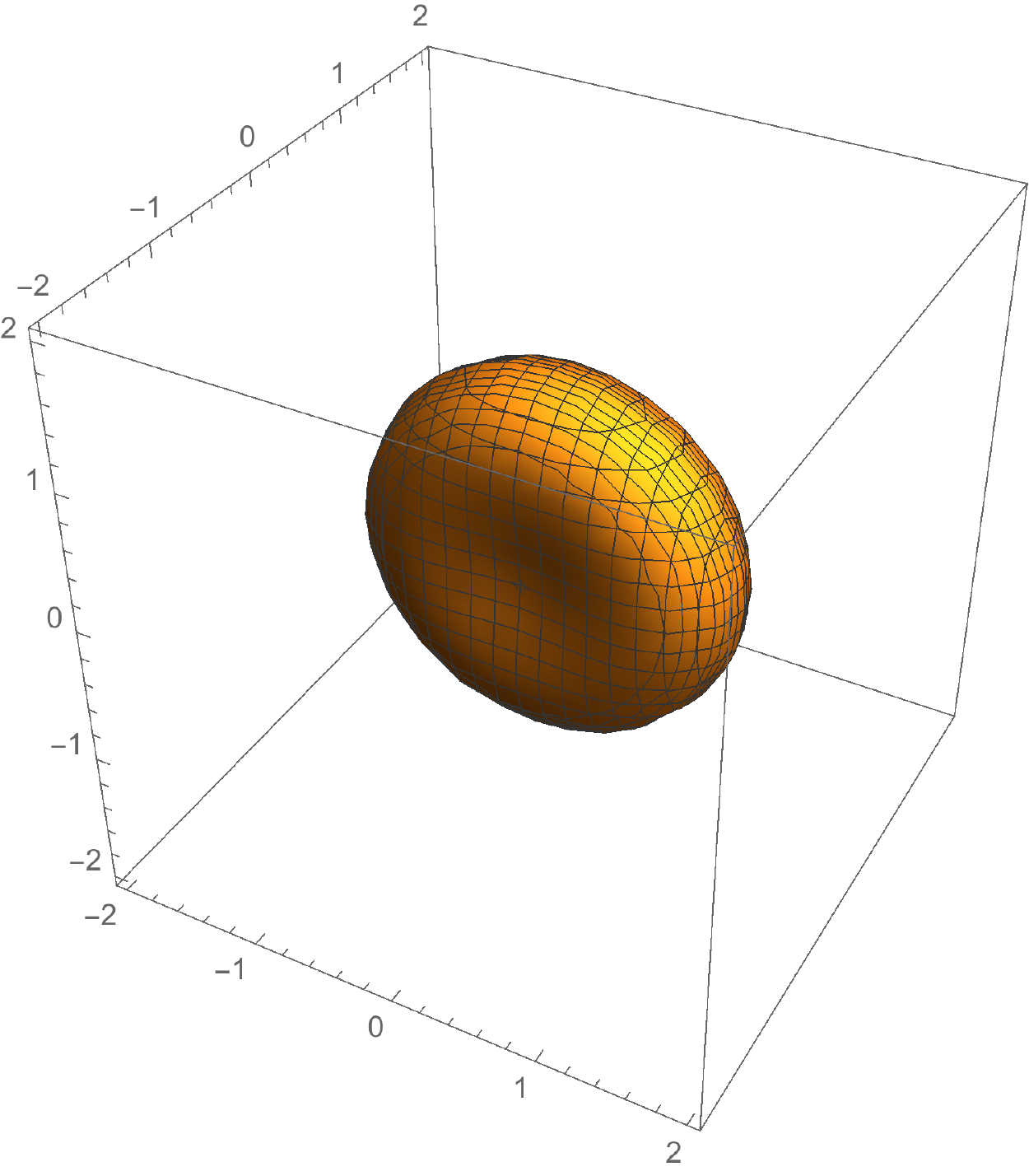}
\includegraphics[width=0.24\textwidth]{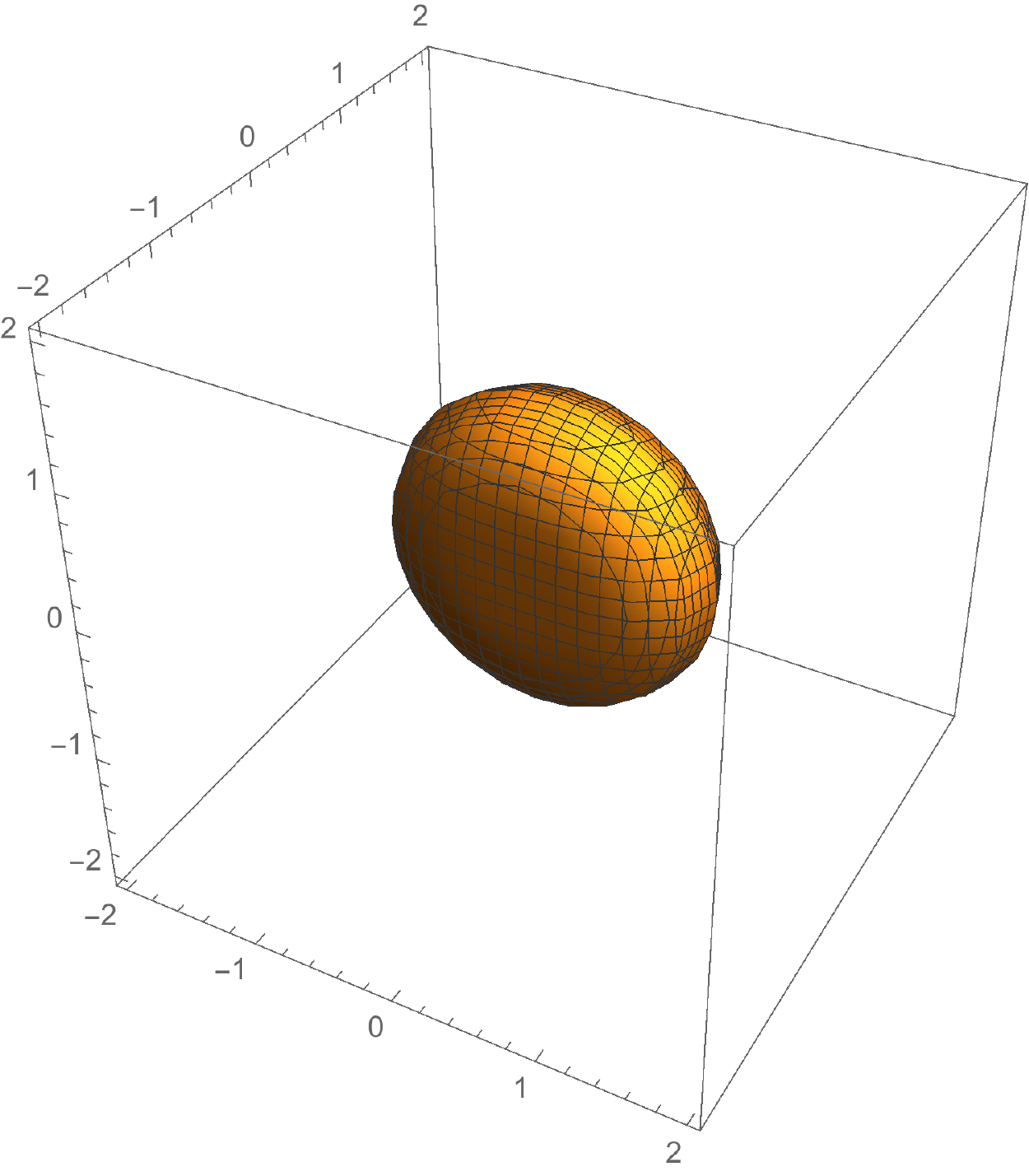}
\caption{Building a continuous family of Riemann surfaces with varying genus: we start on the left with a torus, which then fattens into a sphere. This construction can be extended to build more general interpolating profiles.}
\label{fig:interpolator}
\end{figure}

As an example of this parametrization of Riemann surfaces, we can define a family of tori given with parametrization variable $w$ as:\footnote{We thank R. Donagi for pointing out this construction to us.}
\begin{equation}
  Q = \{ (x^2+y^2+z^2+w^2+R^2-r^2)^2=4R^2(x^2+y^2+w^2) \},
\end{equation}\label{eq:interpolator}
where for $w=0$, $R$ and $r$ are the ``major'' and ``minor'' radii of the torus respectively. We then vary the parameter between $0$ and $R$, noting that at $w=R$ the Riemann surface described now turns into a two-sphere. This is illustrated in figure \ref{fig:interpolator} where we see a torus transform into a sphere as the parameter $w$ increases from $0$ to $R$. As a result, by compactifying on $Q$ with $w$ starting at $w<R$ (in the middle), and reaching $w=R$ as $|x_\bot|\rightarrow \infty$, we obtain families of 4D theories compactified on genus zero surfaces on the left and right, but compactified on a genus one surface in the middle, thus realizing two $S^2$'s cut out of $\Sigma T^2$. Note that once we transition to a genus zero Riemann surface, we can then consider further motion in the moduli space $\overline{\mathcal{M}}_{0,n}$. We can use this to also rotate the phases of ``mass parameters'' on the two sides of the thickened interface. Note that we can also extend this construction to produce interpolating profiles between different genus Riemann surfaces.

We can also consider interpolating profiles for flavor symmetry fluxes. The possibilities for the background gauge field that couples to the flavor current are: a non-trivial monodromy, a flux for an abelian portion, or a 't Hooft flux for a non-simply connected flavor group. We can build an interface that interpolates between any two pairs of monodromies since for the cobordism $Q=\Sigma C_g \backslash (C^R_{g_R}\sqcup C^L_{g_L})$, one is free to chose the monodromy around the cycles. Note also that these interfaces allow for the added possibility of monodromy associated only to the $\Sigma C_g$ cycles and not to either $C^R_{g_R}$ or $C^L_{g_L}$. For the flux cases, the relevant cobordism groups to look at are:
\begin{align}
 \Omega^{SO}_2(BU(1)) =\mathbb{Z} \; \; \; \; &(\textnormal{abelian flux})\\
 \Omega^{SO}_2(BG) = \pi_1 (G) \; \; \; \; &(\textnormal{'t Hooft flux})
\end{align}
where $G$ is the flavor group in question, and $BG$ denotes its classifying space. These express total abelian flavor and 't Hooft charge conversation and follow from an application of Stokes' theorem (along with the universal coefficient theorem for the 't Hooft case) to the cobordism with the assumptions $dF_{U(1)}=0$ and $\delta F_{\text{'t Hooft}}=0\in H^3(Q,\pi_1(G))$ (where here $\delta$ is the coboundary operator).

One can study more general interfaces by adding extra codimension-three defect operators with localized flux in the cobordism leading to the relation:
\begin{align}
\underset{C^L}{\int} c_1(F) & = \underset{C^R}{\int} c_1(F) + \text{monopoles} \\
\underset{C^L}{\int} w_2(F) & = \underset{C^R}{\int} w_2(F) + \text{twists}
\end{align}
where ``monopoles'' and ``twists'' refers to pointlike singular field configurations in the three-manifold.

\subsection{Hypermultiplet Example}

With these general considerations in place, we now turn to a concrete example of 6D hypermultiplets which, when suitably compactified, produces a 4D theory with a thickened 3D interface. This 6D theory arises from the theory of a single M5-brane probing an A-type singularity $\mathbb{C}^{2} / \mathbb{Z}_{k}$. Strictly speaking, this does not produce an interacting fixed point, but it will be adequate for the main ideas we wish to consider. In field theory terms, we have a theory of hypermultiplets in the bifundamental representation of $SU(k) \times SU(k)$.\footnote{The actual flavor symmetry in this case is $SU(2k)$} We will be interested in building an interpolating profile with modes trapped along a 3D interface. We review the case of a position dependent mass term for a Weyl fermion in Appendix \ref{app:WEYL}.

To begin, we consider the compactification of this theory on a genus $g$ Riemann surface $C$. We also consider switching on abelian fluxes in a subgroup $H \subset SU(2k)$ of the flavor symmetry. For ease of exposition, we concentrate on the case of a single $U(1)$ factor, and consider the mass spectrum for states of charge $\pm q$ under this $U(1)$ factor. We leave implicit the representation content under the commutant flavor symmetry. Letting $\mathcal{L}$ denote the line bundle associated with switching on this background flux, the zero mode content on the curve consists of 4D $\mathcal{N} = 1$ chiral multiplets of charge $+q$ and $-q$ under this $U(1)$. The 6D fermion obeys a Dirac equation of the form:
\begin{equation}
\Gamma_{6D} \cdot D_{6D} \Psi_{6D} = 0.
\end{equation}
We expand the 6D fermion in terms of a basis of 4D Weyl fermions and chiral modes on the curve $C$ via:
\begin{equation}
\Psi_{6D} = \sum_{a} \psi^{(a)}_{4D} \otimes \chi^{(a)}_{C}.
\end{equation}
The Dirac equation then takes the form:
\begin{equation}
(\gamma_{4D} \cdot D_{4D} + \gamma_{C} \cdot D_{C}) \sum_{a} \psi^{(a)}_{4D} \otimes \chi^{(a)}_{C} = 0.
\end{equation}
Consequently, the Dirac operator on $C$ controls the spectrum of zero modes and massive modes in the theory. More precisely,
in the expansion of $(\gamma_{C} \cdot D_{C})^{2}$, we see the appearance of the curvature in the spin connection and the gauge field flux.

The number of zero modes is controlled by the cohomology groups (see e.g. \cite{Beasley:2008dc}):
\begin{align}
\#_{+q} & = h^{0}(C, K_{C}^{1/2} \otimes \mathcal{L}^{+q})\\
\#_{-q} & = h^{0}(C, K_{C}^{1/2} \otimes \mathcal{L}^{-q}).
\end{align}
where here, $K_{C}$ denotes the canonical bundle and we need to specify a choice of spin structure, i.e. a choice of square root for $K_{C}$.

As an example, we can engineer a theory with no zero modes by considering the special case of $C$ a $\mathbb{CP}^1$ with $\mathcal{L} = \mathcal{O}$. We can view this as a situation in which all the modes of the 6D hypermultiplet have a Kaluza-Klein scale mass. As an example where we get a single chiral multiplet, we could consider switching on $\mathcal{L} = \mathcal{O}(1)$ on a $\mathbb{CP}^{1}$,
which includes a 4D Weyl fermion and a complex scalar, both of charge $+q$. Finally, we can also produce an example with a 4D Dirac fermion and its superpartners by compactifying on a $T^{2}$, with no fluxes switched on.

\subsection{Strongly Coupled Examples}

We now generalize the above considerations to consider compactifications of 6D SCFTs on three-manifolds with boundary. Our primary
interest will be in localizing states along a thickened 3D interface. To track the appearance of localized degrees of freedom, we consider the 4D anomaly polynomial obtained from compactification of a 6D theory on a curve $C$ with some background fluxes switched on. Recall that the general
form of the anomaly polynomial for a 6D SCFT takes the form:
\begin{align}
I_{8}  &  = \alpha c_{2}(R)^{2} + \beta c_{2}(R) p_{1}(T) + \gamma
p_{1}(T)^{2} + \delta p_{2}(T)\nonumber\\
&  + \sum_{i} \left[  \mu_{i} \, \mathrm{Tr} F_{i}^{4}
+ \, \mathrm{Tr} F_{i}^{2} \left(  \rho_{i}
p_{1}(T) + \sigma_{i} c_{2}(R) + \sum_{j} \eta_{ij} \, \mathrm{Tr} F_{j}^{2}
\right)  \right]  . \label{eq:anomalypoly}%
\end{align}
Here, $c_{2}(R)$ is the second Chern class of the $SU(2)_{R}$ symmetry,
$p_{1}(T)$ is the first Pontryagin class of the tangent bundle, $p_{2}(T)$ is
the second Pontryagin class of the tangent bundle, and $F_{i}$ is the field
strength of the $i$th symmetry, where the sum on $i$ and $j$ runs over the
global symmetries of the theory. In the case where we have sufficiently generic curvatures switched on, we can extract the anomalies of the 4D theory which are inherited from six dimensions by integrating this formal eight-form over a curve $C$ (see e.g. \cite{Benini:2009mz, Razamat:2016dpl, Apruzzi:2018oge}):
\begin{equation}
I_{6} = \underset{C}{\int} I_{8}.
\end{equation}
This, in tandem with $a$-maximization \cite{Intriligator:2003jj} makes it possible to extract the values of the conformal anomalies $a$ and $c$ (see e.g. \cite{Razamat:2016dpl, Apruzzi:2018oge}), which provides a crude ``count'' of the number of degrees of freedom in the 4D theory.

To generate examples of trapped matter, we can attempt to mimic our discussion of the 6D hypermultiplet. In particular, we can engineer examples where the anomalies split up as:
\begin{align}
  a_L \quad & \vert \quad a_{\text{mid}} \quad \vert \quad a_R, \,\,\, \text{with} \,\,\, a_{L},a_{R} < a_{\text{mid}}\\
  c_L \quad & \vert \quad c_{\text{mid}} \quad \vert \quad c_R, \,\,\, \text{with} \,\,\, c_{L},c_{R} < c_{\text{mid}}.
\end{align}
Of course, the anomalies provide only partial information on the structure of localized states, so a priori, it could happen that in each region, there are massless states present which are missing from the other regions. Though we cannot prove it in full generality, we expect that regions with higher $a$ and $c$ are typically the places which have more states as is expected by RG flow.

To illustrate this, consider the case of 6D SCFTs as generated by M5-branes probing an ADE singularity \cite{DelZotto:2014hpa}. In reference \cite{Ohmori:2014kda} the 6D anomalies for these theories were computed, and the anomalies of the 4D theories resulting from compactification were computed in \cite{Razamat:2016dpl, Apruzzi:2018oge}. For example, from compactification on a curve of genus $g \geq 1$ and in the absence of flavor symmetry fluxes, the values of $a$ and $c$ are both proportional to $(g-1)$. In the case of compactification on a genus one curve, one instead gets a 4D $\mathcal{N} = 2$ theory , and in the case of a genus zero curve (with no punctures), the resulting 4D system produces a trivial fixed point \cite{Apruzzi:2018oge}. When fluxes are switched on, the central charges become algebraic numbers, as determined by $a$-maximization. The general feature of $a$ and $c$ increasing with genus still holds in these cases \cite{Razamat:2016dpl, Apruzzi:2018oge}.

\subsection{Generating Thin Interfaces}

The construction we have provided generates a thickened 3D interface. This is simply because the ``middle region'' can also be thought of as compactification of a 6D theory on a Riemann surface which is then further compactified on an finite length interval. In the limit where the size of this interval collapses to zero size, this leads to a thin interface. What we would like to understand is whether the resulting construction still produces localized states.

Returning to the example of the 6D hypermultiplet, we can see some potential issues with such a procedure. For example, in the case of a 4D Dirac fermion with a position dependent mass, the appearance of a localized state in the thin wall limit relies on having a sign flip in the mass term, relating to the two time-reversal invariant values of $\theta$ at weak coupling. From the perspective of our compactification of a 6D anti-chiral two-form, this involves a bordism between two elliptic curves with different values of the complex structure moduli. In the example of a 6D hypermultiplet, we can arrange something similar since the spin connection and gauge field connection implicitly depend on the complex structure of the compactification curve. Working with curves with real coefficients, we can again enforce the appearance of a sign flip in the mass spectrum of Kaluza-Klein modes, thus ensuring that the trapped states ``in the middle'' do not disappear in the zero thickness limit. The same logic also applies in more general compactifications of 6D SCFTs. One reason is that a large number of such examples can be interpreted as 4D $\mathcal{N} = 1$ theories in which marginal couplings have been formally tuned to extremely large values \cite{Razamat:2019vfd}. From this perspective, we can impose a further condition that we restrict to time-reversal invariant values of these marginal couplings, thus providing a way to ``protect'' localized states in this more general setting.

\newpage

\section{Conclusions} \label{sec:CONC}

Interfaces generated by position dependent couplings provide a general way to access non-perturbative structure in quantum field theories.
In this paper we have investigated 3D interfaces generated from 4D theories at strong coupling. In the case of 4D $U(1)$ gauge theories we showed that the appearance of a finite index duality group $\Gamma \subset SL(2,\mathbb{Z})$, in tandem with the condition of time-reversal invariance leads to a rich phase structure for possible interfaces, as captured by the real component of a modular curve $X(\Gamma)_{\mathbb{R}}$. We have also seen that a more general starting point based on compactifications of 6D SCFTs on three-manifolds with boundary leads to a broad class of thickened 3D interfaces with states trapped in an interior region. In the remainder of this section we discuss some avenues for future investigation.

Throughout this paper we have operated under the assumption that time-reversal invariance is preserved by the system, even as we vary
the parameters of the theory. Of course, this is not always the case, and in some cases there is good evidence that time-reversal invariance
is actually spontaneously broken (see for example \cite{Gaiotto:2017tne}). Given the strong constraints on the real component of a modular curve, it would be interesting to study these assumptions in more detail.

One of the outcomes of our analysis is the prediction that in some $U(1)$ gauge theories with duality group $\Gamma \subset SL(2,\mathbb{Z})$,
there are 3D interfaces which are inherently at strong coupling, namely, the resulting parameters are on a different component of $X(\Gamma)_{\mathbb{R}}$ from the one connected to the point of weak coupling. As a further generalization, it is natural to ask whether quantum transitions between these different phases could be activated by adding small time-reversal breaking couplings to the system. Calculating these transition rates would be very interesting in its own right, and would likely shed additional light on the non-perturbative structure of such theories.

The geometry of modular curves also suggests additional ways in which strong coupling phenomena may enter such setups. For example, for suitable duality groups, the modular curve $X(\Gamma)$ can have genus $g > 0$. This in turn means that there are one-cycles which can be traversed by a motion through parameter space. Compactifying our 4D theory on a circle, a non-zero winding number in moving through such a one-cycle of $X(\Gamma)$ suggests another way to produce features protected by topology.

It is also interesting to ask whether coupling such systems to gravity imposes any restrictions. At least in the context of F-theory constructions, there appear to be sharp constraints on the possible torsional structures which can be realized in UV complete models, see e.g.\ \cite{Aspinwall:1998xj, Hajouji:2019vxs}. More generally, Swampland type considerations suggest the possible existence of a sharp upper bound on the genus of the associated modular curves (perhaps they are always genus zero). Determining such bounds would be quite illuminating.

From a mathematical point of view, our study of the real components of the modular curve $X(\Gamma)$ has centered on a particular notion of conjugation given by $\tau \mapsto - \overline{\tau}$, which has a clear physical interpretation in terms of time-reversal. On the other hand, reference \cite{snowden2011real} considers another conjugation operation given by $\tau \mapsto 1 / \overline{\tau}$, and this choice also leads to a rather rich set of conjugation invariant components of the modular curve. This can be thought of
as the composition of time-reversal conjugation with an S-duality transformation. It would be very interesting to develop a physical interpretation of this case as well.

Much of our analysis has focused on the special case of 4D $U(1)$ gauge theory. When additional $U(1)$'s are present, there is again a
fundamental domain of possible couplings as swept out by a congruent subgroup of $Sp(2r, \mathbb{Z})$ acting on the Siegel upper half-space.
In this case, less is known about the analog of modular curves, let alone their real components, but it would nevertheless be interesting to study the phase structure of cusps in this setting.

The main thrust of our analysis has focused on formal aspects of 3D interfaces in 4D systems. One could envision applying these insights to specific concrete condensed matter systems. Additionally, in cases with additional $U(1)$ factors, one might consider scenarios in which a visible sector $U(1)$ kinetically mixes with a dark $U(1)$. The phenomenology of axionic domain walls leads to a rather rich set of signatures \cite{Sikivie:1984yz}, so it would be interesting to investigate the related class of questions for axionic domain walls charged under one of these hidden $U(1)$ factors.

Our analysis was inspired by string compactification considerations, though we have mainly focused on field-theory
considerations. In a related development, M-theory on non-compact $Spin(7)$ backgrounds can sometimes be interpreted as generating interpolating profiles between 4D M- and F-theory vacua \cite{Cvetic:2020piw}. It would be very interesting to study time-reversal invariant configurations engineered from this starting point.

In the same vein, we note that some of the techniques considered use supersymmetry only sparingly. It is therefore tempting to ask whether these considerations could be used to build non-supersymmetric brane configurations which are protected by topological structures. We leave an analysis of this exciting possibility for future work.

\section*{Acknowledgments}

We thank R. Donagi, C.L. Kane and G. Zoccarato for helpful discussions.
The work of M.D. is supported by the individual DFG grant DI 2527/1-1.
The work of JJH is supported by NSF CAREER grant PHY-1756996
and a University Research Foundation grant at the University
of Pennsylvania. ET is supported by a University of Pennsylvania
Fontaine Fellowship.

\newpage

\appendix

\section{Aspects of Elliptic Curves} \label{app:ELLIPTIC}

In this Appendix we review some aspects of the geometry of elliptic curves used in this paper. In
normal Weierstrass form, an elliptic curve can be presented as the hypersurface cut out by the equation:
\begin{align}
y^2 = x^3 + f x z^4 + g z^6 \,,
\end{align}
with complex coefficients $f$ and $g$ and $(x,y,z)$ inhomogeneous coordinates on the weighted projective space $\mathbb{CP}^{2}_{[2,3,1]}$. In the patch $z \neq 0$ one can rescale $z$ to $1$ via the $\mathbb{C}^*$ rescaling leading to the more standard form
\begin{align}\label{eq:WEIER}
y^2 = x^3 + f x + g \,,
\end{align}
which has to be supplemented by the ``point at infinity'' given by $[x,y,z] = [1,1,0]$.
Expressing the cubic equation according to its roots $e_i$ one can write
\begin{align}
y^2 = (x - e_1) (x - e_2) (x - e_3) \,,
\end{align}
and one has
\begin{align}
e_1 + e_2 + e_3 = 0 \,, \quad f = e_1 e_2 + e_2 e_3 + e_3 e_1 \,, \quad g = - e_1  e_2 e_3
\end{align}
The discriminant is given by:
\begin{align}
D_{\mathrm{disc}} = \prod_{i < j} (e_i - e_j)^2 = -( 4f^3 + 27 g^2) \equiv - \Delta \,.
\end{align}
In what follows we follow F-theory conventions and refer to $\Delta = 4 f^3 + 27 g^2$ as the discriminant.

We can define the modular $\lambda$ function knowing the position of the branch cuts $e_i$.
In the Weierstrass form, where one of the roots is at infinity it is given by:
\begin{align}
\lambda = \frac{e_3 - e_2}{e_1 - e_2} \,,
\end{align}
In terms of this, the $j$-function can be expressed as
\begin{align}
j (\tau) = \frac{256 (1 - \lambda - \lambda^2)^3}{\lambda^2 (1 - \lambda)^2} \,.
\end{align}

One can also work in terms of a presentation such as:
\begin{equation}
x^2 = P_4(z) = (z - z_1)(z - z_2)(z - z_3) (z - z_4)
\end{equation}
in which all four roots are at finite values. In this case, the modular $\lambda$
function is defined by the conformal cross ratio
\begin{align}
\lambda = \frac{(z_2 - z_3) (z_1 - z_4)}{(z_1 - z_3) (z_2 - z_4)} \,,
\end{align}
where the branch cuts are chosen between $z_2$ and $z_3$ and $z_1$ and $z_4$.
One can also consider the elliptic curve defined by the equation:
\begin{align}
x^2 = \frac{P_4 (z)}{(z - 1)^2 (z - q)^2} \,,
\end{align}
as is the case for the Seiberg-Witten curve with $N_f = 4$. In this case, we can clear denominators and perform blowups at $z = 1$ and $z = q$
to get an elliptic curve. In this case one can identify the branch points at the zeros of $P_4$ and plug
them into the formula for $\lambda$ which in turn can be used to compute $j (\tau)$.

Let us analyze the behavior of $j(\tau)$ in terms of the cross ratio $\lambda$. Clearly, $j (\tau)$ diverges for the three cases
\begin{align}
\lambda \rightarrow 0 \,, \quad \lambda \rightarrow 1 \,, \quad \lambda \rightarrow \infty \,.
\end{align}
In these limits the branch point at $\lambda$ collides with one of the other three branch points.

\subsection{Phase Structure for Real Elliptic Curves}

Having discussed the general structure of roots in an elliptic curve, we now specialize further, taking $f,g \in \mathbb{R}$.
In section \ref{sec:DUALITY} we argued that the time-reversal invariant components of the fundamental domain of $SL(2,\mathbb{Z})$
split up into three distinct phases based on singularities in the elliptic curve, as dictated by the vanishing of $f,g$ and $\Delta$.
Here we provide some complementary details.

Going back to the description in terms of the explicit branch
points we find that up to a
permutation of indices one has the following two possibilities.
\begin{equation}
\begin{split}
\text{Case I}:& \quad e_1, e_2, e_3 \in \mathbb{R} \,, \\
\text{Case II}:& \quad e_1 \in \mathbb{R} \,, \enspace e_2 = \bar{e}_3 \,.
\end{split}
\label{eq:rootcases}
\end{equation}
Next, we want to relate the different configurations of the branch
points to the regions of $\tau$ given in \eqref{eq:tauregions}
that describe the distinct time-reversal invariant phases of the
abelian gauge theory. For that we hold the root $e_1$ fixed at negative real value.

For Case I in \eqref{eq:rootcases} we can parametrize the two other roots as
\begin{align}
e_2 = - \tfrac{1}{2} e_1 + \delta \,, \quad e_3 = - \tfrac{1}{2} e_1 - \delta \,,
\end{align}
with $\delta \in \mathbb{R}$. In terms of the variable $\delta$ the Weierstrass coefficients and discriminant read
\begin{align}
f = - \tfrac{3}{4} e_1^2 - \delta^2 \,, \quad g = - e_1 \big( \tfrac{1}{4} e_1^2 - \delta^2 \big) \,, \quad \Delta = - \tfrac{1}{4} \delta^2 (9 e_1^2 - 4 \delta^2)^2 \,.
\end{align}
The discriminant vanishes for $\delta = 0$ and $\delta = \pm e_1$, and as expected these points are associated to the collision of two of the branch points. Note also that all the coefficients are invariant with respect to $\delta \rightarrow - \delta$, which corresponds to an exchange of $e_2$ and $e_3$.
\begin{figure}[t!]
\centering
\includegraphics[width=\textwidth]{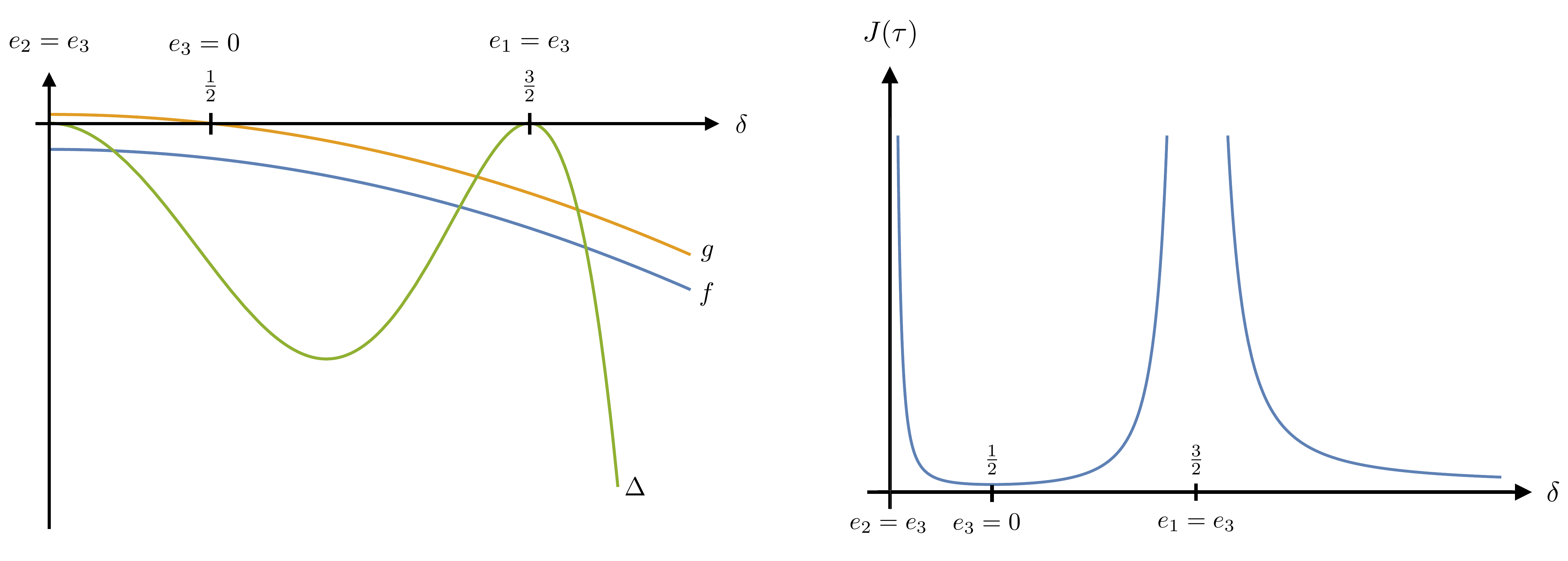}
\caption{The parameters $f$, $g$, and $\Delta$, as well as the $J$-function for all three branch points on the real axis (here: $e_1 = - 1$).}
\label{fig:Jreal}
\end{figure}
The $J$-function is then given by:
\begin{align}
J (\tau) = \frac{(3 e_1^2 + 4 \delta^2)^3}{4 \delta^2 (9 e_1^2 - 4 \delta^2)^2} \,.
\end{align}
Together with $f$, $g$, and $\Delta$ it is depicted in figure \ref{fig:Jreal}. We find that $J(\tau) \geq 1$, which means that all the configurations translate to the trivial phase with $\theta = 0$ and varying gauge coupling. At the collision of two branch points, which happens at $\delta = 0$ and $\delta = - \tfrac{3}{2} e_1$ the $J$-function diverges $J \rightarrow + \infty$. For the special values $\delta = - \tfrac{1}{2} e_1$ and $\delta \rightarrow \infty$ the $J$-function goes to $1$, which means that $\tau$ approaches the strong coupling point $\tau = i$.

For Case II in \eqref{eq:rootcases}, we use the following parametrization:
\begin{align}
e_2 = - \tfrac{1}{2} e_1 + i \tilde{\delta} \,, \quad e_3 = - \tfrac{1}{2} e_1 - i \tilde{\delta} \,,
\end{align}
with $\tilde{\delta} \in \mathbb{R}$. The Weierstrass coefficients and discriminant are given by
\begin{align}
f = - \tfrac{3}{4} e_1^2 + \tilde{\delta}^2 \,, \quad g = - e_1 \big( \tfrac{1}{4} e_1^2 + \tilde{\delta}^2 \big) \,, \quad \Delta = \tfrac{1}{4} \tilde{\delta}^2 (9 e_1^2 + 4 \tilde{\delta}^2)^2 \,.
\end{align}
The discriminant only vanishes at $\tilde{\delta} = 0$, when the two branch points collide on the real axis. Again, we find the symmetry $\tilde{\delta} \rightarrow - \tilde{\delta}$ which exchanges $e_2$ and $e_3$.
\begin{figure}[t!]
\centering
\includegraphics[width=\textwidth]{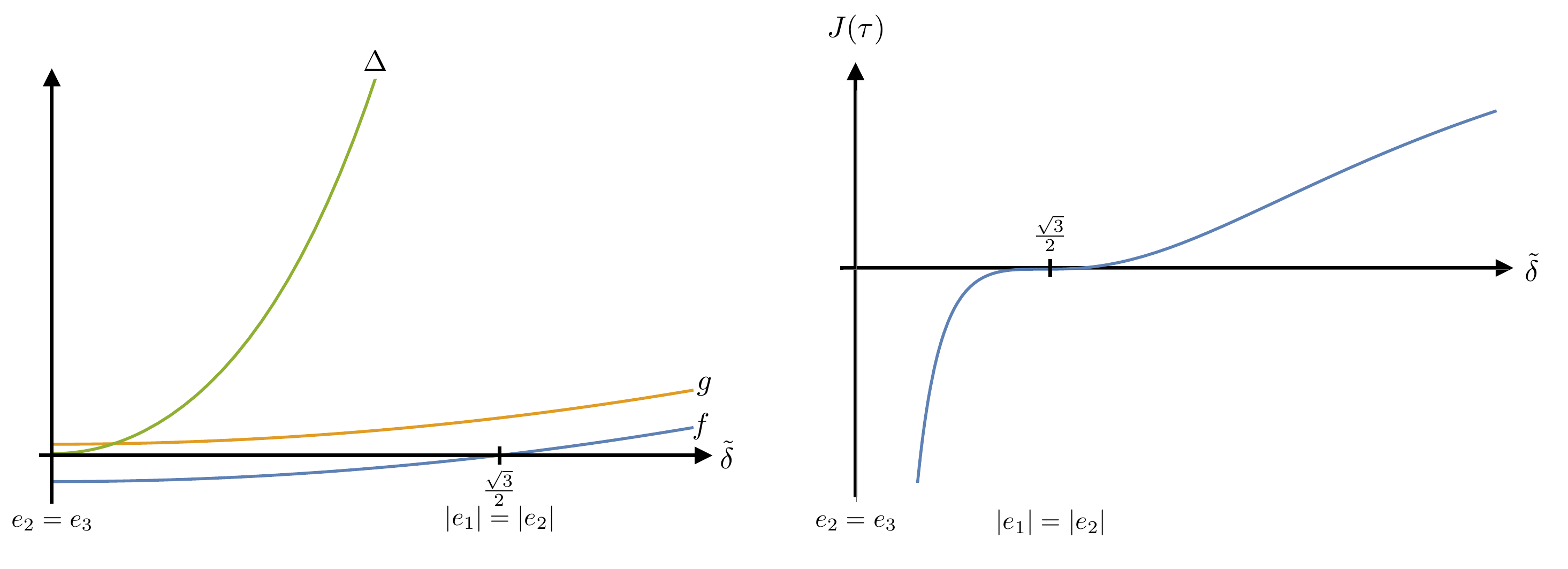}
\caption{The parameters $f$, $g$, and $\Delta$, as well as the $J$-function for two complex conjugate branch points (here: $e_1 = - 1$).}
\label{fig:Jcomp}
\end{figure}
The $J$-function is given by:
\begin{align}
J (\tau) = \frac{(4 \tilde{\delta}^2 - 3 e_1^2)^3}{4 \tilde{\delta}^2 (9 e_1^2 + 4 \tilde{\delta}^2)^2}
\end{align}
and is depicted in figure \ref{fig:Jcomp}. We find that $J(\tau) < 0$ for $\tilde{\delta} \in \big(- \tfrac{\sqrt{3}}{2} |e_1|, \tfrac{\sqrt{3}}{2} |e_1| \big)$, with $J (\tau) \rightarrow - \infty$ for $\tilde{\delta} \rightarrow 0$. This is the region where, $\theta = \pi$ and the gauge coupling varies. Finally, for $|\tilde{\delta}| > \tfrac{\sqrt{3}}{2} |e_1|$ one has $J (\tau) \in (0, 1)$ which indicates the strong coupling region with $|\tau| = 1$.

\begin{figure}[t!]
\centering
\includegraphics[width=\textwidth]{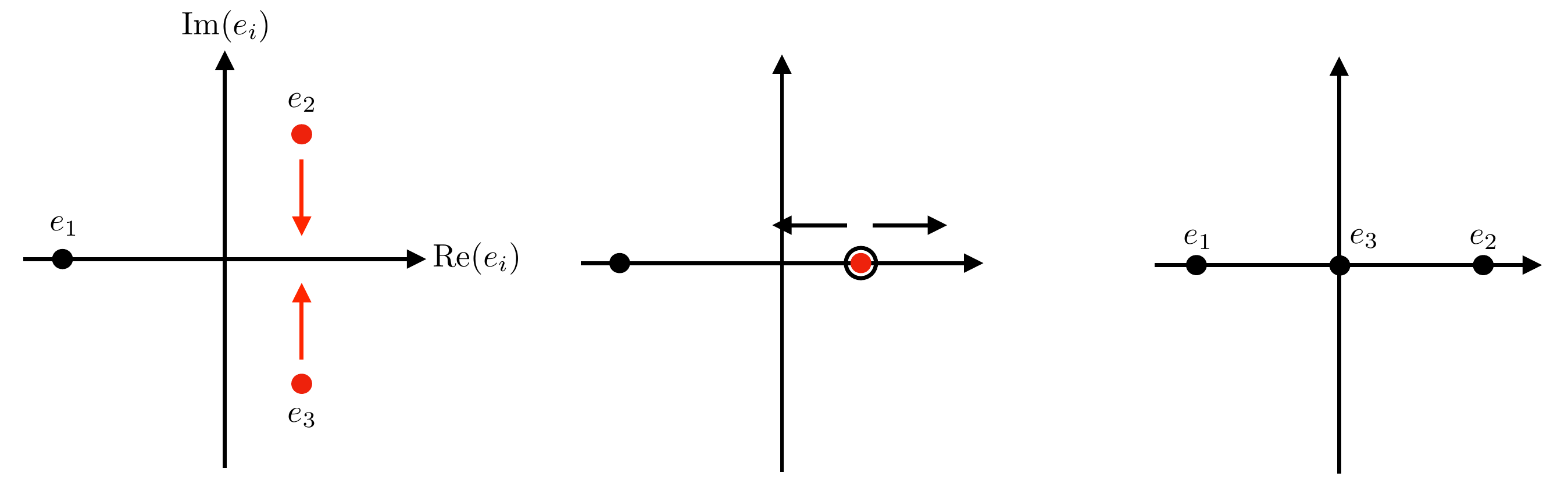}
\caption{The roots in the Weierstrass equation along the considered paths.}
\label{fig:rootpath}
\end{figure}
We see that by considering the configuration above, and depicted in figure \ref{fig:rootpath}, we can scan the full set of real $J (\tau)$ and therefore all the time-reversal invariant values of the complexified coupling constant $\tau$.

To summarize, the three different phases of the time-reversal invariant contour are specified by the following parameters:
\begin{itemize}
\item Trivial Phase: $J > 1 \Leftrightarrow \theta=0$ and $\tau = i \beta$ for $\beta>1$. There we have $\Delta < 0$, $f < 0$ and the roots $e_1 < e_3 < e_2$ are all real. The contours encircle $e_1$ to $e_3$ for $\gamma_B$ and $e_2$ to $e_3$ for $\gamma_A$.
\item Topological Insulator Phase: $J < 0 \Leftrightarrow \theta=\pi$. There we have $\Delta > 0$, $f < 0$ and the roots are such that $e_1 \in \mathbb{R}$, $e_2 = \bar{e}_3$, $\mathrm{Im}(e_2) > 0$. The contours encircle $e_1$ to $e_3$ for $\gamma_B$ and $e_2$ to $e_3$ for $\gamma_A$.
\item Strongly Coupled Phase: $0 \leq J \leq 1 \Leftrightarrow 0 \leq \theta \leq \pi$, $|\tau| = 1$. There we have $\Delta > 0$, $f \geq 0$ and the roots again satisfy $e_1 \in \mathbb{R}$, $e_2 = \bar{e}_3$, $\mathrm{Im}(e_2) > 0$. The contours encircle $e_1$ to $e_2$ for $\gamma_B$ and $e_1$ to $e_3$ for $\gamma_A$.
\end{itemize}
The different time-reversal invariant regions together with the signs of $f$, $g$, $\Delta$ are also indicated in figure \ref{fig:phasessigns}.

\section{Congruence Subgroups and Torsion Points} \label{app:CONG}

In section \ref{sec:MOREDUAL} we showed that compactifying the 6D theory of an anti-chiral two-form on an elliptic curve
can generate 4D $U(1)$ gauge theories with duality group given by a congruence subgroup $\Gamma \subset SL(2,\mathbb{Z})$.
In this Appendix we discuss in greater detail the relation between these congruence subgroups and torsion points.
As a point of notation, in the main text these torsion points are elements of $\widetilde{E}$, the Jacobian of the elliptic curve $E$
on which the 6D theory is compactified. To avoid cluttering the notation, we shall simply discuss an elliptic curve $E$ with torsion points.
The two characterizations are related by the Abel-Jacobi map, so we will not belabor this point in what follows.

We now consider the action of the congruence subgroups on the $N$-torsion points of an elliptic curve $E$,
denoted by $E(N)$, see e.g.\ \cite{diamond2006first}.
\begin{figure}[t!]
\centering
\includegraphics[width=0.7\textwidth]{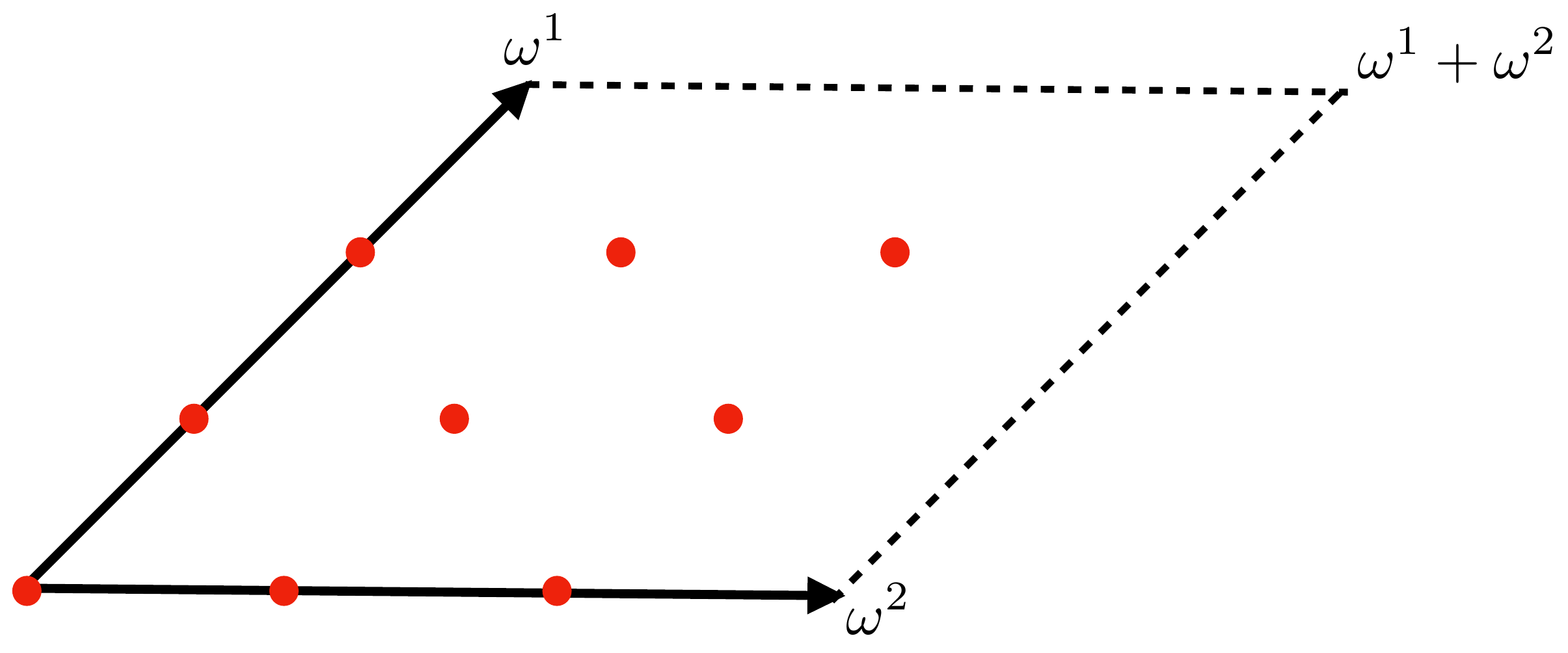}
\caption{Set of 3-torsion points $E(3)$ in the torus fundamental domain spanned by $\omega^1$ and $\omega^2$.}
\label{fig:3tors}
\end{figure}
When we describe $E$ as the quotient of the complex numbers $\mathbb{C}$ by a lattice $\Lambda = \omega^1 \mathbb{Z} \oplus \omega^2 \mathbb{Z}$, these torsion points are simply given by (see figure \ref{fig:3tors}):
\begin{align}
E(N) = \big\{ P \in E: \enspace P = \tfrac{m}{N} \omega^1 + \tfrac{n}{N} \omega^2 \,, \enspace m,n \in \{ 0, 1, \dots, N - 1 \} \big\} \,.
\end{align}
We see that the torsion points generate a subgroup of $E$ isomorphic to $\mathbb{Z} / N \mathbb{Z} \times \mathbb{Z} / N \mathbb{Z}$
with respect to the natural addition on the elliptic curve. An $N$-torsion point $P$ satisfies the condition:
\begin{align}
N P = P + P + \dots + P \in \Lambda \,,
\end{align}
i.e., the point $N P$ it is a lattice vector $k \omega^1 + l \omega^2$ with $k,l \in \mathbb{Z}$. This means that the full $N$-torsion subgroup is generated by two elements. We can choose $\omega^1 = \tau$ and $\omega^2 = 1$ on which a general $SL(2,\mathbb{Z})$ element acts as
\begin{align}
\begin{pmatrix} a & b \\ c & d \end{pmatrix} \begin{pmatrix} \tau \\ 1 \end{pmatrix} = \begin{pmatrix} a \tau + b \\ c \tau + d \end{pmatrix} \sim \begin{pmatrix} \tfrac{a \tau + b}{c \tau + d} \\ 1 \end{pmatrix}
\end{align}

The congruence subgroup $\Gamma (N)$ preserves two $N$-torsion points $P$ and $Q$ which generate the torsion subgroup $E(N)$ and have a Weil pairing given by $e_N (P,Q) = e^{2 \pi i / N}$. For two $N$-torsion points $P$ and $Q$ the Weil pairing is defined by
\begin{align}
e_N (P, Q) = e^{2 \pi i \det \alpha / N} \,,
\end{align}
where $\alpha$ is the matrix with entries in $\mathbb{Z}_N$, which maps $\big( \tfrac{1}{N} \omega^1, \tfrac{1}{N} \omega^2 \big)$ to $(P,Q)$ up to lattice vectors. Therefore, the subgroup $\Gamma (N)$ preserves all $N^2$ torsion points individually.

The congruence subgroup $\Gamma_1 (N)$ preserves a specific $N$-torsion point $P$ and consequently its multiples. This is, it fixes all elements in a $\mathbb{Z}_N$ subgroup of $E(N)$ individually. Note, that by an $SL(2,\mathbb{Z})$ transformation all such points can be mapped to e.g.\ $\tfrac{1}{N} \omega^2$. Conversely, starting from $\tfrac{1}{N} \omega^2$ we can generate all possible choices of the $N$-torsion element by the action of elements in $SL(2,\mathbb{Z}) / \Gamma_1(N)$, i.e.\ by the coset representatives.

Finally, the subgroup $\Gamma_0 (N)$ also preserves a $\mathbb{Z} / N \mathbb{Z}$ subgroup of $E(N)$, but it does not fix the individual elements, which can be mapped to one another in the process. As for $\Gamma_1(N)$ different choices of the $\mathbb{Z} / N \mathbb{Z}$ subgroup are related by a coset representative in $SL(2,\mathbb{Z}) / \Gamma_0(N)$.

Note that some of these congruence subgroups also appear in F-theory models with non-trivial Mordell-Weil torsion \cite{Aspinwall:1998xj, Hajouji:2019vxs}, see also \cite{Mayrhofer:2014opa, Kimura:2016crs, Baume:2017hxm, Cvetic:2017epq, Kimura:2018oze}. These models contain extra torsional sections, which can constrain the global realization of the gauge groups.

Let us illustrate the correspondence between $E(N)$ and the congruence subgroups for the case $N = 3$. We will use the description in terms of $\Lambda = \omega^1 \mathbb{Z} \oplus \omega^2 \mathbb{Z}$.

\subsection{$\Gamma(3)$}

The congruence subgroup $\Gamma(3)$ is generated by the elements
\begin{align}
\gamma_1 = \begin{pmatrix} 1 & 3 \\ 0 & 1 \end{pmatrix} \,, \quad \gamma_2 = \begin{pmatrix} -8 & 3 \\ -3 & 1 \end{pmatrix} \,, \quad \gamma_3 = \begin{pmatrix} 4 & -3 \\ 3 & -2 \end{pmatrix} \,.
\end{align}
A general point $P = (x,y)$ in $E$ is acted on by $SL(2, \mathbb{Z})$ as follows:
\begin{align}
\begin{pmatrix} x \\ y \end{pmatrix} \mapsto \begin{pmatrix} a & b \\ c & d \end{pmatrix} \begin{pmatrix} x \\ y  \end{pmatrix} = \begin{pmatrix} a x + b y \\ c x + d y \end{pmatrix} \,.
\end{align}
Furthermore, we use that the lattice $\Lambda$ is simply given by $\mathbb{Z} \oplus \mathbb{Z}$ and thus all points are understood modulo an integer. Since a point is invariant under the full group if it is invariant with respect to a set of generators, we check which points are invariant with respect to the action of $\gamma_1$, $\gamma_2$, and $\gamma_3$.

The first generator yields
\begin{align}
\begin{pmatrix} x \\ y \end{pmatrix} \mapsto \gamma_1 \begin{pmatrix} x \\ y \end{pmatrix} = \begin{pmatrix} x + 3 y \\ y \end{pmatrix} \sim \begin{pmatrix} x \\ y \end{pmatrix} \,,
\end{align}
which demands that $3 y$ is a lattice vector, or in other words $y \in \big\{ 0, \tfrac{1}{3}, \tfrac{2}{3} \big\}$. For the second generator one finds
\begin{align}
\begin{pmatrix} x \\ y \end{pmatrix} \mapsto \gamma_2 \begin{pmatrix} x \\ y \end{pmatrix} = \begin{pmatrix} - 8 x + 3 y \\ - 3 x + y \end{pmatrix} \,,
\end{align}
telling us that also $x \in \big\{ 0, \tfrac{1}{3}, \tfrac{2}{3} \big\}$. The last generator does not lead to any new constraints and one concludes that the set of invariant points is given by
\begin{align}
\big\{ \tfrac{m}{3} \omega^1 + \tfrac{n}{3} \omega^2 \,, \enspace m,n \in \{ 0, 1, 2\} \big\} = E(3) \,,
\end{align}
as desired.

\subsection{$\Gamma_1 (3)$}

The congruence subgroup $\Gamma_1 (3)$ is generated by the elements
\begin{align}
\widetilde{\gamma}_1 = \begin{pmatrix} 1 & 1 \\ 0 & 1 \end{pmatrix} \,, \quad \widetilde{\gamma}_2 = \begin{pmatrix} 1 & -1 \\ 3 & -2 \end{pmatrix} \,.
\end{align}
From the action of the two generators
\begin{align}
\begin{pmatrix} x \\ y \end{pmatrix} \mapsto \widetilde{\gamma}_1 \begin{pmatrix} x \\ y \end{pmatrix} = \begin{pmatrix} x + y \\ y \end{pmatrix} \,, \quad \begin{pmatrix} x \\ y \end{pmatrix} \mapsto \widetilde{\gamma}_2 \begin{pmatrix} x \\ y \end{pmatrix} = \begin{pmatrix} x - y \\ 3x - 2 y \end{pmatrix} \,,
\end{align}
one concludes that the only invariant points are given by
\begin{align}
\big\{ \tfrac{m}{3} \omega^1 \,, \enspace m \in \{ 0, 1, 2 \} \big\} \subset E(3) \,.
\end{align}
This fixes the elements of a $\mathbb{Z} / 3 \mathbb{Z}$ subgroup of the full torsion subset $E(3)$. Using a coset representative of $\Gamma_1 (3)$ with respect to $SL(2, \mathbb{Z})$, one can also generate different $\mathbb{Z} / 3 \mathbb{Z}$ subgroups which are preserved on the level of the individual elements.

\subsection{$\Gamma_0 (3)$}

The congruence subgroup $\Gamma_0 (3)$ is generated by the elements
\begin{align}
\gamma'_1 = \begin{pmatrix} 1 & 1 \\ 0 & 1 \end{pmatrix} \,, \quad \gamma'_2 = \begin{pmatrix} -1 & 1 \\ -3 & 2 \end{pmatrix} \,.
\end{align}
The action of the generators on points in $E$ is given by
\begin{equation}
\begin{split}
\begin{pmatrix} x \\ y \end{pmatrix} &\mapsto \gamma'_1 \begin{pmatrix} x \\ y \end{pmatrix} = \begin{pmatrix} x + y \\ y \end{pmatrix} \,, \\
\begin{pmatrix} x \\ y \end{pmatrix} &\mapsto \gamma'_2 \begin{pmatrix} x \\ y \end{pmatrix} = \begin{pmatrix} -x + y \\ - 3x + 2 y \end{pmatrix} \,,
\end{split}
\end{equation}
and no point beside the origin is kept fixed. However, the full set
\begin{align}
\big\{ \tfrac{m}{3} \omega^1 \,, \enspace m \in \{ 0, 1, 2 \} \big\} \subset E(N) \,.
\end{align}
is fixed under this group action. The individual elements are mapped to each other as follows
\begin{align}
\tfrac{0}{3} \omega^1 \mapsto \tfrac{0}{3} \omega^1 \,, \quad \tfrac{1}{3} \omega^1 \mapsto - \tfrac{1}{3} \omega^1 \,, \quad \tfrac{2}{3} \omega^1 \mapsto - \tfrac{2}{3} \omega^1 \,.
\end{align}
Again, we can use a coset representatives with respect to $SL(2, \mathbb{Z})$ in order to generate different $\mathbb{Z} / 3 \mathbb{Z}$ subgroups that are fixed by $\Gamma_0 (3)$ as a set but not element by element.

\section{4D $\mathcal{N} = 2$ Gauge Theory with Four Flavors} \label{app:FLAVA}

In this Appendix we discuss in greater detail some aspects of 4D $\mathcal{N} = 2$ gauge theory with gauge group $SU(2)$ and four hypermultiplets in the fundamental representation of $SU(2)$, as studied in reference \cite{Seiberg:1994aj}.
This theory leads to a 4D $\mathcal{N} = 2$ SCFT with flavor symmetry $SO(8)$. Our plan will be to first review some general aspects of
the $\mathcal{N} = 2$ curve in this setting. We then fix a choice of Coulomb branch parameter and vary the mass parameters of the theory
under the condition that the IR theory is time-reversal invariant, and that the mass parameters and Coulomb branch scalar vev preserve
time-reversal invariance.

\subsection{General $\mathcal{N} = 2$ Considerations}
We begin by stating some general considerations about $\mathcal{N} = 2$ theories.
For a state of charge $(q_e, q_m, q_f)$ under the electric, magnetic and flavor symmetry $U(1)$'s, this is controlled by the formula:
\begin{equation}
Z = q_e a - q_m a_D + \frac{1}{\sqrt{2}} \sum_{f = 1}^{\mathrm{dim} \mathcal{R}} q_f m^f, \,\,\, \text{with} \,\,\, M = \sqrt{2} \vert Z \vert.
\end{equation}
where here, $a$ denotes a coordinate on the Coulomb branch, $a_D = \partial \mathcal{F} / \partial a$ is a magnetic dual coordinate controlled by the derivative of $\mathcal{F}$, the $\mathcal{N} = 2$ prepotential, $\mathcal{R}$ denotes a representation of the flavor symmetry, and $M$ denotes the mass of the particle. Recall that in terms of the Seiberg-Witten geometry a massless state occurs whenever a one-cycle of the curve collapses. Following \cite{Minahan:1996cj, Minahan:1996fg}, we introduce a fixed representation $\mathcal{R}$ of the flavor symmetry and write the Seiberg-Witten one-form as:
\begin{equation}
\lambda_{\mathcal{R}} = (c_1 u + c_3) \frac{dx}{y} + c_2 \sum_{b} \frac{m_b y_b(u)}{x - x_b(u)}\frac{dx}{y}.
\end{equation}
for some coefficients $c_i$ which depend on the mass parameters. Introducing an A-cycle and a B-cycle
on the elliptic curve, the coordinates $a$ and $a_D$ can be written as:
\begin{equation}
a = \underset{\gamma_A}{\int} \lambda_{\mathcal{R}} \,\,\, \text{and} \,\,\, a_{D} = \underset{\gamma_B}{\int} \lambda_{\mathcal{R}},
\end{equation}
and the complex structure of the curve is encoded in the derivatives:
\begin{equation}
\tau = \frac{\partial a_{D}}{\partial a} = \frac{\partial a_{D} / \partial u}{\partial a / \partial u}.
\end{equation}

\subsubsection{Seiberg-Witten Curve}

Let us now turn to the Seiberg-Witten curve for the case of $SU(2)$ gauge theory with four flavors. This was originally considered in \cite{Seiberg:1994rs}, and was also presented in a different parametrization in reference \cite{Gaiotto:2009we}.

One way to present the Seiberg-Witten curve is by introducing the 6D SCFT with $\mathcal{N}=(2,0)$ of $A_1$-type, namely the one coming from
the worldvolume of two M5-branes. Wrapping the M5-branes on a $\mathbb{CP}^1$ with four marked points, the moduli space of $\mathcal{N} = 2$ vacua is controlled by the moduli space of the $SU(2)$ Hitchin system on this curve. At a generic point of the moduli space, we obtain a branched double cover of this genus zero curve, namely the ``IR curve'' or Seiberg-Witten curve as obtained from the spectral equation for the Higgs field:
\begin{align}
\lambda^2 - \phi_2 = 0 \,,
\end{align}
with Seiberg-Witten differential $\lambda = x dz/ z$ and $\phi_2$ the quadratic Casimir of the Hitchin system Higgs field
given by:
\begin{align}
\phi_2 =  \frac{P_4 (z)}{(z - 1)^2 (z - q)^2} \frac{dz^2}{z^2} \,.
\end{align}
In the above, $z$ is an affine coordinate on the $\mathbb{CP}^1$. Here, $q$ encodes the UV coupling constant $\tau_{\text{UV}}$ of the $SU(2)$ gauge theory via $q = e^{2 \pi i \tau_{\text{UV}}}$ and $P_4 (z)$ is a fourth order polynomial in $z$ whose coefficients determine the position of the four branch points on the $\mathbb{CP}^1$. Note that the differential on the lefthand side has double poles at $z = 0, 1, \infty$, and $q$. Clearing denominators, we can write this as a hypersurface equation inside $T^{\ast} \mathbb{CP}^{1}$ given by:
\begin{align}
x^2 (z - 1)^2 (z - q)^2 = P_4 (z) \,.
\end{align}
Since we have quadratic order terms on the left-hand side, we can blowup at these zeros, and instead consider the hypersurface equation:\
\begin{equation}
x^2 = P_4 (z),
\end{equation}
which we recognize as the equation of an elliptic curve. To pass to the Weierstrass form, we can use the general prescription given in Appendix \ref{app:ELLIPTIC} to first compute the conformal cross ratio in the roots of $P_4$, and from this extract the $J$-function for the elliptic curve. Next, apply a Moebius transformation on $z$
\begin{align}
z \rightarrow \frac{a z + b}{c z + d} \,, \quad dz \rightarrow \frac{a d - b c}{(c z + d)^2} \, dz = \frac{1}{(c z + d)^2} \, dz \,,
\end{align}
which can be understood as $x \rightarrow (c z + d)^{-2} x$ on the coordinate on the fiber of the cotangent bundle. This can be used to map three marked points to fixed positions, and recover the desired form of the Weierstrass model.

We now use the parametrization of the Seiberg-Witten curve in Weierstrass form as obtained from a D3-brane probe of an $SO(8)$ seven-brane.
From reference \cite{Noguchi:1999xq}, we have:
\begin{equation}
  f = u^2 + \widetilde{w}_4, \quad
  g = w_2 u^2 + w_4 u + w_6.
\end{equation}
The Casimir invariants are given by (equations (2.12)-(2.15) of \cite{Noguchi:1999xq}):
\begin{align}
  u_2 = -\sum_a m_a^2, \qquad & u_4 = \sum_{a<b} m_a^2 m_b^2, \nonumber \\
  u_6= -\sum_{a<b<c} m_a^2 m_b^2 m_c^2, \qquad & \widetilde{u}_4 = -2im_1m_2m_3m_4.
\end{align}
\begin{align}
  u_2=-3w_2, \qquad & u_4=\widetilde{w}_4+3 w_2^2, \nonumber \\
  u_6 = w_6 -w_2 \widetilde{w}_4 -w_2^3, \qquad & \widetilde{u}_4 = w_4.
\end{align}

To simplify we can set all the mass parameters equal to $m$ so that the computations only depend on two parameters.
Furthermore, the Coulomb branch is parameterized by $\widetilde{u}=i u$, and is taken to be real.

\begin{figure}[t!]
\centering
\begin{overpic}[width=0.45\textwidth]{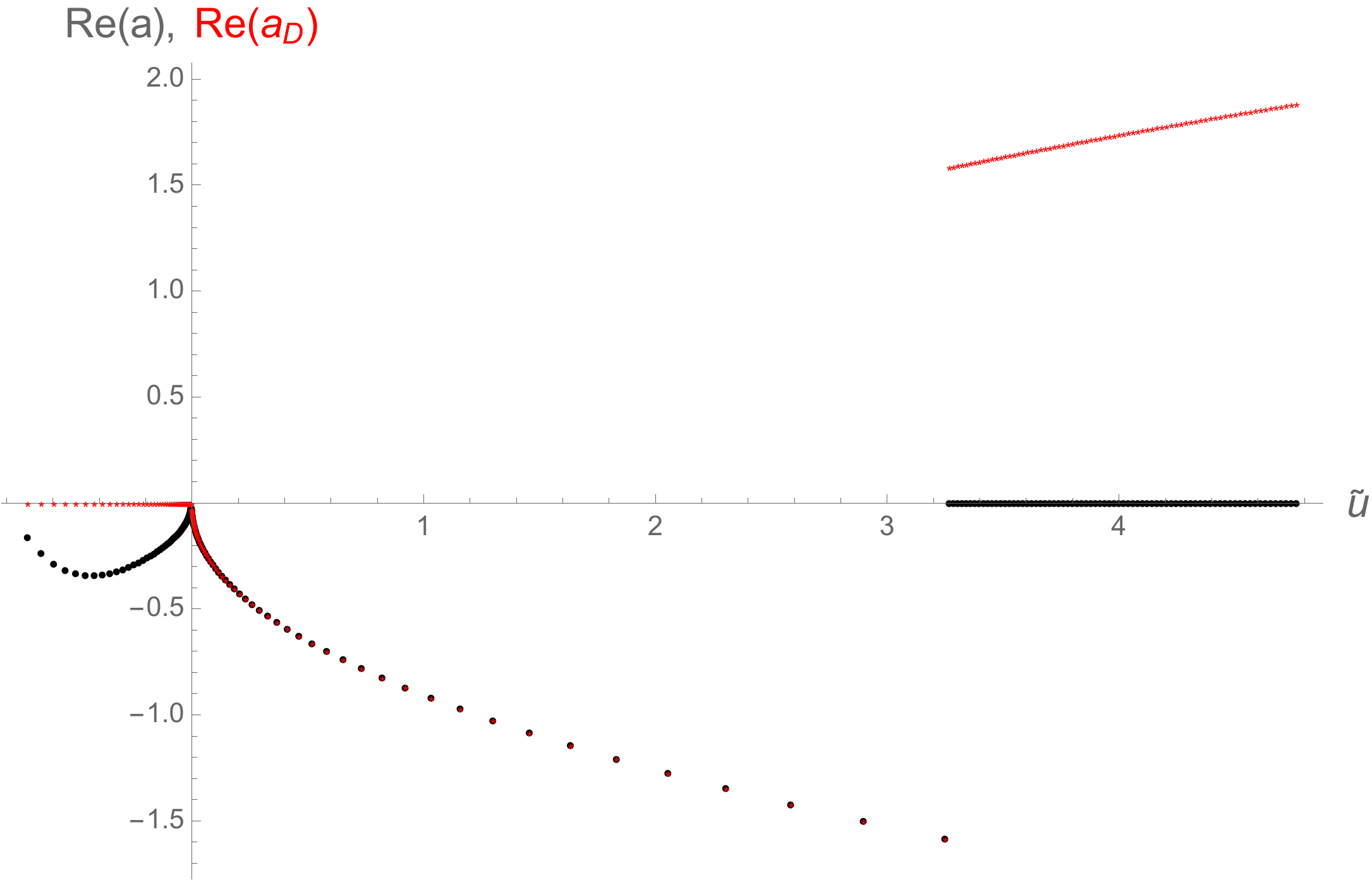}
  \put (1,40) {\tiny\textbf{$\theta=0$}}
  \put (40,40) {\tiny\textbf{$|\tau|=1$}}
  \put (80,40) {\tiny\textbf{$\theta=\pi$}}
\end{overpic}\hspace{.5cm}
\begin{overpic}[width=0.45\textwidth]{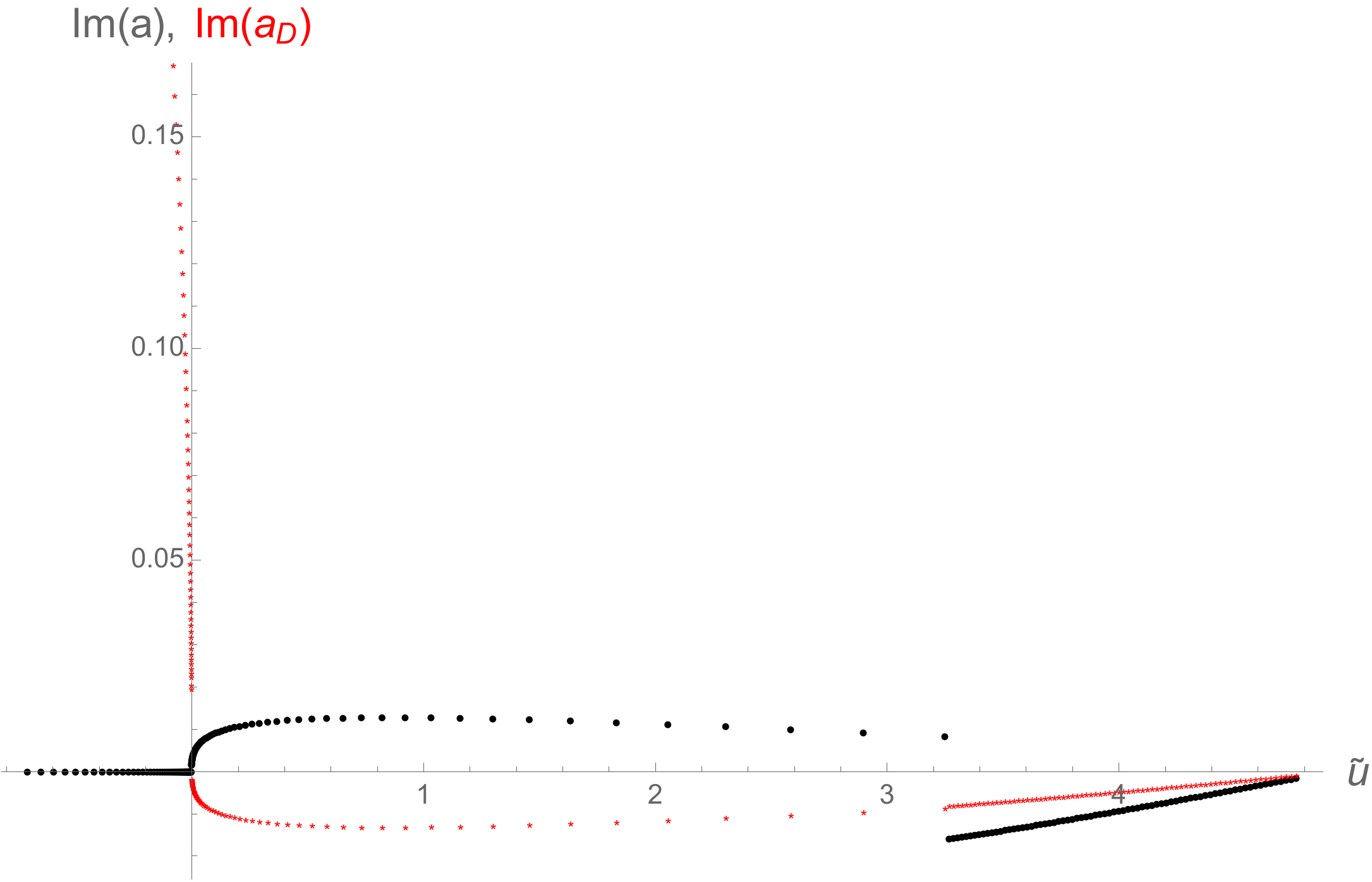}
  \put (1,40) {\tiny\textbf{$\theta=0$}}
  \put (40,40) {\tiny\textbf{$|\tau|=1$}}
  \put (80,40) {\tiny\textbf{$\theta=\pi$}}
\end{overpic}\vspace{.5cm}

\includegraphics[width=0.45\textwidth]{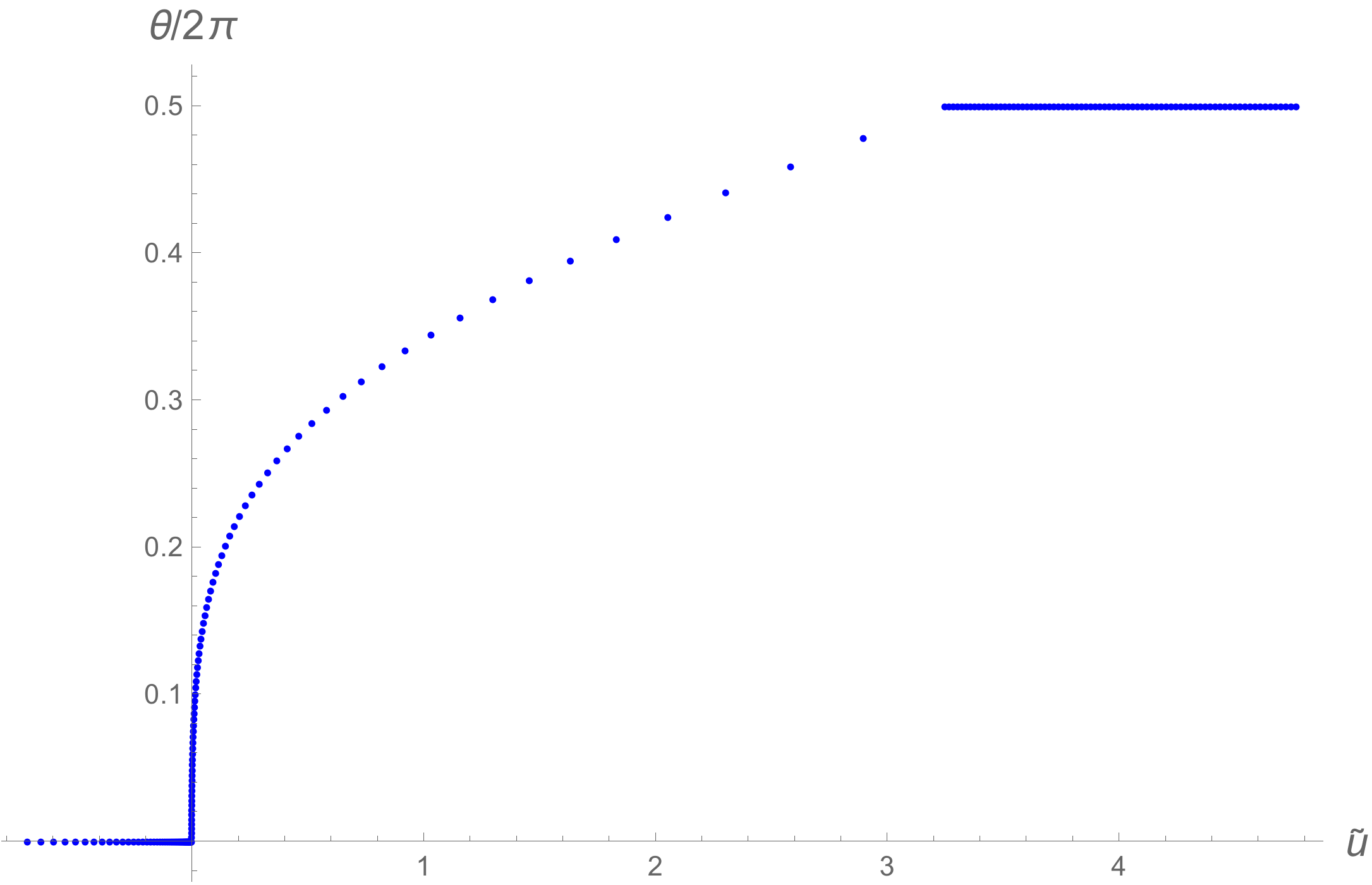} \hspace{.5cm}
\includegraphics[width=0.45\textwidth]{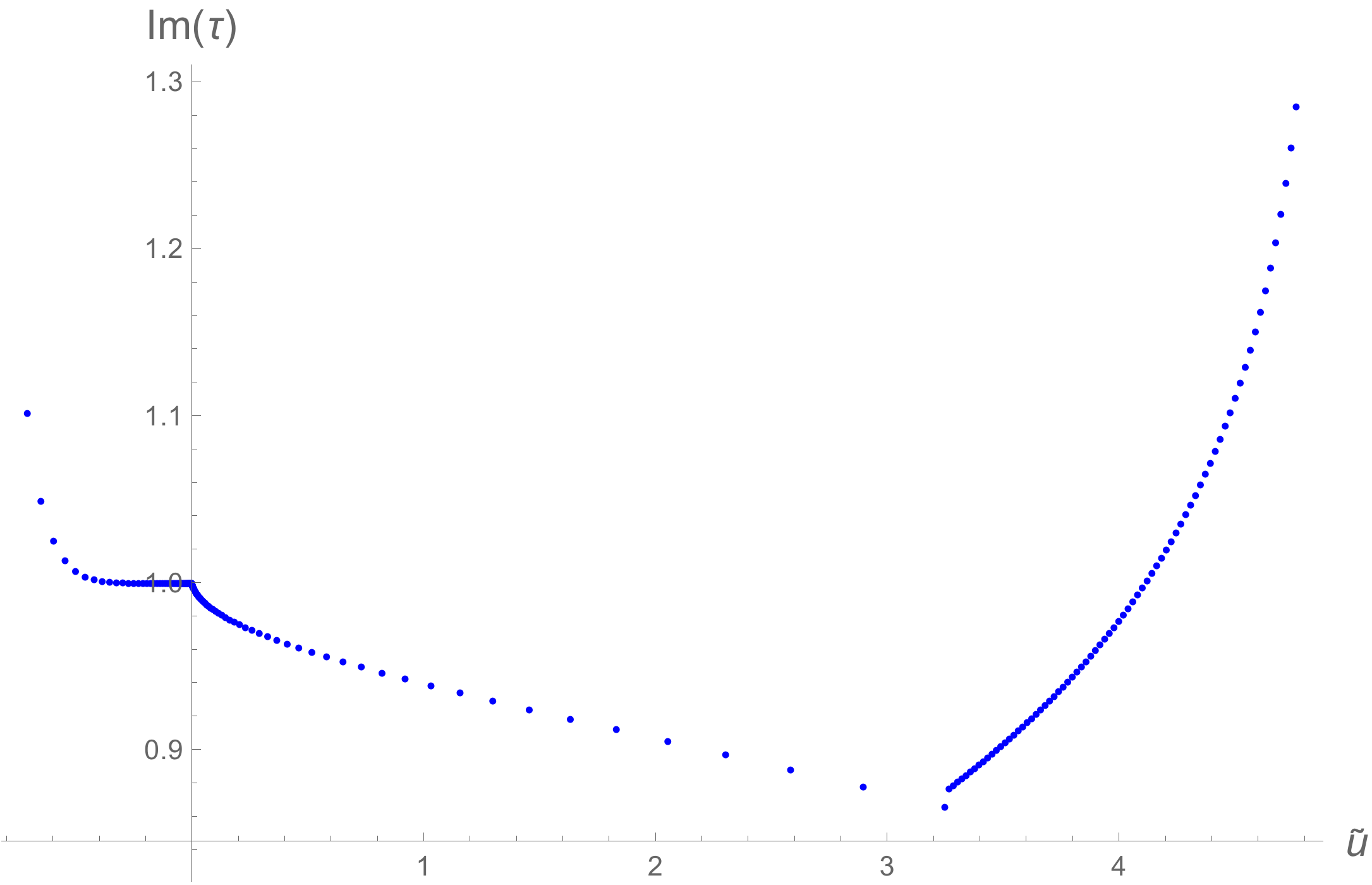}
\caption{The period integrals $a$ (black) and $a_D$ (red) plotted against the Coulomb branch parameter $\tilde{u}$ across the three different phases. The top panel gives the periods while the bottom shows the coupling $\tau$. The left-hand side gives the real part while the right-hand side shows the imaginary piece.
We start off in the trivial phase ($\theta=0$), then transition at $\tau=i$ into the strongly coupled phase $|\tau|=1$. The topological insulator phase ($\theta= \pi$) is then reached at $\tau=e^{\pi i/3}$. Finally, going to the weak coupling limit ($\tau = i \infty$) we can go back into the trivial ($\theta=0)$ phase. Note that the mass parameter $m$, while not plotted, also varies.}%
\label{fig:4flavNumeric}%
\end{figure}

Thus,
\begin{equation}
  f = -\tilde{u}^2 + \frac{2}{3}m^4, \quad
  g = -\frac{4}{3}m^2 \widetilde{u}^2 + 2m^4 \widetilde{u} -\frac{20}{27}m^6.
\end{equation}
And the Seiberg-Witten differential is:
\begin{align}
  \lambda_{8_v} &= \frac{\sqrt{2}}{8\pi i} \left(2u \frac{dx}{y} + \sum_{a=1}^{4} \frac{m_a^2u + w_4/2}{x-m_a^2+w_2}\frac{dx}{y} \right) \nonumber \\
  &= \frac{\sqrt{2}}{8\pi i} \left(2u \frac{dx}{y} + 4m^2 \frac{u-im^2}{x+m^2/3}\frac{dx}{y} \right) \nonumber \\
  &= \frac{\sqrt{2}}{8\pi} \left(2 \widetilde{u} \frac{dx}{y} + 4m^2 (\widetilde{u}-m^2)\frac{dx}{y(x+m^2/3)} \right).
\end{align}
In figure \ref{fig:4flavNumeric} we then plot the result of those computations. We give the period integrals $a$ and $a_D$ across all three possible regions in which $\tau$ belongs to the real component of $X(\Gamma)_{\mathbb{R}}$ for $\Gamma = SL(2,\mathbb{Z})$. We note that as one moves around in the moduli space, the value of $\tau = \partial a_{D} / \partial a$ might move outside the fundamental domain. When this occurs, we perform a change in the ordering of roots $e_i$ appearing in the elliptic curve. This in turn leads to a jump in the values of the periods $a$ and $a_D$, as occurs by applying an $SL(2,\mathbb{Z})$ transformation. In our analysis, it proves convenient to use a slightly different convention from the rest of the paper. So, in this Appendix we take $e_1 > e_2 > e_3$ in the trivial phase, $e_2 \in \mathbb{R}$, Im$(e_1)>\mathrm{Im}(e_3)$ in the strongly coupled phase, and $e_1 \in \mathbb{R}$, Im$(e_3)>\mathrm{Im}(e_2)$ in the topological insulator phase.

In each of the different phases, we observe (from figure \ref{fig:4flavNumeric}) that:
\begin{itemize}
\item $\theta=0$:   $\Delta < 0$, $f < 0$ gives $a \in \mathbb{R}$, $a_D \in i\mathbb{R}$.
\item $|\tau|=1$: $\Delta > 0$, $f \geq 0$ gives $a_D = a^\dagger$.
\item $\theta=\pi$:  $\Delta > 0$, $f < 0$ gives $a \in i \mathbb{R}$, $\mathrm{Im}(a_D) = \mathrm{Im(a)}/2$.
\end{itemize}
Furthermore, both periods vanish at the transition point $\tau=i$, while only $a$ goes to zero at the weak coupling limit $\tau=i\infty$.

\subsection{Elliptic Integrals and Relations Between $a$ and $a_D$}\label{app:contour}
We now derive some of the reality conditions for contour integrals in the three different phases.
\begin{figure}[t!]
\centering
\includegraphics[width=0.7\textwidth]{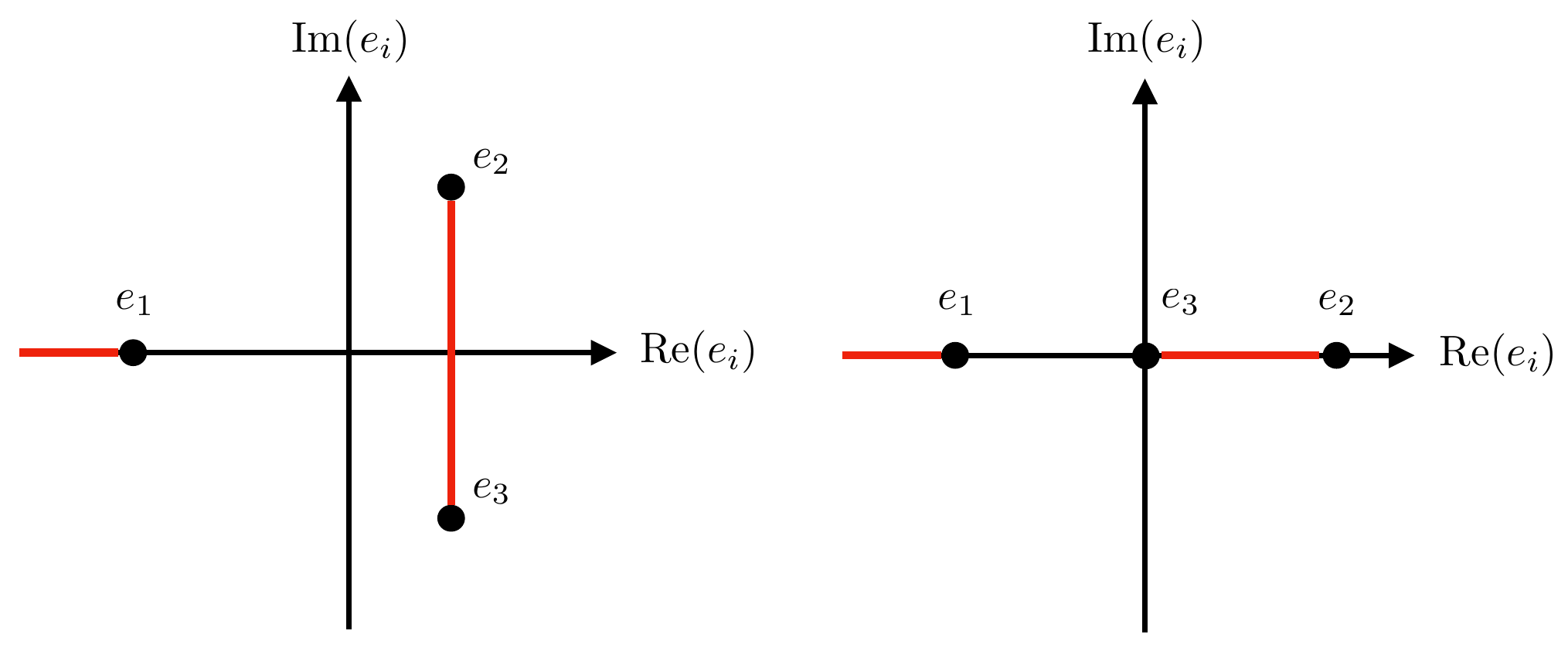}
\caption{Choice of branch cuts in the two cases \eqref{eq:rootcases}.}
\label{fig:branchcuts}
\end{figure}
We choose the distribution of branch cuts as depicted in figure \ref{fig:branchcuts} with contour integrals given in figure \ref{fig:contours}.
In order to prove the various relations between $a$ and $a_D$ we must first take a closer look at the elliptic integrals and fix some conventions about branch cuts.
We want to investigate the properties of the following integrals
\begin{equation}
\int_{e_a}^{e_b} \frac{dx}{y} \,, \quad \text{and} \quad  \int_{e_a}^{e_b} \frac{dx}{y(x-c)} \,,
\end{equation}
where $y$ on the chosen branch is given by $+ \sqrt{x^3 + f x + g}$.

\subsubsection{Proof that $I_A \in i \mathbb{R} $ and $I_B \in \mathbb{R}$ in Phase I (Trivial Phase)}
We fix the real roots such that $e_1 < e_3 < e_2$. Following the same notation as in \cite{Bilal:1997st, DelZotto:2016fju} we want to investigate the integrals:
\begin{align}
  I_A^{(1)} &= \int_{e_3}^{e_2} \frac{dx}{y}, \\
  I_A^{(3)} &= \int_{e_3}^{e_2} \frac{dx}{y(x-c)}, \\
  I_B^{(1)} &= \int_{e_1}^{e_3} \frac{dx}{y}, \\
  I_B^{(3)} &= \int_{e_1}^{e_3} \frac{dx}{y(x-c)},
\end{align}
where $y = \sqrt{(x-e_1)(x-e_2)(x-e_3)}$, so that $y$ is purely imaginary for $e_3 \leq x \leq e_2$, but $y$ is real for $e_1 \leq x \leq e_3$.

Therefore,
\begin{equation}
  I_A \in i \mathbb{R}, \quad I_B \in \mathbb{R}.
\end{equation}

\subsubsection{Proof that $I_A \in i \mathbb{R}$ and $\mathrm{Im}(I_B) = \mathrm{Im}(I_A)/2$ in Phase II (Topological Insulator)}
Let $e_2 = r+i\alpha$ so that $e_3 = r-i\alpha$ and $e_1=-2r$.
First of all we prove that $I_A \in i \mathbb{R}$ by noting that:
\begin{align}
  \begin{split}
    I_A^{(1)} &= \int_{e_2}^{e_3} \frac{dx}{y} \\
    &= \int_0^1 \frac{-2i\alpha}{y}dt, \quad x=(e_3-e_2)t+e_2 = -2i\alpha t + r + i\alpha \\
    &= \int_0^1 \frac{-2i\alpha}{\sqrt{4\alpha^2t\left(-3r+i\alpha(2t-1)\right)(t-1)}}dt \\
    &= \int_0^1 \frac{-i}{\sqrt{s(t)}}dt, \\
  \end{split}
\end{align}
where
\begin{equation}
  s(t) = t(t-1)\left(-3r+i\alpha(2t-1)\right).
\end{equation}
We observe that:
\begin{equation}
  s(t) = \overline{s(1-t)}.
\end{equation}
As a result,
\begin{align}
  \begin{split}
    I_A^{(1)} &= \int_0^1 \frac{-i dt}{\sqrt{s(t)}} \\
    &= \int_0^{\frac{1}{2}} \frac{-i dt}{\sqrt{s(t)}} + \int_{\frac{1}{2}}^{1} \frac{-i dt}{\sqrt{s(t)}} \\
    &= \int_0^{\frac{1}{2}} \frac{-i dt}{\sqrt{s(t)}} + \int_{\frac{1}{2}}^0 \frac{i dt'}{\sqrt{s(1-t')}} \\
    &= \int_0^{\frac{1}{2}} \frac{-i dt}{\sqrt{s(t)}} + \int_{\frac{1}{2}}^0 \frac{i dt'}{\sqrt{\overline{s(t')}}} \,, \\
    I_A^{(1)} &= -i\int_0^{\frac{1}{2}}dt\left(\frac{1}{\sqrt{s(t)}}+\frac{1}{\sqrt{\overline{s(t)}}}\right) \,, \\
  \end{split}
\end{align}
which implies  $I_A^{(1)} \in i \mathbb{R}$.

Furthermore, we have
\begin{equation}
  x(t) = -2i\alpha t + i\alpha + r = \overline{x(1-t)}.
\end{equation}
Thus the same reasoning applies to
\begin{equation}
  I_A^{(3)} = \int_{e_2}^{e_3} \frac{dx}{y(x-c)}.
\end{equation}
This concludes the proof that $I_A \in i \mathbb{R}$.

Next we note that
\begin{equation}
  I_B^{(1)} = \int_{e_1}^{e_3} \frac{dx}{y} = \overline{\int_{e_1}^{e_2} \frac{dx}{y}},
\end{equation}
which implies that
\begin{align}
  2i \mathrm{\, Im}I_B^{(1)} &= \int_{e_1}^{e_3} \frac{dx}{y} - \int_{e_1}^{e_2} \frac{dx}{y} \\
  &= \int_{e_1}^{e_3} \frac{dx}{y} + \int_{e_2}^{e_1} \frac{dx}{y} \\
  &= \int_{e_2}^{e_3} \frac{dx}{y} \\
  &= I_A^{(1)}.
\end{align}
And similarly we have $2i \mathrm{\, Im}I_B^{(3)} = I_A^{(3)}$. So that indeed, $\mathrm{Im}(I_B) = \mathrm{Im}(I_A)/2$

\subsubsection{Proof that $I_B= \bar{I}_A$ in Phase III (Strongly Coupled Phase)}
In this phase, we note that in order for $\tau$ to be in the fundamental domain, the roots are chosen so that $e_1 \in \mathbb{R}$, $e_2 = \bar{e}_3$, and the period integrals given by:
\begin{align}
  I_A^{(1)}= \int_{e_1}^{e_3} \frac{dx}{y}, && I_A^{(3)} = \int_{e_1}^{e_3} \frac{dx}{y(x-c)}, \nonumber \\
  I_B^{(1)}= \int_{e_1}^{e_2} \frac{dx}{y}, && I_B^{(3)} = \int_{e_1}^{e_2} \frac{dx}{y(x-c)}.
\end{align}
Therefore, $I_B^{(1)} = \overline{I_A^{(1)}}$ and $I_B^{(3)} = \overline{I_A^{(3)}}$.

\section{Localizing a 4D Weyl Fermion} \label{app:WEYL}

In this Appendix we consider the localization of a 4D Weyl fermion
$\chi_{\alpha}$ with a position dependent mass term on a thin wall. We will be
specifically interested in the case where the mass is non-zero outside some
finite size interval, but vanishes inside this interval. We take
\textquotedblleft particle physics conventions\textquotedblright\ and work in
signature $(+,-,-,-)$. We consider a position dependent mass term in the
spatial direction $x_{\bot}=x^{3}\equiv z$ given by:
\begin{equation}
m=m_{L}\Theta(-z)+m_{R}\Theta(z-h),
\end{equation}
where $\Theta$ denotes the Heaviside step function and $m_{L}=\left\vert
m_{L}\right\vert e^{i\phi_{L}}$ and $m_{R}=\left\vert m_{R}\right\vert
e^{i\phi_{R}}$ are non-zero complex numbers. The massless region runs from
$z=0$ to $z=h$, and would describe a thick interface. We will be interested in
the special case where $h\rightarrow0$. We will also need the derivative of
the mass term:%
\begin{equation}
\partial_{z}m=m_{R}\,\delta(z-h)-m_{L}\,\delta(z).
\end{equation}

Our 4D Weyl fermion satisfies the equation of motion:
\begin{equation}
i\left(  \overline{\sigma}^{\mu}\right)  ^{\dot{\alpha}\beta}\partial_{\mu}%
\chi_{\beta}=m(z)\left(  \chi^{\dagger}\right)  ^{\dot{\alpha}}\,,
\end{equation}
We will be interested in explicit solutions to this equation, so we write out
the form of the Dirac equation equation in terms of the two component
doublet:
\begin{equation}
\chi_{\beta}=%
\begin{pmatrix}
a\\
b
\end{pmatrix}
\quad\text{and}\quad\left(  \chi^{\dagger}\right)  ^{\dot{\alpha}}=%
\begin{pmatrix}
-b^{\dagger}\\
a^{\dagger}%
\end{pmatrix}
.
\end{equation}
From there, our Dirac equation can be simplified into a pair of differential
equations:
\begin{align}
(\partial_{\mathrm{4D}}^{2}+|m|^{2})a &  =i(\partial_{z}m^{\dagger}%
)b^{\dagger}\\
(\partial_{\mathrm{4D}}^{2}+|m|^{2})b &  =i(\partial_{z}m^{\dagger}%
)a^{\dagger}.
\end{align}
where the 4D$\ $D'Alembertian $\partial_{\mathrm{4D}}^{2}$ can be further
expanded as:%
\begin{equation}
\partial_{\mathrm{4D}}^{2}=\partial_{\mathrm{3D}}^{2}-\partial_{z}^{2},
\end{equation}
with $\partial_{\mathrm{3D}}^{2}$ the 3D D'Alembertian in the directions
transverse to the $z$-direction. We will mainly be interested in modes which
are exactly massless on a thin 3D slice, so we impose the condition that
$\partial_{\mathrm{3D}}^{2}$ annihilates all functions. We note that in the
case of a thick interface, this condition is not quite appropriate because we
really have a 4D Weyl fermion on an interval (in the interior region).

Focussing now on the case where $h\rightarrow0$, it is enough to consider
just the $z$-dependence of our solutions so we can now write our
differential equation as:%
\begin{align}
\left(  -\partial_{z}^{2}+|m|^{2}\right)  a &  =i(\partial_{z}m^{\dagger
})b^{\dagger}\\
\left(  -\partial_{z}^{2}+|m|^{2}\right)  b &  =i(\partial_{z}m^{\dagger
})a^{\dagger},
\end{align}

We now turn to the solutions of this differential equation. This is
essentially an exercise of the form found in introductory quantum mechanics
textbooks, but we include some general comments for completeness. In the thin
wall limit, the solution splits up into a piecewise smooth function. In the
$z<0$ region we have:
\begin{align}
z &  <0\\
a_{L} &  =A_{L}\exp(+\left\vert m_{L}\right\vert z)\\
b_{R} &  =B_{L}\exp(+\left\vert m_{L}\right\vert z).
\end{align}
for some as yet unfixed coefficients $A_{L}$ and $B_{L}$. Consider next the
solution in the region $z>0$. In this case we have:%
\begin{align}
z &  >0\\
a_{R} &  =A_{R}\exp(-|m_{R}|z)\\
b_{R} &  =B_{R}\exp(-|m_{R}|z).
\end{align}

Next, we need to match the form of our solutions across the three regions.
First, we impose continuity. This leads to the conditions:%
\begin{equation}
A_{L}=A_{R}=A\text{ \ \ and \ \ }B_{L}=B_{R}=B.
\end{equation}
Next, we integrate our differential equation across the interfaces. This
yields the conditions:%
\begin{align}
(\left\vert m_{R}\right\vert +\left\vert m_{L}\right\vert )A &  =i(m_{R}%
^{\dag}-m_{L}^{\dag})B^{\dag}\\
(\left\vert m_{R}\right\vert +\left\vert m_{L}\right\vert )B &  =i(m_{R}%
^{\dag}-m_{L}^{\dag})A^{\dag}.
\end{align}
so we get the condition:%
\begin{equation}
\left\vert m_{R}-m_{L}\right\vert ^{2}=\left\vert \left\vert m_{R}\right\vert
+\left\vert m_{L}\right\vert \right\vert ^{2}.
\end{equation}
To get a localized mode we therefore need to set $e^{i(\phi_{L}-\phi_{R})}%
=-1$, namely the mass term is rotated by a phase of exactly $\pi$ in passing
from the left to the right side of the thin interface. Note that
we also get a non-trivial constraint on the relative phases of $A$ and
$B$. Indeed, we have:
\begin{equation}
A=ie^{-i\phi_{R}}B^{\dag}.
\end{equation}
Consequently, we learn that out of the original two-dimensional complex
doublet of $\mathfrak{spin}(3,1)$, we only retain a single real doublet of
$\mathfrak{spin}(2,1)$ on the wall.

Returning to the more general setting where we have a thick interface, in this
case we should really include non-zero values of the three-momentum. We should
then consider a more general differential equation:%
\begin{align}
\left(  -\partial_{z}^{2}+\Delta\right)  a &  =i(\partial_{z}m^{\dagger
})b^{\dagger}\\
\left(  -\partial_{z}^{2}+\Delta\right)  b &  =i(\partial_{z}m^{\dagger
})a^{\dagger},
\end{align}
with:%
\begin{equation}
\Delta= \partial_{3D}^{2} + \left\vert m\right\vert ^{2} .
\end{equation}
In a thick interior region we have a standard
4D\ wave equation. Switching on specific phases for the mass
terms outside this region amounts to setting a boundary condition on the left $(z=0)$ and right
$(z=h)$ of the middle region. Note that this also leads to an oscillatory
behavior in the middle region. In the thin interface limit, the boundary
conditions on the left and right become correlated, and this imposes a further
condition on the zero modes (as we have seen).

\newpage
\bibliographystyle{utphys}
\bibliography{3DInterface}

\end{document}